\documentclass[twocolumn]{emulateapj}
\usepackage{graphicx}
\usepackage{subfigure}
\usepackage{amssymb,amsmath}

\graphicspath{{figsastroph/}}

\def\snrsurvey{121}
\def\snrdetectedlevela{39}

\begin{document}
	
\title{The MIPSGAL view of supernova remnants in the Galactic Plane}

   \author{D. Pinheiro Gon\c calves\altaffilmark{1}, A. Noriega-Crespo\altaffilmark{2}, R. Paladini\altaffilmark{2}, P.G. Martin\altaffilmark{3}, S. J. Carey\altaffilmark{2}}
    
      \affil{1. Department of Astronomy and Astrophysics, University of Toronto, 50 St. George street, Toronto, ON, M5S 3H4, Canada \\
      email: goncalves@astro.utoronto.ca}
      \affil{2. Spitzer Science Center, California Institute of Technology, 1200 East California Boulevard, Pasadena, CA 91125, USA}     
      \affil{3. CITA, University of Toronto, 60 St. George street, Toronto, ON, M5S 3H4, Canada }

\begin{abstract}
  
We report the detection of Galactic supernova remnants (SNRs) in the mid-infrared (at 24 and 70 $\micron$), in the coordinate ranges $10^\circ< l <65^\circ$  and $285^\circ<l<350^\circ$, $| b |<1^\circ$, using the Multiband Imaging Photometer (MIPS) aboard the Spitzer Space Telescope. We search for infrared counterparts to SNRs in Green's catalog and identify 39 out of \snrsurvey, i.e., a detection rate of about 32\%. Such a relatively low detection fraction is mainly due to confusion with nearby foreground/background sources and diffuse emission. 
The SNRs in our sample show a linear trend in [F$_{8}$/$F_{24}$] versus [F$_{70}$/$F_{24}$]. We compare their infrared fluxes with their corresponding radio flux at 1.4 GHz and find that most remnants have ratios of 70 $\micron$ to 1.4 GHz characteristic of SNRs (with the exception of a few which have ratios closer to those of \ion{H}{2} regions).  Furthermore, we retrieve a slope close to unity when correlating infrared (24 and 70 $\micron$) with 1.4 GHz emission.  Our survey is more successful in detecting remnants with bright X-ray emission, which we find is well correlated with the 24 $\micron$ morphology. Moreover, by comparing the power emitted in X-ray, infrared and radio, we conclude that the energy released in the infrared is comparable to the cooling in the X-ray range.
 
\end{abstract}

   \keywords{infrared: ISM --- shock waves --- supernova remnants}
 
   \maketitle
 
\section{Introduction}

With the help of Galactic surveys (e.g., \citealp{1989ApJS...70..181A}, \citealp{1992ApJS...81..715S}), our understanding of the infrared (IR) energetics, morphology and evolution of supernova remnants (SNRs) has increased substantially in the past decades. However, even at \emph{IRAS} (Infrared Astronomical Satellite) resolution, confusion with other IR sources in the Galactic plane is a limiting factor that can prevent a clear picture from emerging. Such confusion, for example, can make the correct assessment of ejecta dust masses for Type II SNRs difficult, a key factor for interpreting high redshift dust in the Universe (\citealp{2010ApJ...719.1553S}, and references therein). Infrared emission from SNRs can also provide insight into the radio-IR correlation of external galaxies \citep{1985ApJ...298L...7H}. Additionally, the study of dust re-emission is relevant for determining the cooling rate of SNRs, an important quantity that shapes their evolution (\citealp{1987ApJ...322..812D}).

 In this paper, we search for counterparts of SNRs in the mid-infrared (at 24 and 70 $\micron$) using the higher resolution \emph{Spitzer} data obtained with the Multiband Imaging Photometer (MIPS), in the coordinate ranges $10^\circ< l <65^\circ$  and $285^\circ<l<350^\circ$, $| b |<1^\circ$, complementing the work of \cite{2006AJ....131.1479R}. We also briefly explore the relation of SNRs IR fluxes with other wavelength ranges, such as the radio and the X-ray.

The structure of this paper is as follows: In \S\ref{presurv}, we review previous Galactic infrared surveys and their detection rates. In \S\ref{emission}, we discuss SNR emission mechanisms which can contribute in the mid-infrared regime. In \S\ref{sec:data}, we describe the mid-infrared, radio and X-ray data used in this work. In \S\ref{detec}, we report the MIPS detections. A discussion of SNR identifications, derived color ratios and radio and X-ray to infrared ratios follows in \S\ref{discussion}. Finally, in \S\ref{con}, we summarize our findings. Details on each detected SNR can be found in the appendix.

\subsection{Previous Galactic Infrared Surveys \label{presurv}}

Using the Galactic Legacy Infrared Mid-Plane Survey Extraordinaire, GLIMPSE (\citealp{2003PASP..115..953B}; \citealp{2009PASP..121..213C}),  \cite{2006AJ....131.1479R} searched the Galactic Plane for infrared counterparts to known SNRs. Out of 95 objects, 18 were clearly detected, a detection rate of about 20\%. Previous searches were conducted using the \emph{IRAS} all-sky survey.  Within the GLIMPSE surveyed area, \cite{1989ApJS...70..181A} obtained a detection rate of 17\% and \cite{1992ApJS...81..715S} 18\% (\citealp{2006AJ....131.1479R}). In the present study, we identify 39 out of \snrsurvey, i.e., a detection rate of about 32\%.

\subsection{SNR infrared emission mechanisms \label{emission}}

The IR part of the spectral energy distribution (SED) of SNRs can result from multiple sources ranging from dust (either stochastically or thermally heated), atomic/molecular line emission, PAH (polycyclic aromatic hydrocarbon) emission, and even synchrotron emission. In fact, the amount of the latter emission can vary substantially and is dependent on the type of remnant. In plerion remnants such as the Crab Nebula \citep{1992Natur.358..654S}, it accounts for about 90\% of the mid-infrared flux density (at 25 $\micron$). In contrast, about 2\% of the mid-infrared flux density (at around 25 $\micron$) in Cas A is due to synchrotron emission \citep{2010ApJ...719.1553S}.

 Dust emission is expected to be a substantial component of the IR emission of SNRs since their
spectra is well fitted by one or more thermal dust populations, as revealed by previous infrared surveys (IRAS; \citealp{1989ApJS...70..181A}, \citealp{1992ApJS...81..715S}), or by stochastic heating of the grains, as proposed by \cite{2004ApJS..154..290H} to explain the mid-infrared emission from Cas A.
 The dust is heated by charged particles in hot plasma generated by shocks (e.g., \citealp{1981ApJ...248..138D}, \citealp{1987ApJ...322..812D}, \citealp{1992ARA&A..30...11D}). Direct evidence for the shock-heated plasma is provided by X-ray observations in the continuum, and X-ray lines are a good diagnostic of high energy interactions and abundances (\citealp{2004NuPhS.132...21V}). While for the majority of SNRs, the observed dust emission is thought to originate from the interaction of the shockwave with the surrounding interstellar medium (ISM; e.g., \citealp{2001A&A...373..281D}, \citealp{2006ApJ...642L.141B}, \citealp{2006ApJ...652L..33W}, \citealp{2007ApJ...662..998B}), evidence of SNRs with ejecta dust is scarce (e.g., \citealp{1999ApJ...521..234A}) even with the help of higher resolution IR telescopes such as \emph{ISO} and \emph{Spitzer}. However, observations with the \emph{Spitzer Infrared Spectrograph} (IRS) \citep{2008ApJ...673..271R} of the young remnant Cas A have pointed to the presence of a large hot dust continuum peaking at 21 $\micron$, from ejecta dust
formed through condensation of freshly synthesized heavy elements. This component has temperatures ranging from 60 to 120 K. Furthermore,
recent far-infrared and sub-millimeter work
(\citealp{2010ApJ...719.1553S}) on the same remnant, also suggested
the existence of a `tepid' ($\sim 35$ K) central dust component in the ejected material (see also \citealp{2010A&A...518L.138B}).

 Observational evidence for the importance of IR lines has been provided
 by multiple SNRs spectroscopic studies which identified a plethora of
 lines from elements such as Fe and O (e.g.,
 \citealp{1999A&A...343..943O}; \citealp{2002ApJ...564..302R}). 
 Much of the infrared line emission seen in SNRs results from the interaction of the blastwave with the surrounding environment where they are born as the shocked gas cools down. For example, IC443 is encountering an atomic region at the northeast side and a molecular cloud at the southern border (e.g., \citealp{2001ApJ...547..885R}). Spectral observations by \cite{1999A&A...341L..75O} on the north of the remnant revealed that most of the emission at 12 and 25 $\micron$, previously surveyed with IRAS, was in fact due to ionized line emission (e.g., [Fe II]) with only a small contribution from dust. Because optical spectra also showed strong collisionally excited fine structure atomic lines (\citealp{1980ApJ...242.1023F}), the infrared line emission is not unexpected. Likewise, for the 'optically bright regions' in N49, an old remnant in the Large Magellanic Cloud (LMC), \cite{2006AJ....132.1877W} found that IR line emission can be a substantial fraction of the total IR flux, up to 80\% at 24 $\micron$.
	On the south of IC443, however, the spectrum is dominated by the H$_2$ pure rotational lines  S(2) through S(7) (\citealp{1999A&A...348..945C}), once again with a faint dust continuum. Moreover, \cite{2006ApJ...649..258T} found a good agreement between the soft X-ray emitting plasma and the radio/optical structure on the northeastern part of the remnant where the plasma density is highest. Such energetic encounters change the morphology of the neighboring ISM and of the remnant itself thereby creating prominent emission lines (e.g., molecular hydrogen) which can be detected in the infrared. OH masers (in particular, the 1720 MHz line) are also good tracers of the remnant encounter with dense molecular regions with 10\% of Galactic SNRs showing associated masers (\citealp{2002Sci...296.2350W} and references therein).

 The existence of infrared emission associated with PAHs in SNRs was first observed in a LMC remnant by \cite{2006ApJ...653..267T}. Plus, evidence for such emission in some Galactic SNRs has also been shown through near-infrared color-color ratios by \cite{2006AJ....131.1479R}. This is likely excitation of PAHs from the ISM, rather than PAHs from supernova dust.

\section{Data Used}\label{sec:data}

\subsection{Infrared Data}

Infrared data used in this paper originates from two Spitzer surveys: GLIMPSE using the Infrared Array Camera (IRAC) and MIPSGAL (survey of the inner Galactic plane using MIPS, \citealp{2009PASP..121...76C}). Both surveys cover the Galactic coordinates  $10^\circ< l <65^\circ$  and $285^\circ<l<350^\circ$, $| b |<1^\circ$.

%%%%%%%%%%%%%%%%%%%%%%%%%%%%%%%%%%%%%%%%%
\subsubsection{MIPSGAL}

\begin{figure*}
\centering
\includegraphics[width=1.\textwidth]{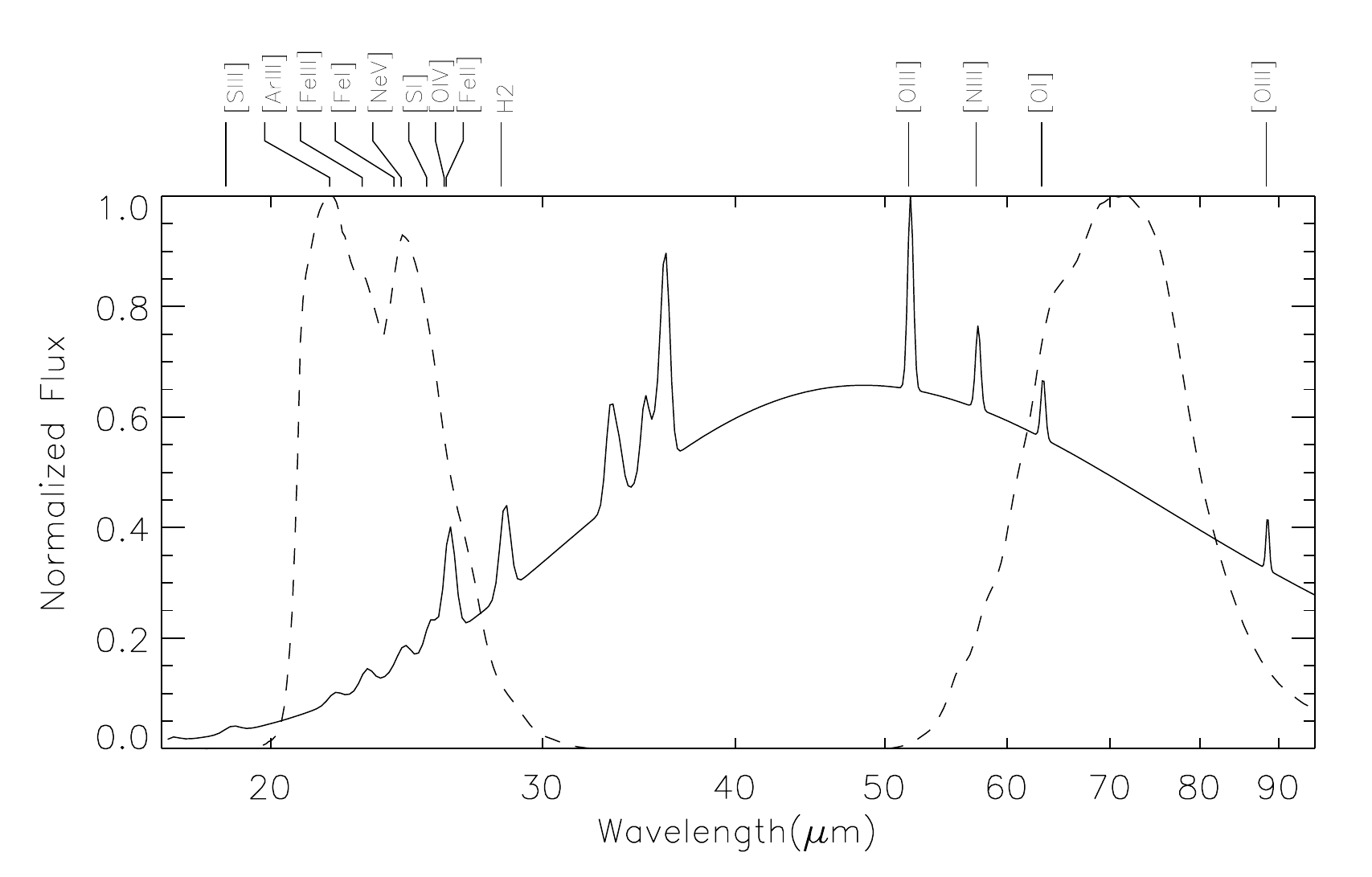}
\caption{A sketch of a dusty SNR spectrum made with a blackbody at 60 K and superimposed the most important fine structure and ionic emission lines (mimicking ISO Short and Long wavelength spectrometer observations of SNR RCW103). Emission lines within the wavelength range of the MIPS filters are annotated. Overplotted (dashed lines) are the normalized response functions of the MIPS filters at 24 and 70 $\micron$. This shows that the contribution of line emission to the MIPS filters is relatively small when a dust continuum is present.}
\label{mipsfilters}
\end{figure*}

The MIPS wavelength coverage is approximately 20 to 30 $\micron$ at ch 1 (24 $\micron$) and 50 to 100 $\micron$ at ch 2 (70 $\micron$), thus covering a potentially rich set of diagnostic lines and dust emission features (see Fig.~\ref{mipsfilters}). In the 24 $\micron$ bandpass, the classic [O IV] 25.9 and [Fe II] 26.0 $\micron$ tracers can be quite strong (e.g., \citealp{1999A&A...341L..75O}; \citealp{2009eimw.confE..46N}), and likewise [O~I] 63 and [O~III] 88 $\micron$ (e.g., \citealp{2001ApJ...547..885R}) within the 70 $\micron$ bandpass. Other standard shock excitation tracers, like [Ne~II] 12.81 or [Si II] 34.8 $\micron$, are unfortunately not included within the IRAC or MIPS bandpasses. Nevertheless, when dust is present, it is the dominant component of the IR radiation at 24 and 70 $\micron$ (see Fig.~\ref{mipsfilters}).

Although the 24 $\micron$ data delivered by the Spitzer Science Center in their Post-BCD products are of a very high quality, the MIPSGAL pipeline was designed to enhance the data a step further. The details are described in \cite{2008PASP..120.1028M}, but briefly the pipeline: (1) carefully deals with saturated and non-saturated data, a key issue in the Galactic Plane; (2) removes a series of artifacts, the most common being a column-to-column jailbar striping, plus (3) that of short-duration afterimage latencies, (4) long-duration responsivity variations, and (5) background mismatches. 
The MIPSGAL observations were designed to be a 24 $\micron$ survey as the high background levels of the Galactic plane at longer wavelengths saturated the 160 $\micron$ detectors and placed the 70 $\micron$ at a fluence level not well characterized. Since MIPS takes data simultaneously at 70 and 160 $\micron$, and that at 70 $\micron$ at first sight looks reasonable, then a concerted effort has been invested by the MIPSGAL team to improve its quality. The specific steps of the 70 $\micron$ pipeline are described in {Paladini et al.\ (\it{in preparation})}, but the main issue is to mitigate the effect of the non-linear response of the Ge:Ga photoconducting detectors. Their standard behavior leads to striping of the images, sudden jumps in brightness and a large uncertainty in the calibration for bright sources and regions.
The MIPSGAL 70 $\micron$ pipeline reduces these effects and decreases the calibration uncertainties to a level of $\sim$15$\%$. This uncertainty on the absolute calibration of the extended emission is consistent with that determined by the MIPS team in their calibration at 24 and 70 $\micron$ (\citealp{2007PASP..119..994E}; \citealp{2007PASP..119.1019G}).
The adopted uncertainty in calibration for MIPS 24 $\micron$ is 10\% . For the 70 $\micron$  fluxes, the main errors come from uncertainties in the background estimation and calibration errors (we adopt a conservative value of 20\%). Moreover, we have used the 24 $\micron$ point source subtracted data. For details in the point source removal procedure see {Shenoy et al.\ (\it{in preparation})}. Point source sensitivities are 2 and 75 mJy (3$\sigma$) at 24 and 70 $\micron$, respectively.
 The resolution of the 24 $\micron$ data is $6^{\prime\prime}$ while for 70 $\micron$ it is $18^{\prime\prime}$.

\subsubsection{GLIMPSE} \label{glimpse}

At the IRAC bands (\citealp{2006AJ....131.1479R}, in particular Fig.\ 2) one expects to find a wealth of atomic lines, including Br$\alpha$ (4.05 $\micron$), Pf$\beta$ (4.65 $\micron$), [Fe~II] (5.34 $\micron$), and [Ar~II] (8.99 $\micron$), among others, plus the molecular S(13) through S(4) H${_2}$ rotational lines, and CO transitions at 4-5 $\micron$. This is very much the case in IC 443 (\citealp{2008ApJ...678..974N}; \citealp{1999A&A...348..945C}).
We use primarily the 8 $\micron$ images from IRAC. Occasionally, other channels are also used in order to constrain the possible emission mechanisms. The 4.5 $\micron$ channel provides a valuable diagnostic of shocked molecular gas (e.g., \citealp{2006AJ....131.1479R}). At these wavelengths, our measured flux represents an upper limit since star contamination is present. We adopt a conservative value of 10\% for the calibration error given the vagaries in color correction plus measurement errors. The IRAC sampling is $1.2^{\prime\prime}$ but the 8 $\micron$ resolution is about $2^{\prime\prime}$.

\subsection{Ancillary Data}\label{otherdata}

Most of the supernova remnants in Green's catalog \citep{2009BASI...37...45G} are discovered using radio observations. By comparing the emission in at least two different radio wavelengths, one can calculate the spectral index and thus distinguish the kind of emission produced (thermal or non-thermal). In the case of SNRs, we expect to find a spectral index which is characteristic of synchrotron radiation.  In contrast, \ion{H}{2} regions show a flat spectrum (when optically thin) which is indicative of free-free emission. The SNR radio emission is associated with regions where shock-induced particle acceleration take place within the source.

 In addition to data in Green's catalog, the following radio data sets are used for comparison with the infrared images: VLA (Very Large Array) archival data at 20 and 90 cm, available from the Multi-Array Galactic Plane Imaging Survey (MAGPIS; \citealp{2006AJ....131.2525H}) website\footnote[1]{http://third.ucllnl.org/gps/index.html}, VGPS (VLA Galactic Plane Survey; \citealp{2006AJ....132.1158S})\footnote[2]{http://www.ras.ucalgary.ca/VGPS/VGPS-data.html} at 21 cm and MOST (Mongolo Observatory Synthesis Telescope; \citealp{1996A&AS..118..329W})\footnote[3]{http://www.physics.usyd.edu.au/sifa/Main/MSC} at about 36 cm. 

Wide band (300-10000 eV) X-ray images from the \emph{Chandra} Supernova Remnant Catalog\footnote[4] {http://hea-www.harvard.edu/ChandraSNR/index.html} archive are also used for comparison. X-ray emission locates the shock heated plasma, including the important reverse shock in the interior region.

\section{Detections of SNRs}\label{detec}

To describe the amount of (or lack of) infrared emission from SNRs in our sample, we use a similar classification scheme to the one adopted by \citet{2006AJ....131.1479R}. Detection levels are characterized as follows: (1) detected, (2) possible detection, (3) unlikely detection and (4) not detected (see Table \ref{table:level}). Level 1 detections have a good positional match between the infrared contours and those at other energies (radio or X-ray). Level 2 detections display infrared emission within the remnant region but confused with cirrus or nearby \ion{H}{2} regions. Level 3 detections show the presence of some infrared emission which does not seem to be correlated with the remnant, although source confusion prevents a clear assessment of the emission origin. Level 4 is characterized by the absence of infrared emission within or nearby the contours of the radio remnant. 

\begin{figure*}
\centering 
\includegraphics[width=.38\textwidth]{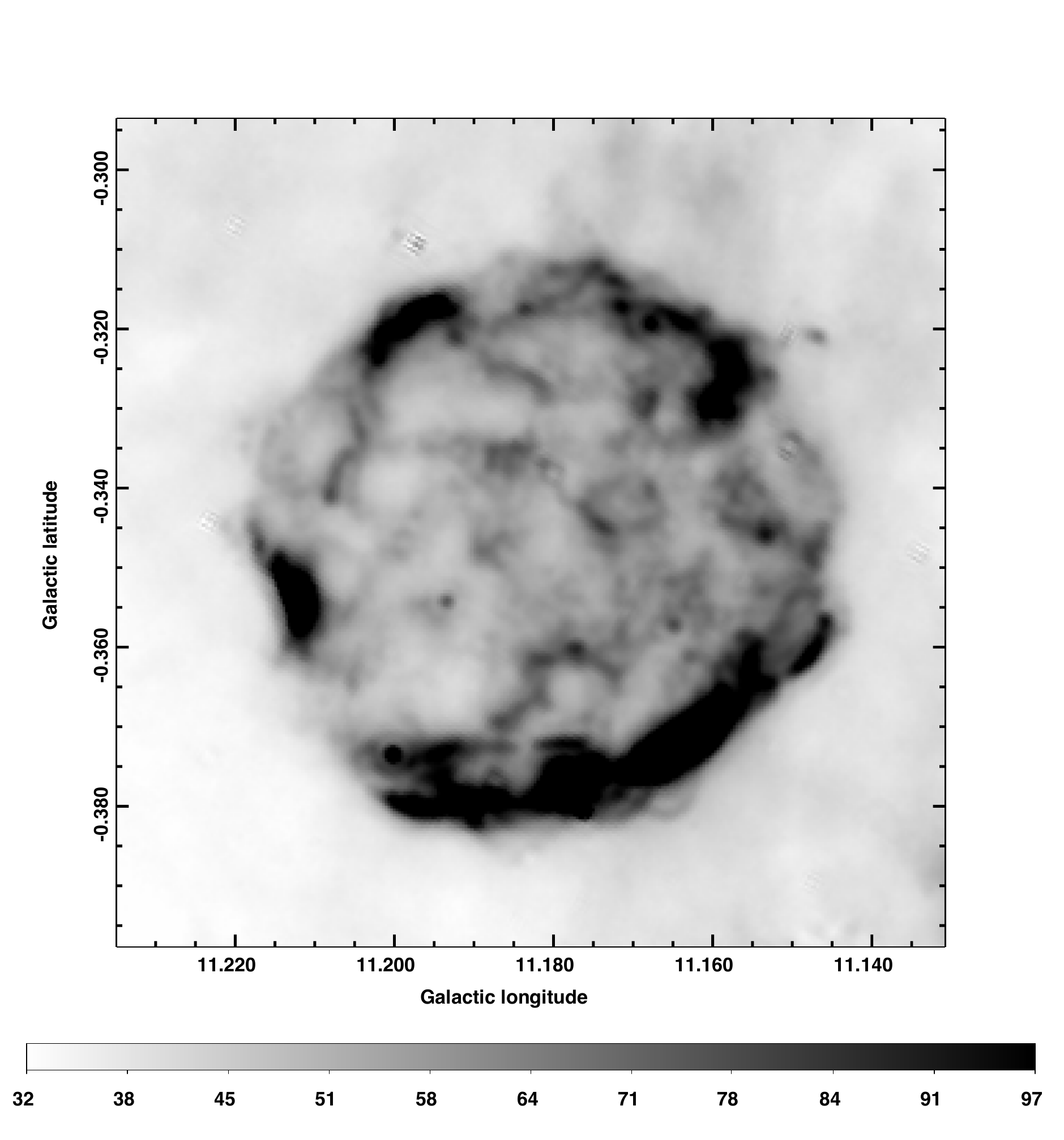}
\includegraphics[width=.38\textwidth]{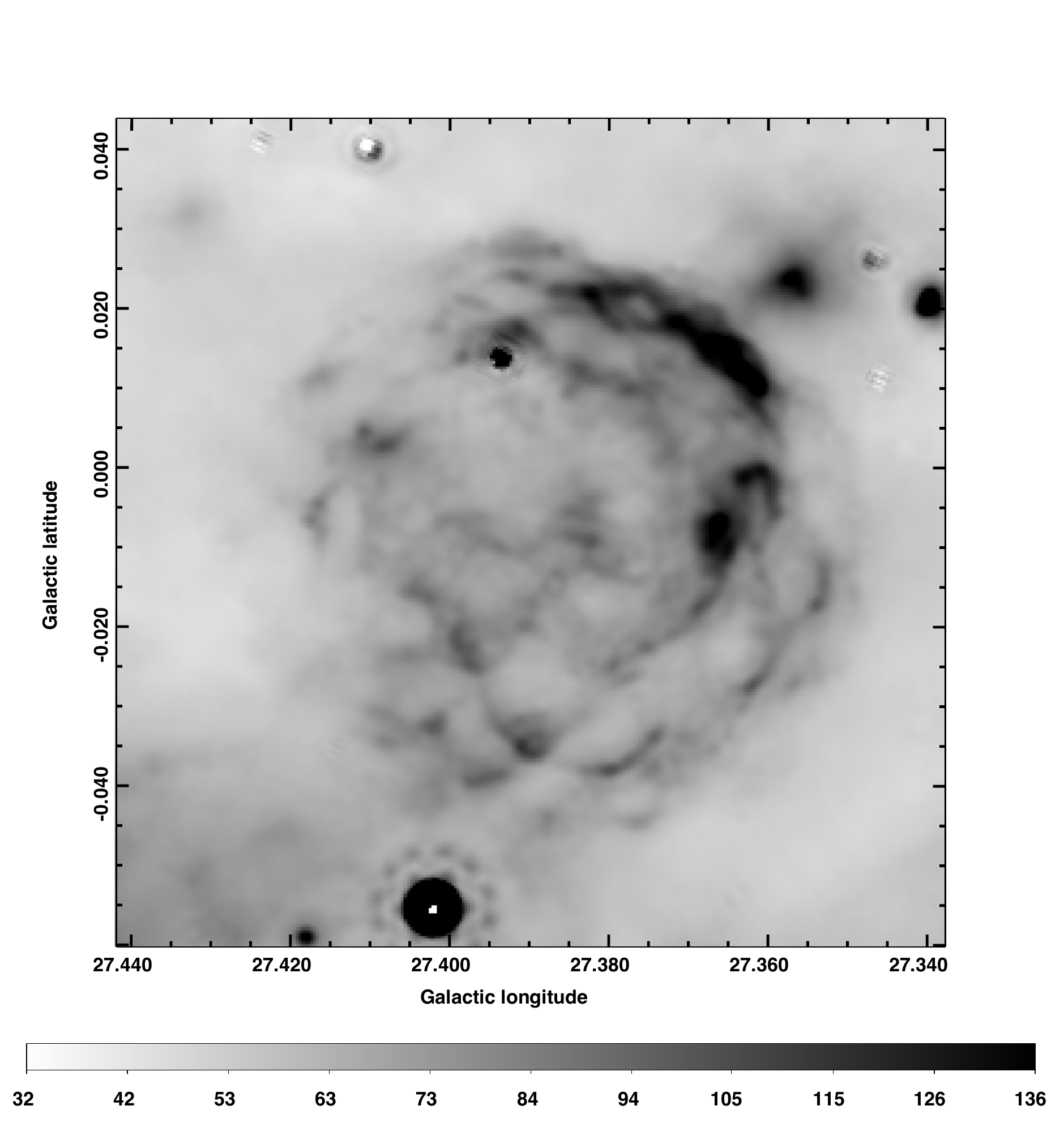}\\
\includegraphics[width=.38\textwidth]{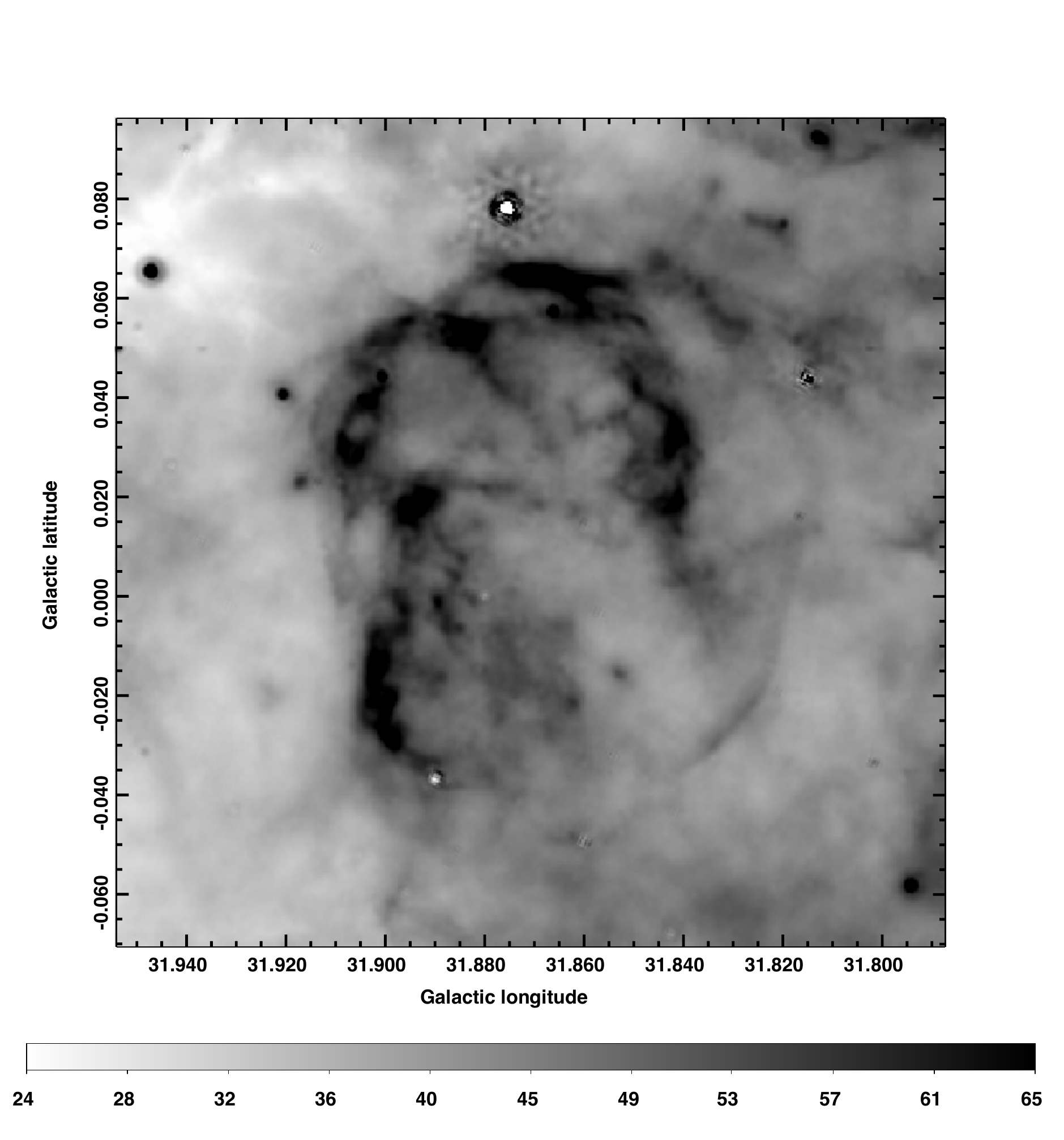}
\includegraphics[width=.38\textwidth]{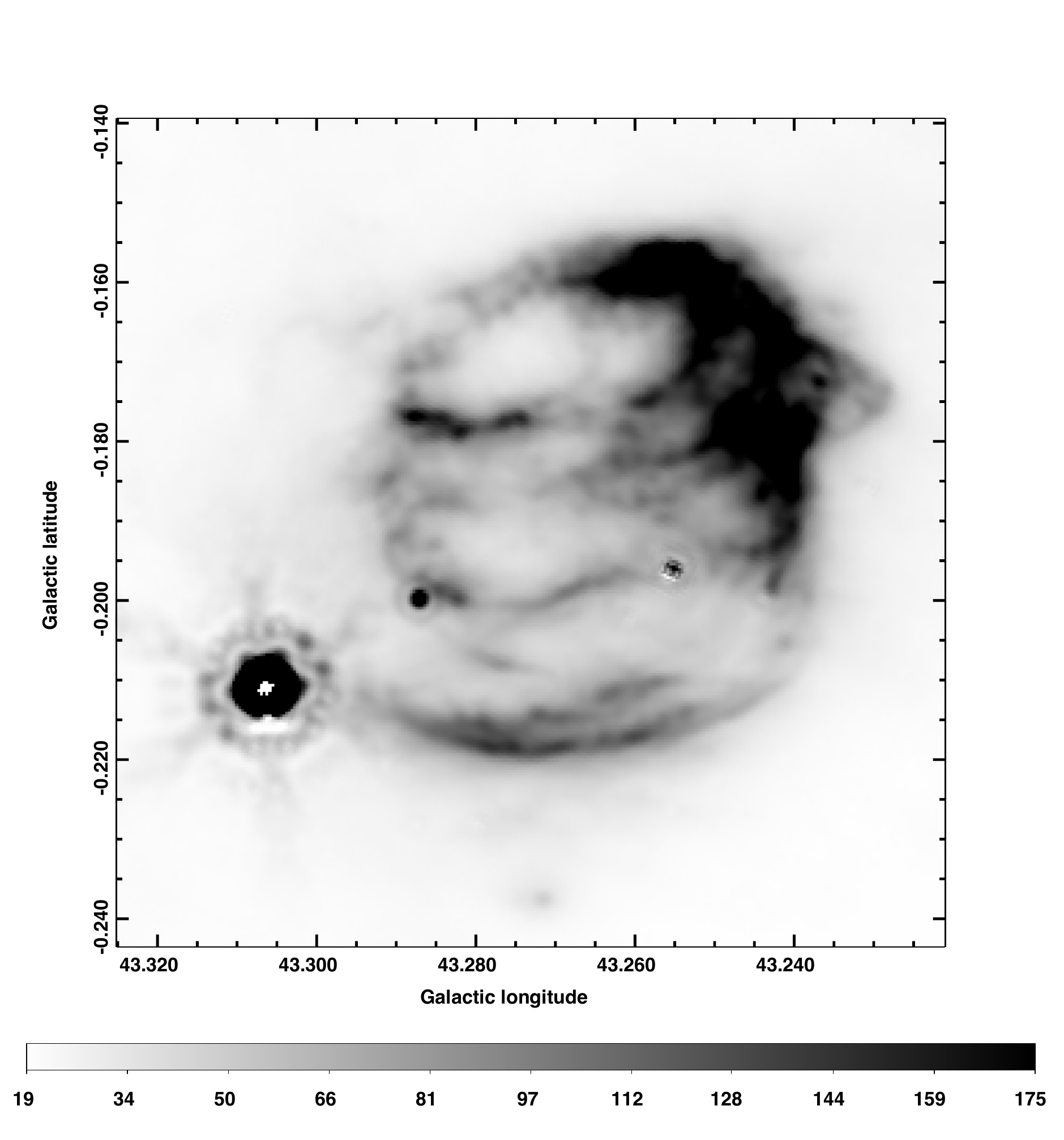}\\
\includegraphics[width=.38\textwidth]{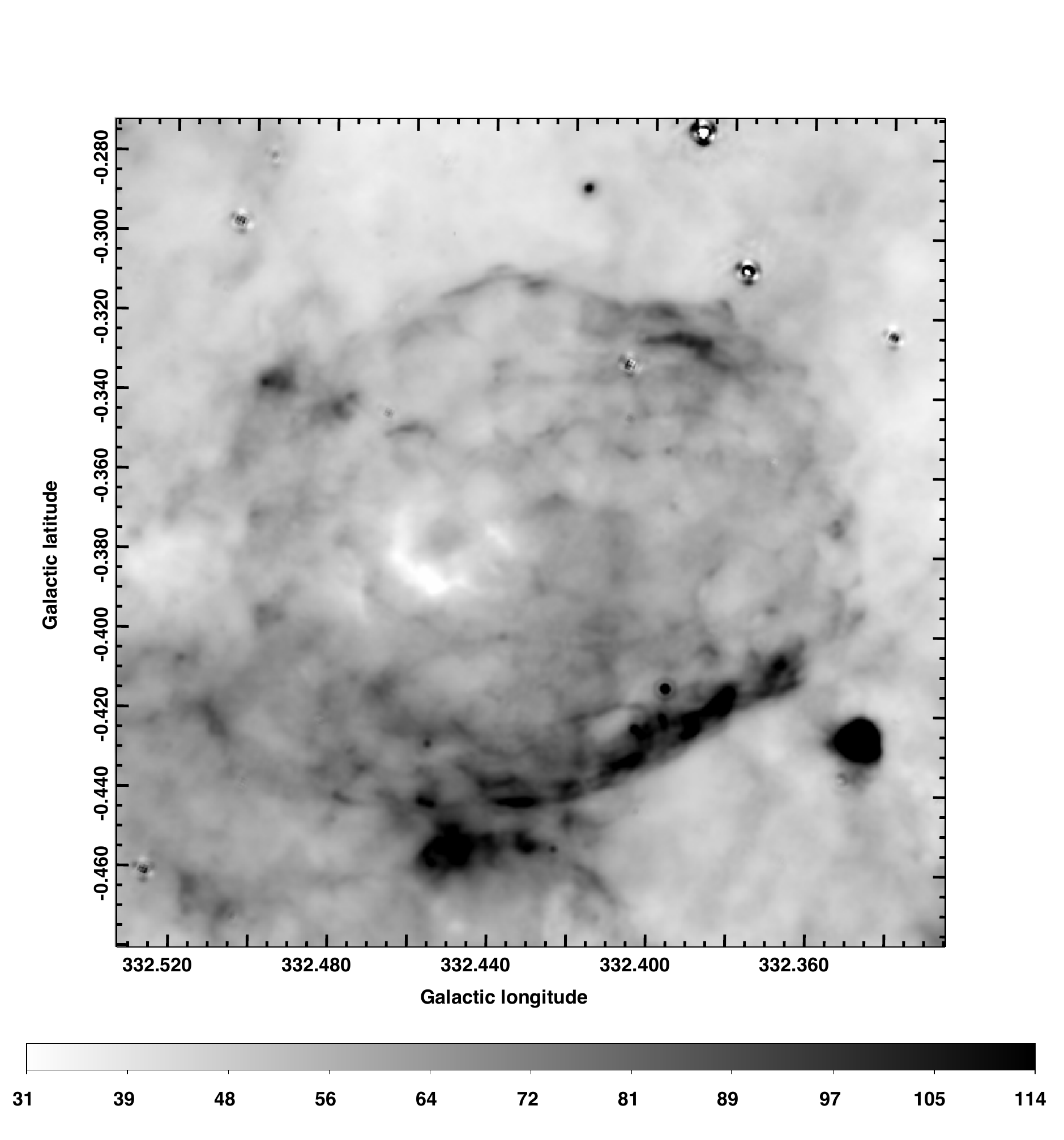}
\includegraphics[width=.38\textwidth]{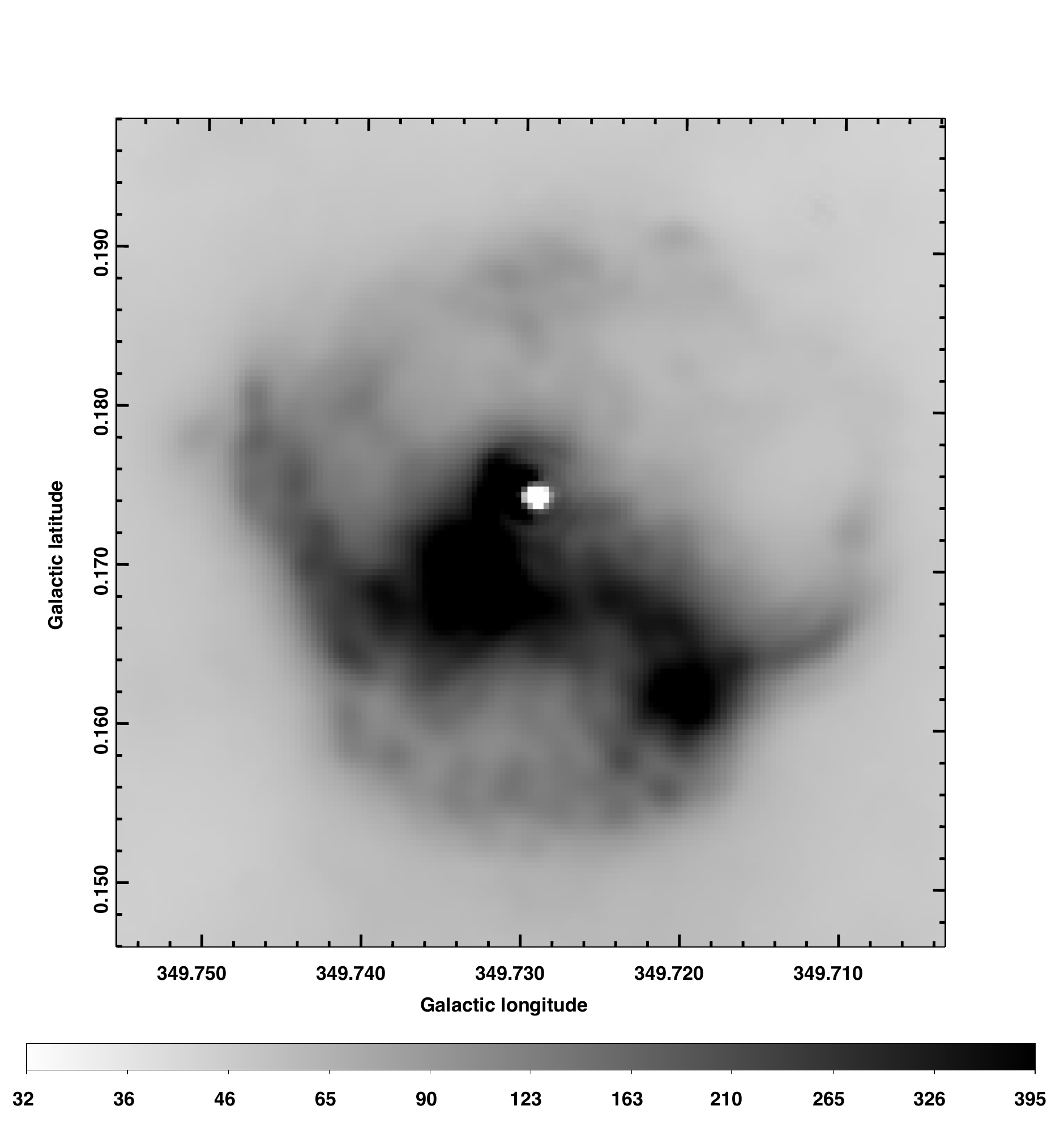}
\caption{Examples of level 1 detections at 24 $\micron$. From left to right, top to bottom: G11.2-0.3, G27.4+0.0 (Kes 73), G31.9+0.0 (3C 391), G43.3-0.2 (W49B), G332.4-0.4 (RCW 103) and G349.7-0.2. The greyscale is linear with the ranges in MJy/sr as displayed in the intensity bars below each image. The angular extents of the images differ.}
\label{good}
\end{figure*}

Out of a sample of \snrsurvey\  possible radio remnants, \snrdetectedlevela\ are level 1 detections, a detection rate of 32\% compared to 18\% with GLIMPSE. A few examples of level 1 detections are shown in Fig.~\ref{good}. The MIPSGAL sensitivity at 24 $\micron$ matches very well that of IRAC at 8 $\micron$ (\citealp{2009PASP..121...76C}; \citealp{2003PASP..115..953B}). The fact that we can identify many more SNRs than with IRAC could potentially be due to an extinction effect (lower at 24 $\micron$ than at 8 $\micron$ by 40$\%$, \citealp{2003ARA&A..41..241D}). However, it is far more plausible that it could simply be the details of how SNRs emit. In the case of SNRs interacting with the ISM where shocks modify the dust grain size distribution (e.g.,~\citealp{2010arXiv1004.0677G}), 24 $\micron$ continuum emission (due to Very Small Grains, VSGs) rises while at 8 $\micron$ the emission by stochastically heated PAHs is not as strong.

We calculate flux densities at 8, 24, and 70 $\micron$ for SNRs which show obvious infrared counterparts (level 1 detections). The aperture size used for the flux derivations is, in some cases, different from the one listed at radio wavelengths (see Table \ref{table:1}). This is done in order to account for the different morphologies a remnant can have in various wavelength ranges. Furthermore, there are cases where the SNR is nearby other unrelated extended Galactic sources whose contamination can lead to an overestimate of the infrared flux (e.g., diffuse Galactic emission and \ion{H}{2} regions).
To avoid such problems, we examined the MIPS images overlaid with contours from other wavelengths to help constrain the location and shape of the SNR infrared emission; however, such discrimination is not always well achieved. Another issue that arises, especially when dealing with Galactic plane images such as these, is the presence of unrelated infrared emission along the line of sight. This confusion can hinder the detection of SNRs as demonstrated by \cite{1989ApJS...70..181A} and \cite{1992ApJS...81..715S}.

Flux densities of the detected remnants were obtained using either a circular or elliptical aperture with a background removed. The background was determined using the median value of the sky brightness around the remnant via two methods. We used an annulus around the source for cases where the external source contamination in the neighborhood is negligible. However, that method was often not feasible due to considerable overlap of sources (extended or point-like) with the periphery of the SNR. For those cases, the sky brightness was estimated with a user-selected region characteristic of the local background emission. In order to quantify the magnitude of background variations, we measured the surface brightness on several regions around one of the faintest 24$\micron$~SNR, G21.8-0.6. We used aperture sizes similar to the one used for the remnant and found that the standard deviation relative to the mean background value across the field, for our three wavelengths are of the order of 12\% (8 $\micron$), 10\% (24 $\micron$) and 20\% (70 $\micron$).

Another consideration for photometry measurements is point source contamination, especially at 8 $\micron$ (as discussed in \S~\ref{glimpse}). Finally, since we are dealing with spatially extended emission, we applied corrections to the 8 $\micron$ estimates to account for internal scattering in the detectors as instructed by the IRAC data handbook\footnote[5]{http://ssc.spitzer.caltech.edu/irac/iracinstrumenthandbook/IRAC-Instrument-Handbook.pdf}. The `infinite' radius aperture is most appropriate for the angular extent of the remnants; we multiplied the surface brightness measurements by 0.74.

\section{Discussion} \label{discussion}

\subsection{Comparison of detections with previous infrared surveys}\label{othersur}
Out of the previous 18 detections (level 1) using GLIMPSE \citep{2006AJ....131.1479R}, we confirm 16 at the same level with MIPSGAL (Table~1). The remaining two remnants are confused with the background and are reported as possible detections in our analysis. Of the other 23 (level 1) infrared counterparts found in this work, 10 are not in the initial sample from Green's catalog used by \citet{2006AJ....131.1479R} and the remaining 13 are mostly bright X-ray remnants which seem to be brighter at 24 $\micron$. Regarding the SNRs identified by \cite{1989ApJS...70..181A}, all of their level 1 detections are also seen in MIPSGAL with the exception of G12.0-0.1 and G54.1+0.3 which are both level 3 in the MIPS data. Comparing with \cite{1992ApJS...81..715S}, six SNRs have been detected in common (level 1). They also reported a level 1 detection of 8 others, which we classify as level 2 or 3. Again, we observe that these cases are mostly regions where there is a large contamination by nearby or overlapping \ion{H}{2} regions making a clear identification challenging even at the superior Spitzer resolution.

\subsection{Lack of infrared signature}
Given the strong ambient emission in the inner Galactic plane, it is not surprising that many SNRs, either too big or too old, and having low infrared surface brightness, do not have an identifiable infrared signature. Also, as noted by \cite{2006AJ....132.1877W}, the environs in which SNRs are encountered need to be sufficiently dense to promote collisional heating of the dust, thus making it distinguishable. Examples of non-detections in this survey are G18.1-0.1 and G299.6-0.5.

\subsection{Color-color Diagrams}

\begin{figure*}
\centering 
\includegraphics[width=1.\textwidth]{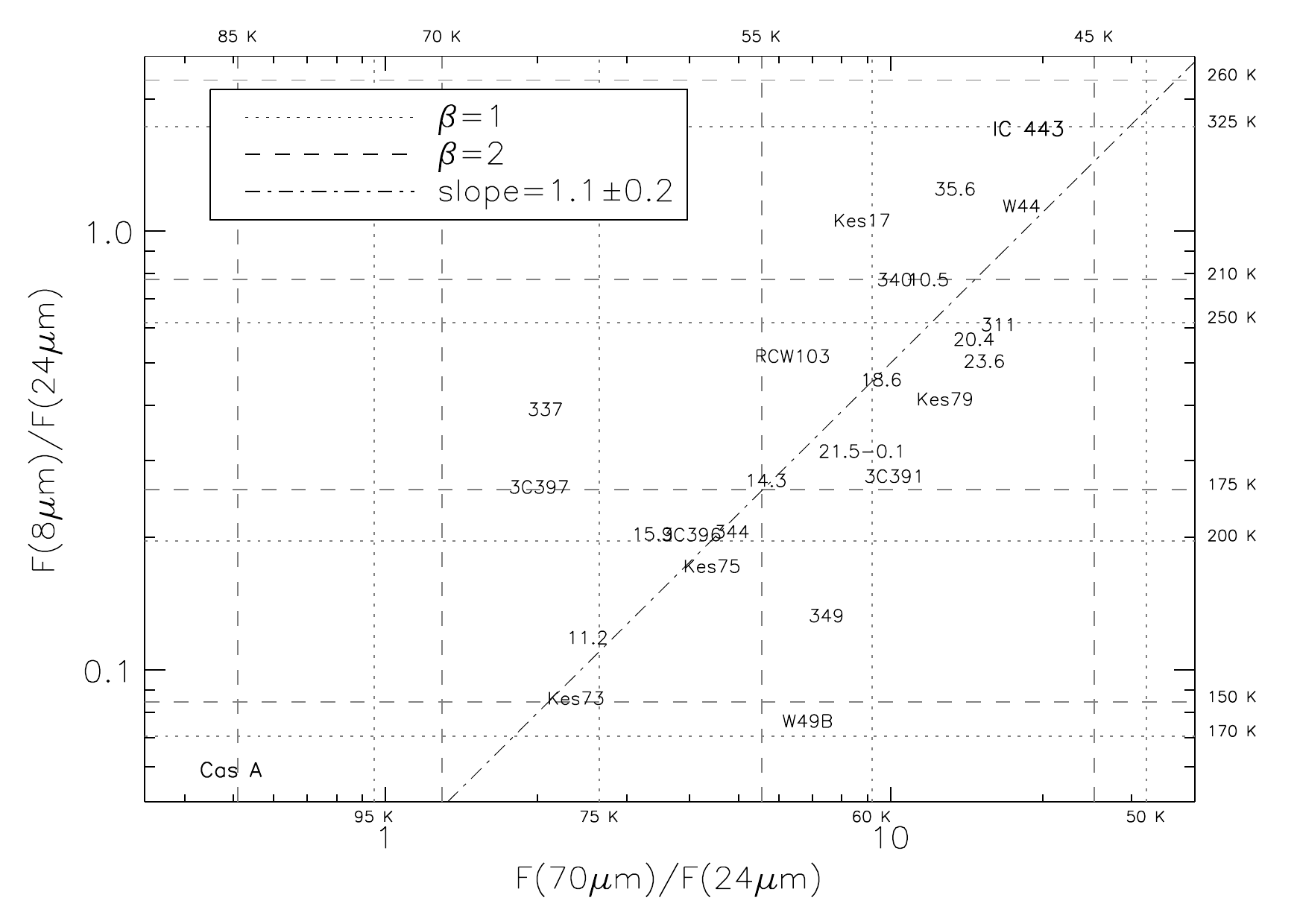}
\caption{Color-color plot of the SNRs in our sample detected and measured as a whole. For comparison, we have also plotted the location of Cas A and IC443 in this diagram. We find that this sample is well fitted by a slope of $1.1\pm0.2$ in this logarithmic representation, showing that 8 and 70 $\micron$ emission are related roughly by a constant while they change dramatically with respect to 24 $\micron$. Vertical lines show the range of expected temperatures for a modified blackbody using only the flux measured at 24 and 70 $\micron$ while for the horizontal lines, the temperatures are calculated using the flux at 8 and 24 $\micron$. Dotted and dashed lines correspond to a $\beta$ of 1 or 2, respectively. If the IR emission at longer wavelengths is mainly due to dust then the range of temperatures for our sample of SNRs is roughly between 50 and 85 K for $\beta$=1 and between 45 and 70 K for $\beta$=2.}
\label{colorplotflux}
\end{figure*}

Table~\ref{table:1} shows flux densities (at 8, 24 and 70 $\micron$) and color ratios for the detected remnants. Figure \ref{colorplotflux} contains a logarithmic plot of [F$_{8}$/F$_{24}$] \emph{versus} [F$_{70}$/$F_{24}$] for which SNRs are detected as a whole (see last column of Table~\ref{table:1}).

Two other SNRs are plotted: for Cas A, values are retrieved from \cite{2004ApJS..154..290H} and for IC443 from Noriega-Crespo et al.\ \emph{(in preparation)}. The data show a linear trend with a slope of $1.1\pm0.2$: remnants with a low [F$_{8}$/F$_{24}$] ratio statistically have a low [F$_{70}$/F$_{24}$] ratio. Furthermore, there seems to exist an age effect, with older remnants (such as IC443 and W44) showing higher [F$_{70}$/F$_{24}$] and [F$_{8}$/F$_{24}$]. The opposite seems to be more characteristic of younger remnants, such as Cas A, Kes 73 and G11.2-0.3.

\begin{figure*}
\centering 
\includegraphics[width=1\textwidth]{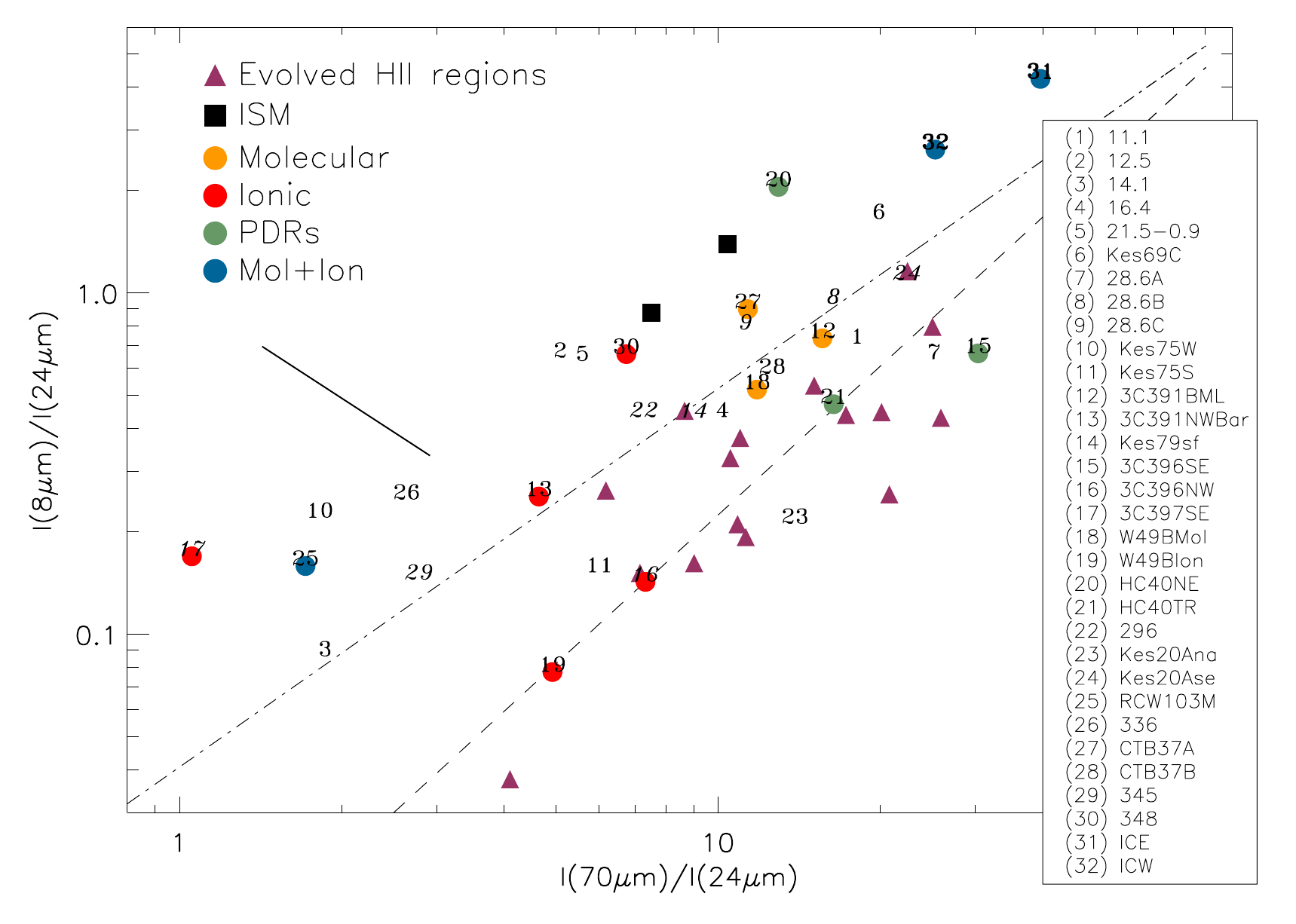}
\caption{Color-color plot based on surface brightnesses of localized regions in SNRs in our sample. Ratios for SNRs known to have molecular interactions, atomic fine-structure line emission or PDRs colors are also shown. ICE and ICW refer to IC443, eastern (strong [O I] emission at 63 $\micron$) and western (interacting with a nearby \ion{H}{2} region) regions, respectively. Colors of pure infrared synchrotron emission are plotted as a straight line. Color ratios for the diffuse ISM \citep{2010arXiv1010.2774C} and evolved \ion{H}{2} regions ({Paladini et al.\ \emph{(in preparation)}}) are also included. Remnants with upper flux limits are represented in italic. For comparison, we plot the slope (dotted-dashed line) of $1.1\pm0.2$ found for the sample of detected remnants and measured as a whole (see Fig.~\ref{colorplotflux}). Also depicted is a dashed line showing a different trend for evolved HII regions.}
\label{colorplot}
\end{figure*}

  In some remnants, infrared emission is only detected in certain parts inside the radio/X-ray contours. Figure~\ref{colorplot} shows surface brightness ratios of these localized regions. To assess the mechanism of the infrared emission, we also calculate surface brightnesses for regions of SNRs known to be either interacting with neighboring molecular regions (such as the BML region in 3C391) and/or that show molecular emission lines (such as W49B) or known to have ionic lines (such as 3C396). For this exercise, we used regions of some remnants for which the main emission mechanism had previously been identified by \cite{2006AJ....131.1479R} (see their Fig. 22). These are also identified in the SNR figures in the appendix. Remnants with colors characteristic of ionic shocks seem to occupy the lower left part of the [I$_{70}$/I$_{24}$] versus [I$_{8}$/I$_{24}$] diagram, while remnants whose colors are consistent with molecular shocks and photodissociation regions are found in the upper right part. However, given the small number of examples and the fact that all three passbands (8, 24 and 70 $\micron$) contain dust emission and lines, we suggest that the [I$_{70}$/I$_{24}$] and [I$_{8}$/I$_{24}$] color ratios are not a completely secure method of distinguishing between different emission mechanisms.

  We plot some other color ratios for comparison. The locus for pure synchrotron is for spectral index of 0.3 to 1 \footnote{$\alpha$ is the spectral index and the radio flux density is defined as $S{_{\nu}}\propto\nu^{-\alpha}$, where $\nu$ is the frequency.}. For the diffuse ISM, we used the color ratios obtained by \cite{2010ApJ...724L..44C} for two regions (with different abundances of very small grains) in the Galactic plane at a longitude of approximately $59^\circ$. We have also included color ratios for evolved \ion{H}{2} regions (found near Galactic longitudes of $35^\circ$; {Paladini et al.\ \emph{(in preparation)}}).
  
 Are the color ratios [F$_{70}$/$F_{24}$] and [F$_{8}$/F$_{24}$] good discriminators between such very different regions? Based on Figure \ref{colorplot}, the plotted colors of individual evolved \ion{H}{2} regions often seem indistinguishable from SNRs using Spitzer colors. However, the overall trend for SNRs is displaced from the \ion{H}{2} regions and covers a broader range of colors. This is similar to what \cite{1989ApJS...70..181A} found for older SNRs and compact \ion{H}{2} regions using IRAS colors.

Although line contamination is significant for radiative remnants, for younger ones the majority of the IR emission should be due to dust grains. For instance, for RCW 103 (age approximately 2000 yr; \citealp{1997PASP..109..990C}) we calculate the ratio of line emission (using ISO data) versus dust continuum within MIPS filters. We find that line emission contributes about 6\% and 3\% of the observed flux at 24 and 70 $\micron$, respectively.

If emission from the remnants comes primarily from shocked hot dust, then we can derive the dust temperature just by assuming a simple modified blackbody emission. If dust has a thermal spectral energy distribution
\begin{equation}
S{_\nu}\propto~B{_\nu}(T)~\nu^{\beta} \, ,
\end{equation}
where $\beta$ is the dust emissivity index (which depends on the dust composition, $\beta\sim1$ to 2), then flux ratios imply certain temperatures. Figure \ref{colorplotflux} shows temperatures based on the [F$_{8}$/$F_{24}$] and [F$_{24}$/$F_{70}$] ratios. The latter flux ratio yields dust temperatures for our SNR sample ranging between 45 and 70 K for $\beta=2$ and from 50 to 85 K for $\beta=1$. This is a very crude estimate of the dust temperature given that we are just considering one population of dust grains in temperature equilibrium when it is far more likely that there are several populations plus non-equilibrium emission. Applying the same simple dust model to the [F$_{8}$/$F_{24}$] ratio, we obtain temperature values higher than 145 K (for $\beta=2$) with younger SNRs having color temperatures lower that older remnants.  This trend is similar to what was obtained by \cite{1989ApJS...70..181A} using color temperatures based on the IRAS 12 to 25 $\micron$ ratio. The infrared colors are not well explained by a single equilibrium dust temperature and atomic and molecular line emission as well as PAHs might significantly contribute to the infrared emission for SNRs.

However, assuming that the infrared emission at 24 and 70 $\micron$ is cospatial, entirely due to dust, and well fitted by a single temperature modified blackbody, then SNR dust masses are given by the following equation:
\begin{equation}
M{_{\mathrm{dust}}} = \frac{d{^2}~F{_{\mathrm{\nu}}}} {{\kappa}{_{\mathrm{\nu}}}~B{_{\mathrm{\nu}}(T{_{\mathrm{dust}}})}} \, ,
\end{equation}
where $d$ is the distance to each SNRs, $F{_{\nu}}$ is the flux density and $\kappa{_{\nu}}$ is the dust mass absorption coefficient. For the dust mass absorption coefficient, we use the diffuse ISM model by \cite{2001ApJ...554..778L} which consists of a mixture of amorphous silicate and carbonaceous grains. Distances are retrieved from Green's catalog. Results are reported in Table~\ref{mass}. Note that these derived dust masses, based on the 24 and 70 $\micron$ fluxes from MIPS, are overestimated given the probable contamination of these fluxes by line emission which can be substantial in radiative remnants. Even in the case where all of the flux in these bands is due to dust, mass estimates are based on a single color temperature (24/70) which can be reasonable for younger remnants but not ideal for older remnants given that their IR SED generally requires more that one dust population (\citealp{1989ApJS...70..181A}; \citealp{1992ApJS...81..715S}).

\subsection{Infrared to radio (1.4 GHz) ratio}
 
 According to \cite{1987Natur.327..211H}, the comparison between mid-infrared (60 $\micron$ from IRAS) and radio continuum (11 cm) emission is a good diagnostic for distinguishing between thermal from non-thermal radio emitters (see  also \citep{1987A&AS...71...63F}  and \citep{1989MNRAS.237..381B}). \ion{H}{2} regions are shown to have infrared to radio ratios of the order of $\geq 500$ while ratios for SNRs are thought to be lower than 20. Moreover, \cite{1996A&AS..118..329W} using MOST found that SNRs have a ratio of infrared (60 $\micron$) to radio (843 MHz or 36 cm) of $\leq 50$ while \ion{H}{2} regions have again ratios of the order of 500 or more. 
 
\begin{figure*}
\centering
\includegraphics[width=1\textwidth]{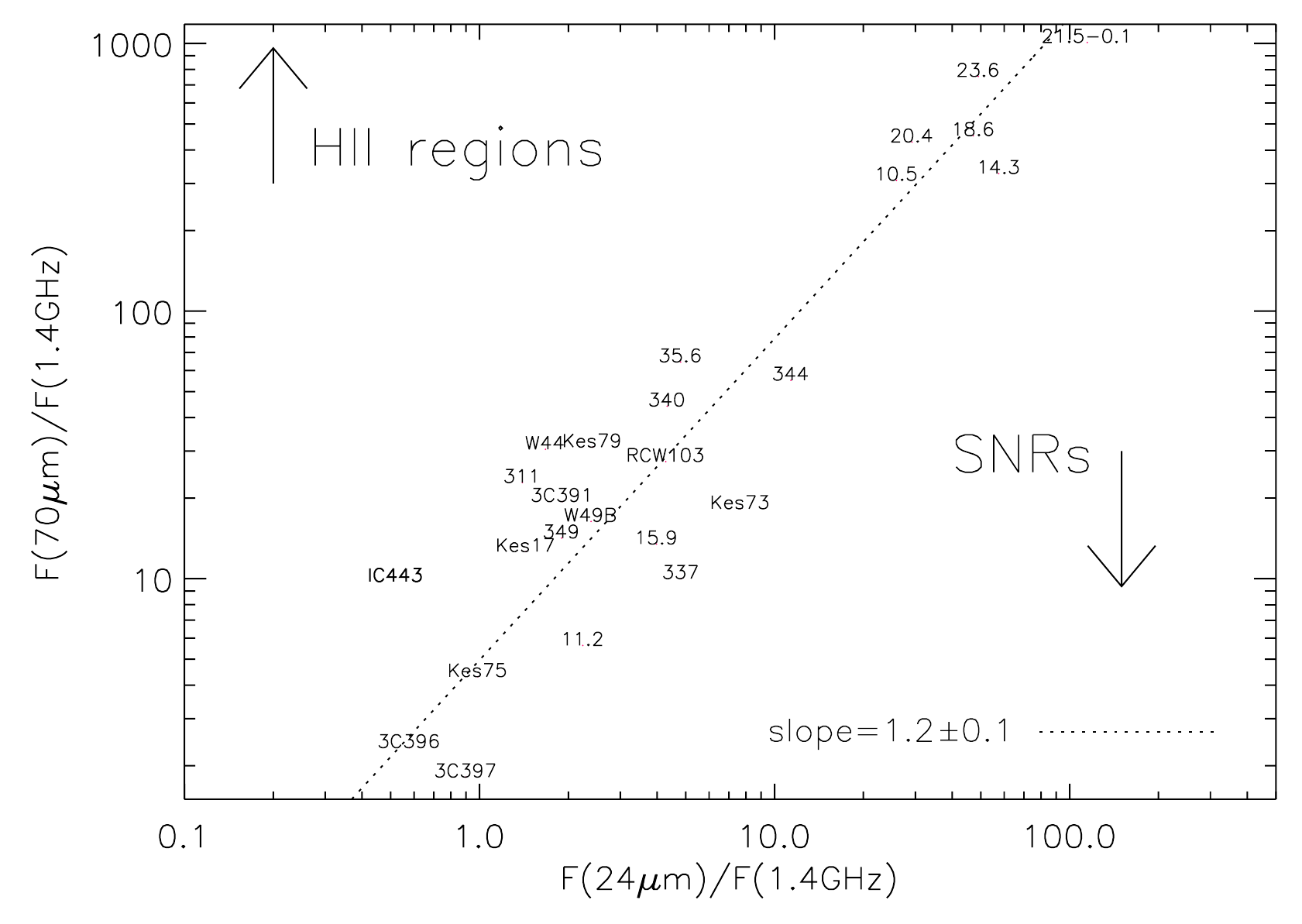}
\caption{Ratio of MIPS 70 $\micron$ and radio continuum at 1.4 GHz ($q{_{\mathrm{70}}}=\log(F{_{\mathrm{70\micron}}}/F{_{\mathrm{1.4GHz}}}$)) versus ratio between 24 $\micron$ and radio again ($q{_{\mathrm{24}}}$). There is a considerable range in the positions of the SNRs. The ratios are correlated and are well fitted by a slope of 1.2. There seems to exist a group of catalogued sources which have ratios more characteristic of \ion{H}{2} regions (upper population in the plot). The infrared counterparts in the lower part of the plot show ratios of infrared (70 $\micron$) to radio comparable to what was found for SNRs in previous studies (e.g., \cite{1989MNRAS.237..381B}) which suggested ratios of infrared (60 $\micron$) to 2.7 GHz lower than 20. The location of IC443 is also shown. Cas A falls off the plot and below the trend with ratios (0.1, 0.05).}
\label{radio_infrared}
\end{figure*}

 We use the 1 GHz flux densities provided in Green's catalog \citep{2009BASI...37...45G}. Those values were converted to 1.4 GHz (21 cm) flux densities using the spectral indices quoted in the same catalog. We compare those to the 24 and 70 $\micron$ emission from MIPSGAL. Figure~\ref{radio_infrared} shows the spread in the ratio of the infrared to radio for the detected SNRs. Note that these ratios are often defined by $q{_{\mathrm{IR\lambda}}}=\log (F{_{\mathrm{IR\lambda}}}/F{_{\mathrm{21cm}}})$ where F${_\mathrm{IR\lambda}}$ and F$_\mathrm{21cm}$ are the flux densities (in Jy) at specific mid-infrared wavelengths and in the radio.

There seems to be a group with ratios (or high $q{_{\mathrm{24}}}$ and $q{_{\mathrm{70}}}$) similar to those of \ion{H}{2} regions.  In particular, we suggest that the morphology of the remnant G23.6+0.3 (and possibly G14.3+0.1) resembles more closely an \ion{H}{2} region and so whether it is a SNR should be re-considered. See also its infrared colors compared to \ion{H}{2} regions (Figs.~\ref{colorplotflux} and \ref{colorplot}). Image in Figure \ref{snrG236-03} (and \ref{snrG14301}) clearly shows a composite infrared emission, with the 24 $\micron$ (probing hot and small particles) matching the radio, while the 8 and 70 $\micron$ emission do not spatially coincide with the 24 $\micron$. In particular, the 8 $\micron$ appears as an outer layer which marks the transition between the ionized region and colder gas. The 70 $\micron$ emission is also seen along the same location as the 8 $\micron$, which is consistent with the premise that we are now probing a colder dust population.

\begin{figure*}
\centering
\includegraphics[width=1.\textwidth]{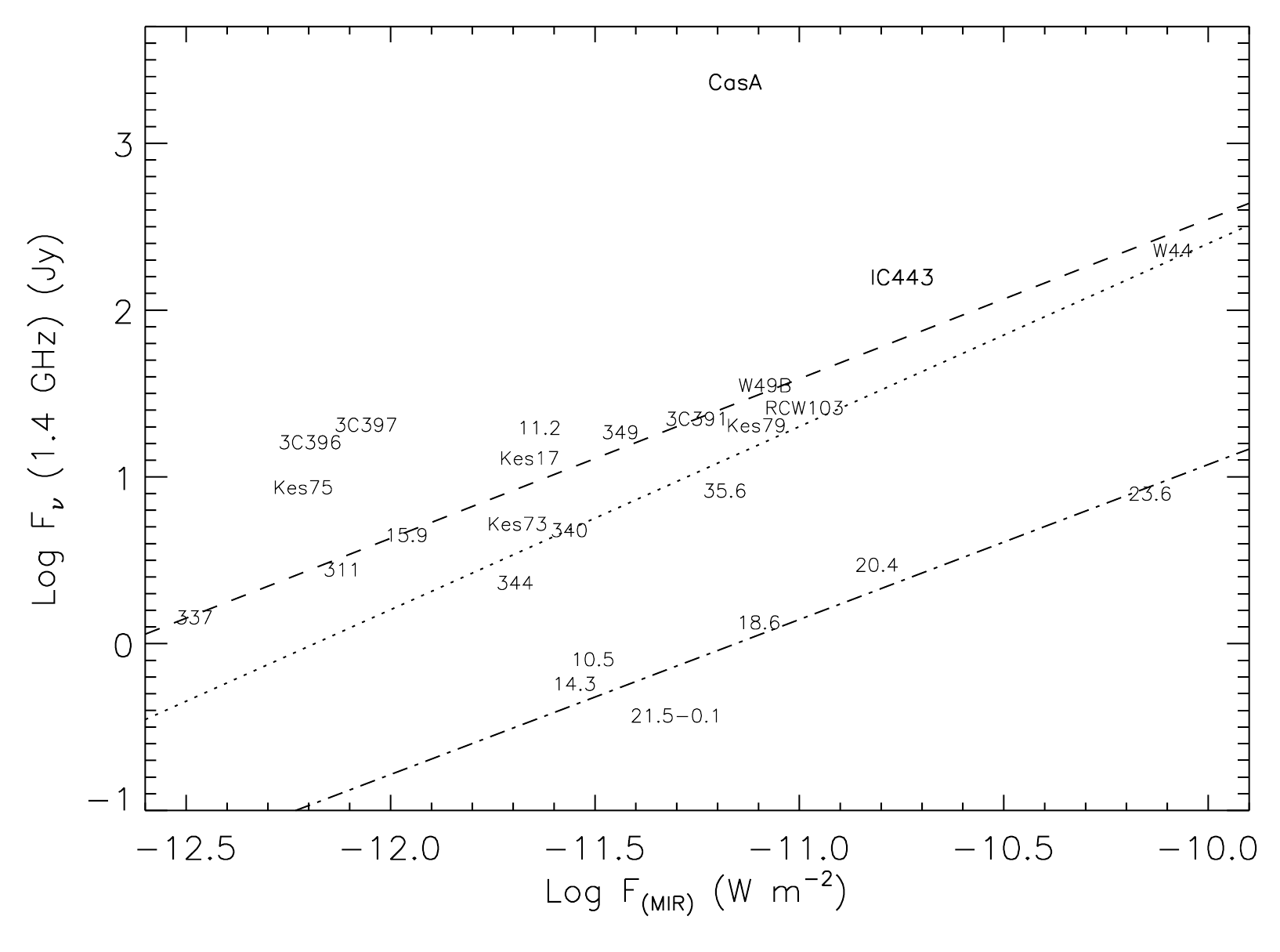}
\caption{Correlation between radio and infrared emission. For the entire population, the slope of the correlation is 1.10 (dotted line). There seem to exist two trends. The upper population is more characteristic of SNRs while the lower population looks more like \ion{H}{2} regions.  The latter are the same objects with the high values of $q{_{\mathrm{24}}}$ and $q{_{\mathrm{70}}}$ in Figure~\ref{radio_infrared}. The slopes for the upper and lower populations are 0.96 (dashed line) and 0.93 (dot-dashed line), respectively, not significantly different. Again, IC443 and Cas A are also included for comparison. Cas A stands out due to its strong radio synchrotron emission.}
\label{helou_82470}
\end{figure*}

There is a strong empirical correlation between infrared and radio emission (at 1.4 GHz) found in spiral galaxies \citep{1985ApJ...298L...7H}. The radio emission in this range is thought to be mainly due to synchrotron emission with only 10\% is attributed to thermal bremsstrahlung produced by \ion{H}{2} regions (\citealp{2009ApJ...706..482M} and references therein). In our Galaxy and other local star forming galaxies, the bulk of the mid and far-infrared arises from warm dust associated with star forming regions. In older systems, things are more complicated due to cold cirrus and evolved stars (e.g., AGBs). We look for a similar correlation using the infrared and radio emission from our SNR sample. We estimate the mid-infrared emission as the sum of the integrated emission in each MIPS passband (24 and 70 $\micron$) and obtain 
an integrated mid-infrared flux F${_{\mathrm{MIR}}}$ in units of W/m$^2$ (see \citealp{1985ApJ...298L...7H}). The F${_{\mathrm{MIR}}}$ is an approximation (underestimate) of the mid-infrared bolometric flux (given that the bandpasses of 24 and 70 $\micron$ filters do not overlap). Figure~\ref{helou_82470} shows the correlation between infrared and 1.4 GHz non-thermal radio flux. The slope of the correlation is $1.10\pm0.13$ when combining all data (plus IC443). Again, two distinct populations seem to exist, an upper main trend and a lower one.  Using only the lower one (i.e., objects which have 70 $\micron$ to 1.4 GHz ratios closer to \ion{H}{2} regions and high values of $q{_{\mathrm{24}}}$ and $q{_{\mathrm{70}}}$ in Fig.~\ref{radio_infrared}), we obtain a slope of $0.93\pm0.13$ while for the upper trend we get $0.96\pm0.09$.

The correlation with the slope fixed to unity leads to the dimensionless parameter  $q{_{\mathrm{MIR}}}$ (\citealp{1985ApJ...298L...7H}) which represents the ratio of mid-infrared to radio and is defined as
\begin{equation}
\begin{small}
q{_{\mathrm{MIR}}}\equiv \mathrm{\log} \Big( \frac{F {_{\mathrm{MIR}}}}{3.75\times10^{12}~\mathrm{Wm^{-2}}} \Big) - ~\mathrm{\log} \Big( \frac{F~{_{\mathrm{1.4~ GHz}}}} {\mathrm{W m^{-2} ~Hz^{-1}}} \Big) \, .
\end{small}
\label{qmir}
\end{equation}

 Our results show that remnants from the lower trend population in Figure~\ref{helou_82470} have  $q{_{\mathrm{MIR}}}$ larger than 2.06 (up to 2.60). The other group has  $q{_{\mathrm{MIR}}}$ ranging from 0.06 and 1.41. Specific $q$ values for each remnant are displayed on Table~\ref{table:qsnr}. For reference, a previous study by \cite{2003ApJ...586..794B} reported a median value of $q{_{\mathrm{TIR}}}$ (i.e., the ratio of the total 8-1000 $\micron$ IR luminosity to the radio power) of $2.64$ for a sample of 162 galaxies. But note this would be higher because it is based on the total IR emission.
 
Table~\ref{table:qlambda} reports on the average and dispersion of the monochromatic $q{_{\mathrm{8}}}$, $q{_{\mathrm{24}}}$ and $q{_{\mathrm{70}}}$ parameters for the two trends. For comparison, a study of extragalactic VLA radio sources by \cite{2004ApJS..154..147A} obtained $q{_{\mathrm{24}}}=0.84$ and $q{_{\mathrm{70}}}=2.15$ while \cite{2007MNRAS.376.1182B} found $q{_{\mathrm{24}}}=1.39$ for somewhat fainter galaxies.
Later on, \cite{2008PASJ...60S.453S} using AKARI data for some LMC SNRs compared the ratio of 24 $\micron$ to radio fluxes; their correlation implies $q{_{\mathrm{24}}}=0$. If SNRs are the sole contributors of synchrotron emission in a star-forming galaxy, then by comparison with the above $q{_{\mathrm{24}}}$'s for galaxies they concluded that about 4 to 14\% of the 24 $\micron$ emission in galaxies is due to SNRs. By doing a similar exercise using the upper trend remnants, for which $q{_{\mathrm{24}}}=0.39$ (see Table~\ref{table:qlambda}), we find that 10 to 35\% of the 24 $\micron$ galactic emission seen would be due to the remnants. Likewise, at 70 $\micron$ that contribution would be 11\%. The rest would be due to dust heated in \ion{H}{2} and photo-dissociation regions, as well as diffuse emission. 

\subsection{High energy emission from SNRs}

In this section, we explore the relationship between the infrared and X-ray energetics of SNRs. Shocks produce hot plasma whose thermal energy is transferred to the dust grains via collisions and then re-emitted in the infrared (\citealp{1973ApJ...184L.113O}; \citealp{1981ApJ...245..880D}; \citealp{1987ApJ...322..812D}).
Heating up the dust grains takes energy from the X-ray gas, thus cooling it and a good tool to measure this transfer of energy is the ratio of infrared to X-ray power, usually referred to as IRX \citep{1987ApJ...320L..27D}, which compares dust cooling (gas-grain collisions) with X-ray cooling (continuum and lines). 
The gas cooling function increases at lower plasma temperatures given that more lines become available due to recombination. On the other hand, at high temperatures of around $10^7$ K, most of the energy is released through the bremsstrahlung continuum (\citealp{1976ApJ...204..290R}). Assuming that line emission in the infrared is not energetically significant compared to the X-ray lines, then most of the line cooling must happen in the X-ray domain. This implies that if the observed powers in the infrared and X-ray are comparable then dust must be the essential contributor to the cooling in the infrared.

\begin{figure*}
\centering
\includegraphics[width=1.\textwidth]{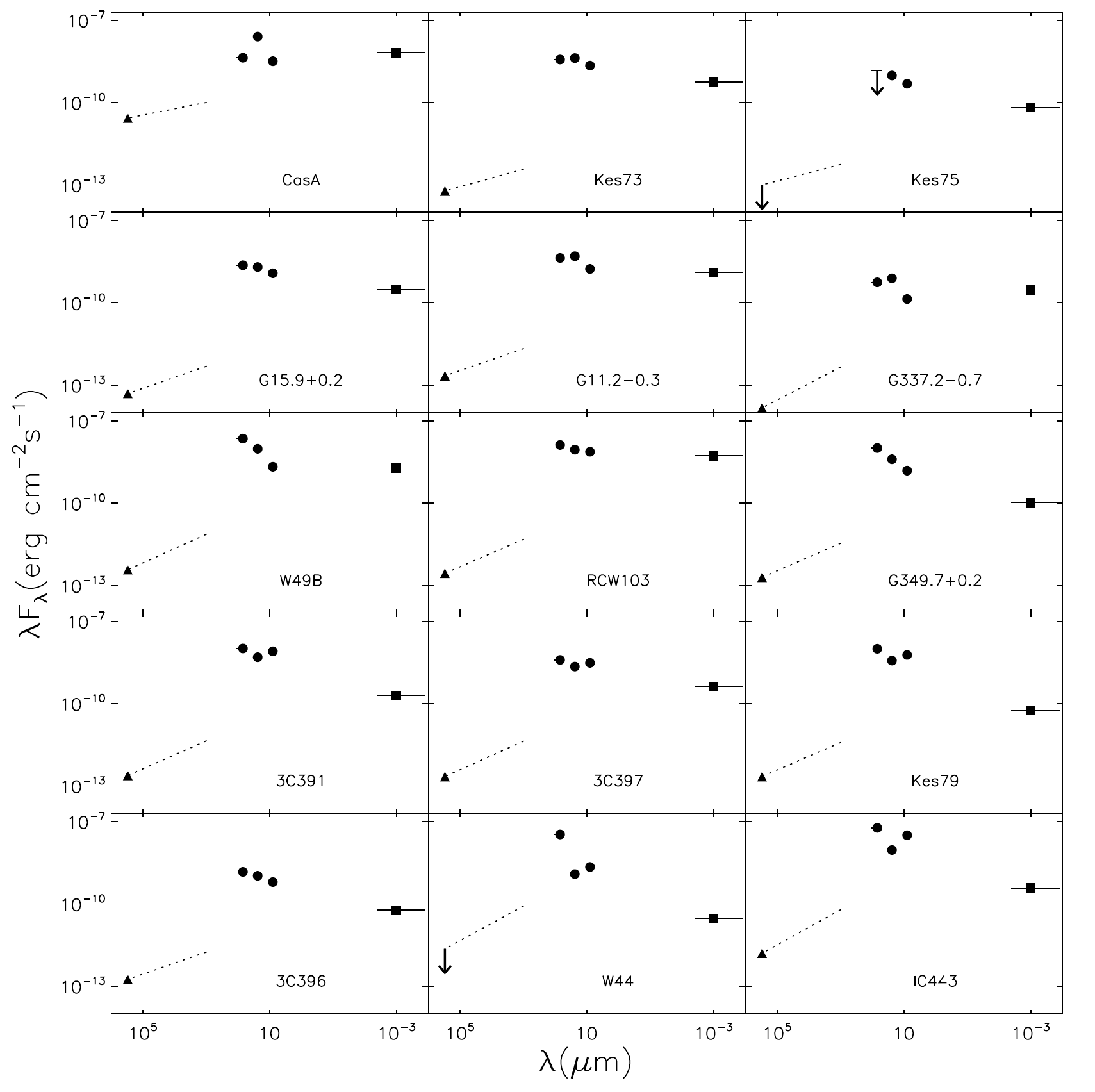}
\caption{SEDs of Chandra SNRs ordered according to approximate remnant's age (increasing left to right, top to bottom). For the radio, the synchrotron spectrum is shown by the dotted line for the appropriate spectral index, anchored in the observed range at 1 GHz (triangles). The circles indicate the power in the infrared at 8, 24 and 70 $\micron$. In the X-ray, the band (0.3-10 keV) of the Chandra images is used (squares).  Horizontal lines represent the width of the bandpass. We also include Cas A and IC443 for comparison. Their infrared flux densities are taken from \cite{2004ApJS..154..290H} and  Noriega-Crespo et al. (\emph{in preparation}) while the X-ray fluxes are from the Chandra Supernova Remnant Catalog and \cite{1987ApJ...320L..27D}, respectively. Again, note that for W44 the infrared and X-ray values apply only to the central region while the radio estimate is for the whole remnant. The integrals over the infrared and X-ray range reveal their contributions to cooling the shocked plasma.}
\label{figxray}
\end{figure*}

Table\ \ref{table:xray} shows the values of IRX for a sample of SNRs. The total X-ray fluxes (with energy range from 0.3 to 10 keV) were retrieved from the \emph{Chandra} Supernova Remnant Catalog\footnote[4] {http://hea-www.harvard.edu/ChandraSNR/index.html}. Infrared flux densities (in Jy) were measured for regions approximately matching those in the X-ray. We obtain the IR flux by integrating under the SED using simple linear interpolation in log space plus $\nu$F$\nu$ constant across the 24 $\micron$ passband. Also, note that the flux densities used in this analysis are not corrected for extinction. For our sample, IRXs range from about 1.6 for remnant RCW103 to 240 for W44. \\
These data are presented as SEDs in Figure~\ref{figxray}, in the logarithmic form $\lambda F{_{\lambda}}$ vs. $\lambda$ which is convenient for assessing the energetics. The radio values are obtained from Green's catalog (\citealp{2009BASI...37...45G}).  We also include Cas A and IC443 for comparison. Their infrared flux densities are taken from \cite{2004ApJS..154..290H} and  Noriega-Crespo et al. (\emph{in preparation}) while the X-ray fluxes are from the Chandra Supernova Remnant Catalog and \cite{1987ApJ...320L..27D}, respectively. In Figure~\ref{figxray} the remnants are sorted according to increasing age (see Table~\ref{table:xray}). Although the ages for some remnants are uncertain, this Figure shows that hot dust grain cooling (24 $\micron$) tends to be the most important contributor in early phases of the remnant, while in the later stage the warm dust traced by the 70 $\micron$ emission is the main contributor to plasma cooling. It also appears that older remnants tend to have greater IRXs. This can potentially be either due to an increase in the dust-to-gas ratio and/or a change in the size distribution favoring smaller sizes.

%_____________________________________________________________

\section{Conclusion}\label{con}
We have compiled a catalog of SNRs detected within the MIPSGAL survey at 24 and 70 $\micron$, with complementary measurements at 8 $\micron$ from the GLIMPSE survey. In order to better assess the nature of the detected infrared emission, we have compared it with radio and X-ray data. Our main findings are the following:

\begin{itemize}
\item{The detection rate of SNRs given the MIPSGAL sensitivity is 32$\%$, 39 out of the 121 candidates from Green's SNR catalog and higher than any previous infrared survey.}
\item{We find a linear trend (slope $=1.1\pm0.2$) in the logarithmic relationship between [F$_{8}$/F$_{24}$] versus [F$_{70}$/$F_{24}$]. If there is indeed an age effect, then the youngest SNRs will have the lowest [F$_{8}$/F$_{24}$] and [F$_{70}$/F$_{24}$] ratios.}
\item{The [I$_{70}$/I$_{24}$] and [I$_{8}$/I$_{24}$] color ratios provide a method of distinguishing between different emission mechanisms. This is not completely secure and the color ratios of some SNRs overlap with those of \ion{H}{2} regions.}
\item{Assuming a simple modified blackbody model (at 24 and 70 $\micron$), we retrieve SNRs dust temperatures which range from 45 to 70 K for a dust emissivity of $\beta=2$.}
\item{Using the previous color temperature (T${_{24/70}}$), we find rough estimates of dust masses ranging from 0.02 to 2.5 M$_{\sun}$. Note that the dust masses obtained here may be overestimated given the possible contribution of line emission to the MIPS fluxes.}
\item{We also compare infrared fluxes with their corresponding radio fluxes at 1.4 GHz and find that most of the remnants have ratios of 70 $\micron$ to 1.4 GHz characteristic of SNRs, although six (about 18\% of the detected sample) have ratios closer to those found for \ion{H}{2} regions.}
\item{The slope of the logarithmic correlation between `total' mid-infrared flux (24 and 70 $\micron$) and the 1.4 GHz non-thermal radio flux is close to unity (1.10) as found for galaxies. $q{_{\mathrm{MIR}}}$ values were calculated for each fully detected remnant and they range between approximately 0.06 and 2.60.}
\item{Whether the strong 24 $\micron$ emission is the result of line emission or hot dust, it is clear that there is a good morphological association of the 24 $\micron$ and X-ray features in bright X-ray remnants. The mechanism for the 24 $\micron$ emission for these remnants is most likely grains heated by collisions in the hot plasma. The morphology of this mid-infrared emission is also generally distinct from the other infrared wavelengths which implies that the emission at 8 and 70 $\micron$ has a different origin. }
\item{We present SEDs (radio, infrared, X-ray) for a sample of remnants and show that the energy released in the infrared is comparable to the cooling in the X-ray range. Moreover, IRX seems to increase with age.}
\end{itemize}

%__________________________________________________________________

\begin{acknowledgements}
This work was based on observations made with the Spitzer Space Telescope, which is operated by the Jet Propulsion Laboratory (JPL), California Institute of Technology under a contract with NASA. Support for this work was provided by NASA in part through an award issued by JPL/Caltech and by the Natural Sciences and Engineering Research Council of Canada. The National Radio Astronomy Observatory is a facility of the National Science Foundation operated under cooperative agreement by Associated Universities, Inc. This research is supported as part of the International Galactic Plane Survey through a Collaborative Research Opportunities grant from the Natural Sciences and Engineering Research Council of Canada.
The MOST is operated by The University of Sydney with support from the Australian Research Council and the Science Foundation for Physics within The University of Sydney. This paper benefited from VLA archival data from The Multi-Array Galactic Plane Imaging Survey (MAGPIS) as well as Chandra archival data which was obtained in the online Chandra Supernova Remnant Catalog and is maintained by Fred Seward (SAO). The authors would like to thank Crystal Brogan for the 20 and 90 cm high resolution VLA images of the SNR G39.2-0.3 and Dae-Sik Moon and Lu\'{\i}s Be\c{}ca for useful discussions. Finally, we acknowledge the referee for valuable comments and corrections which improved this manuscript.
\end{acknowledgements}

%::::::::::::::::::::::::::::::::::::::::::::::::TABLES::::::::::::::::::::::::::::::

\begin{deluxetable}{lccccccclccccccc}
\tabletypesize{\scriptsize}
\tablecaption{Infrared detection classification levels of SNRs from Green's catalog  \label{table:level}}                    
\tablehead{
\colhead{l+b} & \colhead{Name} & \colhead{Size\tablenotemark{a}} &
\colhead{IA\tablenotemark{b}} & \colhead{IS\tablenotemark{c}} & \colhead{GR\tablenotemark{d}} & \colhead{M\tablenotemark{e}}& 
& 
\colhead{l+b} & \colhead{Name} & \colhead{Size} &
\colhead{IA} & \colhead{IS} & \colhead{GR} & \colhead{M}}
\startdata
   10.5-0.0 & & 6 & - & - & - & 1 &&  54.4-0.3 & HC40 & $40$ & 3& 4 & $1$ & 1\\
   11.0-0.0 & & $9\times11$ & - & - & - & 3 &&  55.0+0.3 & & $17$& - & -& $2$ & 2 \\ 
   11.1-0.7 & & $11\times7$ & - & - &- &  3 && 57.2+0.8 & 4C21.53 & $12?$ & 4& 4 & $4$ & 3 \\
   11.1+0.1 & & $12\times10$ & - & -& - & 1 && 59.5+0.1 & & $15$ & - & -& $3$ & 3 \\
   11.2-0.3 & & $4$ & 1 & 2 & 1 & 1 && 296.1-0.5 & & $37\times25$ & 2 & 4 & 3 & 3 \\ 
   11.4-0.1 & & $8$ & 4 & 4 &3 & 3 && 296.8-0.3 & & $20\times14$ & 4 & 4 & 3 & 1 \\
   11.8-0.2 & & 4 & - & -& - & 2 && 298.5-0.3 & & $5?$ & 4 & 4 & 2 & 2\\
   12.0-0.1 & & $7$&  1 & 1 & 3 & 3 && 298.6-0.0 & & $12\times9$ & 3 & 4 & 2 & 2\\
   12.2+0.3 & & $5\times6$ & - & - &  - & 3 && 299.6-0.5 & & 13 & - & - &3 & 4\\
   12.5+0.2 & & $5\times6$ & - &-& - & 1 && 302.3+0.7 & & 17 & 3 & 4 & 3 & 2\\
   12.7-0.0 & & 6 & - & - &-& 3 && 304.6+0.1 & Kes 17 & 8 & 1 & 1 & 1 & 1\\ 
   12.8-0.0 & & 3 & - & - &-& 3 && 308.1-0.7 & & 13 & - & - & 4 & 3\\
   13.5+0.2 & & $5$ & - & -&3 & 3 && 308.8-0.1 & & $30\times20?$ & - & 4 & 2 & 2\\ 
   14.1+0.1 & & $6\times5$ & - &-& - & 1 &&  309.2-0.6 & &$15\times12$ & 4 & 4 & 3 & 3 \\ 
   14.3+0.1 & & $5\times4$ & - & -&- & 1 && 309.8+0.0 & &$25\times19$ & 4 & 4 & 3 & 3 \\
   15.4+0.1 & & $14\times15$ & - &-& - & 3 && 310.6-0.3 & Kes 20B & 8 & - & - & 2 & 3\\
   15.9+0.2 & & $6$ & 4 & 4& 3 & 1 && 310.8-0.4 & Kes 20A & 12 & - & - & 1 & 1 \\
   16.0-0.5 & & $15\times10$ &-& - & - & 3 && 311.5-0.3 & & 5 & 4 & 4  & 1 & 1\\
   16.4-0.5 & & 13 & - & - & -&1 && 312.4-0.4 & & 38 & 2 &  4 & 3 & 3\\
   16.7+0.1 & & $4$ & - & 4 & 3 &  2 && 315.4-0.3 & & $24\times13$ & 2 & 2 & 2 & 2\\
   17.0-0.0 & & 5 & - & - & - & 2 && 315.9-0.0 & & $25\times14$ & - & - & 3 & 3\\
   17.4-0.1 & & 6 & - & - & - & 3 &&  316.3-0.0 & MSH 14-57 & $29\times14$ & 3 & 4 & 3 & 3\\
   18.1-0.1 & & 8 & - & - & -& 4 && 317.3-0.2 & & 11 & - &- &  3 & 2 \\
   18.6-0.2 & & 6 & - & - &-& 1  && 318.2+0.1 & & $40\times35$ &-& -& 3 & 2\\
   18.8+0.3 & Kes 67  & $14$ & 2 & 4 & 3 & 3 && 318.9+0.4 & & $30\times14$ & - & - & 3 & 2\\
   19.1+0.2 & & 27 & - & - & -& 3 && 321.9-0.3 & & $31\times23$ & 4 & 3 & 3 & 3\\
   20.0-0.2 & & $10$ & 4 & 4 & 3 & 3 && 322.5-0.1 & & 15 &-& -& 3 & 3 \\
   20.4+0.1 & & 8 & - & - & -& 1 && 323.5+0.1 & & 13 & 2 & 1 & 2 & 2\\ 
   21.0-0.4 & & $9\times7$ & -& - & - & 3 && 327.2-0.1 & & 5 & - & -& - & 3\\ 
   21.5-0.9 & & $4$ & 4 & 4 & 3 & 1 && 327.4+0.4 & Kes 27 & 21 & 3 & 4 & 2 & 2\\
   21.5-0.1 & & 5 & - & - & -& 1 && 328.4+0.2 &  MSH15-57 & 5 & 4 & 4 & 4 & 3\\ 
   21.8-0.6 & Kes 69 & $20$ & 3 & 4 & 1 & 1 && 329.7+0.4 & & $40\times33$ & - & - & 2 & 2\\
   22.7-0.2 & & $26$ & 3 & 4 &1 & 2 && 332.0+0.2 & & 12 & 3 & 4 & 4 & 3\\
   23.3-0.3 & W41 & $27$ & 3 & 4& 2 & 2 && 332.4-0.4 & RCW 103 & 10 & 4 & 4 & 1 & 1\\
   23.6+0.3 & & $10$&  1 & 2 & 3 & 1 && 332.4+0.1 & Kes32 & 15 & 2& 3 & 2 & 2\\
   24.7-0.6 & & $15$ & 4 & 4 & 4 & 4 && 335.2+0.1 & &21 & 4 & 4 & 2 & 2\\
   24.7+0.6 & & $21$ & 4 & 4 & 3 & 2 && 336.7+0.5 & & $14\times10$ & 4 & 4 & 4 & 1\\
   27.4+0.0 & Kes 73 & $4$ & 3  & 1 & 3 & 1 && 337.0-0.1 & CTB 33 & 1.5 & 4 & 4 & 3 & 3 \\
   27.8+0.6 & & $50\times30$ & 4 & 4 & $3$ & 3 && 337.2-0.7 &  & 6 & 4 & 4 & 4 & 1\\ 
   28.6-0.1 & & $13\times9$ & - & -& $3$ & 1 && 337.2+0.1 & & $3\times2$ & - & - & - & 3\\ 
   29.6+0.1 & & $5$ & - &  -& $4$ & 3 && 337.8-0.1 & Kes 41 & $9\times6$ & 4 & 4 & 2 & 2 \\
   29.7-0.3 & Kes 75 & $3$ & 4& 4 & 3 & 1 && 338.1+0.4 & & $15?$ & 3 & 4 & 4 & 3\\ 
   31.5-0.6 & & $18$ & -&- & $3$ & 2 && 338.3-0.0 & & 8 & 4 & 4 & 3 & 3 \\%
   31.9+0.0 & 3C 391 & $6$ & 1 & 1& $1$ & 1 && 338.5+0.1 & & 9 & 3 & 4 & 3 & 3\\ 
   32.1-0.9 & & $40$ & -& - & $3$ & 3 && 340.4+0.4 & & $10\times7$ & 4& 4 & 4 & 3\\ 
   32.4+0.1 & & 6 & - & - & -& 2 && 340.6+0.3 & & 6 & 1 & 1 & 2 & 1 \\
   32.8-0.1 & Kes 78 & $17$ & 3 & 1 & 3 & 3 && 341.2+0.9 & & $16\times22$ & - & - & 4 & 3 \\
   33.2-0.6 & &$18$ & 2 & 1 & 3 & 2 && 341.9-0.3 & & 7 & 3 & 4 &  4 & 3\\ 
   33.6+0.1 & Kes 79 & $10$ & 2 & 2 & 2 & 1 && 342.0-0.2 & & $12\times9$ & 2 & 4 & 3 & 3\\ 
   34.7-0.4 & W44 & $31$ & 1 & 4 & $1$ & 1 && 342.1+0.9 & & $10\times9$ & - &-& 4 & 3\\
   35.6-0.4 &  & $15\times11$& - & -& - & 1 && 343.1-0.7 & & $27\times21$ & - & -&  3 & 3 \\
   36.6-0.7 & & $25?$ & - & 1 & $2$ & 3 && 344.7-0.1 & & 10 & 3 & 4 & 1 & 1\\ 
   39.2-0.3 & 3C 396 & $7$ & 4 & 4 & $1$ & 1 && 345.7-0.2 & & 6 & - & -& 4 & 1 \\
   40.5-0.5 & & $22$ & 2 & 4 &$4$ & 2 && 346.6-0.2 & & 8 & 4 & 4 & 1 & 2 \\
   41.1-0.3 & 3C 397 & $4.5\times2.5$ & 3 & 4& $1$ & $1$ && 347.3-0.5 & & $65\times55$ & - & -& 3 & 3\\
   42.8+0.6 & & $24$ & - & 4& $4$ & 4 && 348.5-0.0 & & 10 & - &  - & 1 & 1\\
   43.3-0.2 & W49B & $4\times3$ & 1 & 1& $1$ & 1 && 348.5+0.1 & CTB 37A & 15 & 2 & 4 & 1 & 1\\
   45.7-0.4 & & $22$ & - & 4 & $2$ & 2 && 348.7+0.3 & CTB 37B & $17?$ & 1 & 3 & 3 & 1\\ 
   46.8-0.3 & HC30 & $17\times13$ & 3 & 1 & $3$ & 3 && 349.2-0.1 & & $9\times6$ & - & - & 3 & 3\\ 
   49.2-0.7 & W51C & $30$ & 4 & 1 & $3$ & 3 && 349.7+0.2 & & $2.5\times2$ & 1 & 1 &  1 & 1\\ 
   54.1+0.3 & & $1.5$ & 1 & 1& $3$ & 3 \\ 
\enddata
\tablenotetext{a}{The radio sizes are given in arc-minutes and were taken from Green's catalog. When two dimensions are reported they refer to the major and minor axis of the ellipse.}
\tablenotetext{b}{IRAS survey: Arendt 1989 , similar classification scheme as the one employed for this work}
\tablenotetext{c}{IRAS survey: Saken et al. 1992. Detection levels 1 to 4 explained in the text. For the \cite{1992ApJS...81..715S} survey, detection classification is as follows: Y (detected), N (not detected), P (possible detections) and ? (not conclusive). We have translated these to the same classification used in this work for ready comparison (Y:1, P:2, ?:3, N:4).}     
\tablenotetext{d}{GLIMPSE survey: Reach et al. 2006. Similar classification scheme as the one employed for this work}
\tablenotetext{e}{MIPSGAL survey: this paper}
\end{deluxetable}

\begin{deluxetable}{llrrrrrrrrl}
\tabletypesize{\footnotesize}
\tablewidth{0pt}
\tablecaption{Characteristics of detected SNRs in the MIPSGAL survey\tablenotemark{a}\label{table:1}}                    
\tablehead{
\colhead{l+b} & \colhead{Name\tablenotemark{b}} & \colhead{S${_{\mathrm{rad}}}(^{\prime})$} & \colhead{S${_{\mathrm{phot}}}(^{\prime})$} &
\colhead{F$_{8}$(Jy)} & \colhead{F$_{24}$(Jy)} & \colhead{F$_{70}$(Jy)} & 
\colhead{8/24} & \colhead{8/70} & \colhead{24/70} & \colhead{Region\tablenotemark{c}}}  
\startdata
10.5-0.0 & 10.5 & 6 & 4.8 & $14\pm2$ & $19\pm2 $ &  $227\pm47$ & 0.75 & 0.06 & 0.08 & whole \\ 
11.1+0.1 & 11.1 & $12\times10$ & $5.4\times3.8$ & $12\pm2$ & $16\pm2$ & $297\pm60$ & 0.71 & 0.04 & 0.06 & central region\\
11.2-0.3 & 11.2 & $4$ & 4.8 & $4.6\pm1.1$ &  $40\pm5$ & $101\pm23$ & 0.12 & 0.05 & 0.40 & whole \\
12.5+0.2 & 12.5 & $6\times5$ & 1.5 & $0.9\pm0.2$ & $1.5\pm0.2$ & $7.4\pm1.7$ & 0.64 & 0.13 & 0.20 & northeastern region \\ 
14.1-0.1 & 14.1& $6\times5$ & $1$ &  $0.9\pm0.1$ &$ 10\pm1$ &$ 19\pm4$ & 0.09 & 0.54 & 0.05 & northwestern region\\
14.3+0.1 & 14.3 & $5\times4$ & 4 & $7.9\pm1.1$ & $30\pm4$ & $171\pm36$ & 0.26 & 0.05 & 0.18 & whole\\
15.9+0.2 & 15.9 & $7\times5$ & 6.5 & $3.2\pm2.5$ & $16\pm2$ & $55\pm16$ & 0.20 & 0.06 & 0.30 & whole \\
16.4-0.5 & 16.4 & 13 & $6.5\times4.2$ &  $40\pm5$ & $93\pm10$ & $950\pm192$ & 0.43 & 0.04 & 0.10  & northern arc\\ 
18.6-0.2 & 18.6 & 6 & 7 & $26\pm4$ & $58\pm6$ &$ 554\pm115$ & 0.44 & 0.05 & 0.10 & whole\\
20.4+0.1 & 20.4 & 8 & 8.2 &  $43\pm5$ & $79\pm9$ & $1160\pm230$ &0.55 &0.04&  0.07 & whole\\ 
21.5-0.1 & 21.5-0.1 & 5 & 5.2 &$ 12\pm3 $ & $39\pm4$ & $341\pm69$ & 0.31  & 0.04 & 0.11 & whole\\
21.5-0.9 & 21.5-0.9& 4 & 1.6 & $0.4\pm0.1$ & $0.6\pm0.1$ & $3.5\pm0.8$ &0.63 & 0.11 & 0.18 &  central region \\
21.8-0.6 & Kes69C & 20 & $8.9\times6.8$ & $38\pm6$ & $23\pm3$ & $458\pm93$ & 1.6 & 0.08 & 0.05 & central region\\
23.6+0.3 & 23.6 & 10 & $15.6\times6.2$ &$173\pm18$ & $355\pm36$ &$ 5440\pm1090$ & 0.49 & 0.03 &  0.07 & whole\\
27.4+0.0 & Kes 73 & 4 & 4.8 & $3.0\pm1.2$ & $37\pm4$ & $87\pm18$ &0.08 &0.04 & 0.42 & whole\\
28.6-0.1 & 28.6A& & 3.6 &  $7.2\pm1.0$ & $12\pm2$ & $294\pm65$ &  0.62 & 0.03 & 0.04 & nothwestern arc\\
             & 28.6B & & 2.8  &  $\mathrm{<}2.0$ &$ 3.8\pm0.5$ &  $\mathrm{<}35$ & $<$0.53 & - &0.11$>$  & southern filaments\\
             & 28.6C & & 4.1  &  $\mathrm{<}1.9$ &$ 5.7\pm0.8$ &  $\mathrm{<}23$ & $<$0.32 & - & 0.25$>$ & eastern filament\\
29.7-0.3 & Kes75S & 3 & $1.4\times0.9$  &  $0.7\pm0.2$ & $4.8\pm0.5$ & $\mathrm{<}29$ & 0.15 & 0.17$>$ & 0.03$>$ & southern shell\\  
         & Kes75W &   & $0.3\times1.1$  &  $0.6\pm0.1$ & $2.9\pm0.3$ & $5.3\pm1.1$ & 0.22 & 0.55 & 0.12 & western shell\\
         & Kes75  &     &            &  $1.3\pm0.3$ & $7.8\pm0.8$ & $\mathrm{<}34$ & 0.17 & 0.23$>$ & 0.04$>$ & whole (both shells) \\ 
31.9+0.0 & 3C 391 & $7\times5$ & 6.2 & $10\pm3$& $39\pm4$ & $395\pm81$ & 0.27 & 0.03 & 0.10 & whole\\
          & 3C391BML & & 0.8 & $0.6\pm0.1$ & $0.9\pm0.1$ & $13\pm3$ & 0.74 & 0.05&0.06  & BML\\
           & 3C391NWBar && $0.7\times0.2$ & $0.2\pm0.1$ & $0.8\pm0.1$ & $3.9\pm1.0$& 0.25&0.05 &0.22 & northwestern bar\\ 
33.6+0.1 & Kes 79  & 10 &  8.4  &  $18\pm6$ &  $45\pm5 $ &  $577\pm116$ & 0.40 & 0.03 & 0.08 & whole\\
            & Kes79sf & & $1.7\times0.4$ & $\mathrm{<}0.4$ & $1.2\pm0.2$ & $\mathrm{<}10$ & $<$0.36 & - & 0.11$>$& southern filaments\\
34.7-0.4 & W44 & $35\times27$ & $38\times28$ &$384\pm44$ & $348\pm39$& $6330\pm1300$ & 1.10 & 0.06 & 0.06 & whole\\
35.6-0.4 & 35.6 & $15\times11$ & 11 & $44\pm8$ &$ 36\pm5$ & $487\pm99$ &   1.21 & 0.09 & 0.07 & whole\\
39.2-0.3 & 3C 396 &  $8\times6$ & $3.5\times2.5$  & $1.7\pm0.8$ & $8.5\pm1.0$ & $34\pm11$ & 0.20 & 0.05 & 0.25 & whole \\
            & 3C396SE &  & 1.4  & $3.5\pm2.3$ & $5.26\pm0.6$ & $158\pm32$ & 0.66 & 0.02 & 0.03 & southeastern plume\\
            & 3C396NW &  & $1.4\times2.8$ & $0.7\pm0.2$ & $4.9\pm0.6$ & $\mathrm{<}22$ & 0.14 & 0.02$>$ & 0.14$>$ &  northwestern arc\\
41.1-0.3 & 3C 397  & $4.5\times2.5$ & $5\times3.6$  &  $4.3\pm0.8$& $17\pm2$ & $34\pm14$ & 0.25 & 0.13 & 0.50 & whole\\ 
         & 3C397SE & & $0.3$ & $<0.1$ & $0.6\pm0.1$ & $0.6\pm0.1$ & $<0.18$ & $<0.17$ & $0.95$ & ionic shock region\\
43.3-0.2 & W49B  & $4\times3$ & 5  & $5.7\pm0.7$ & $77\pm8$ & $529\pm107$ & 0.07 & 0.01 & 0.15  & whole\\
                 & W49BMol & & $0.6\times0.3$ & $0.3\pm0.1$ & $0.6\pm0.1$ & $6.7\pm1.4$ &  0.52&0.04 &0.08 & molecular shock region\\

        & W49BIon & & $0.6\times0.3$ & $0.2\pm0.1$ & $3.1\pm0.3$ & $4.3\pm0.9$ & 0.09& 0.03 & 0.28& ionic shock region\\ 
54.4-0.3 & HC40NE & 40 & 2.8 & $5.9\pm0.7$ & $2.9\pm0.4$ & $38\pm8$ & 2.04 & 0.16 &0.08 & northeastern region \\
            & HC40TR & 40 & $4.7\times2.1$ & $3.8\pm0.5$& $8.0\pm0.9$ & $131\pm27$ & 0.47 & 0.03 & 0.06 & north top region\\
296.8-0.3 & 296 & $20\times14$ & $4\times1.8$& $\mathrm{<}0.6$ &$ 1.5\pm0.2$ &$\mathrm{<}11$ & $<$0.39 & - & 0.14$>$ & western arc \\
304.6+0.1 & Kes 17 & 8 & 8  & $17\pm2$ &$ 17\pm2$ & $149\pm31$ & 1.02 & 0.12 & 0.11 & whole\\
310.8-0.4 & Kes20Ana & 12 &1.1 &$0.3\pm0.1$ & $1.4\pm0.2$ & $20\pm4$ &0.21 & 0.02 & 0.07 &northern arc\\
                 & Kes20Ase & 12 & 1.2 & $0.8\pm0.1$ & $0.8\pm0.1$ & $\mathrm{<}17$ & 1.1 & 0.05$>$& 0.04$>$ &southeastern ridge\\
311.5-0.3 & 311& 5 & 4 & $2.1\pm0.4$ & $3.5\pm0.5$ & $58\pm14$ & 0.60 & 0.04 & 0.06 & whole \\
332.4-0.4 & RCW 103 & 10 & $10\times8.7$ & $51\pm6$ & $101\pm11$ & $648\pm134$ &  0.50 & 0.08 & 0.16 & whole\\
          & RCW103M & & $1.9\times0.4$ & $0.9\pm0.3$ & $5.9\pm0.6$ & $10\pm3$ & 0.16 & 0.09 & 0.58 & molecular region\\
336.7+0.5 & 336 & $14\times10$ & 2.3  & $0.6\pm0.2$ & $2.4\pm0.3$ & $6.3\pm2.5$ & 0.25 & 0.09 & 0.38 & bow\\ 
337.2-0.7 & 337 & 6 & 5.4 & $2.4\pm0.6$ & $6.3\pm0.8$ &$ 13\pm5$ & 0.38 & 0.18 & 0.48 & whole \\ 
340.6+0.3 & 340 & 6 & 5.4 & $14\pm2$ &  $19\pm2$ & $193\pm40$ & 0.75 & 0.07 & 0.10 & whole\\ 
344.7-0.1 & 344 & 10 & 9 & $4.8\pm1.9$ & $24\pm3$ & $117\pm28$ & 0.20 & 0.04 & 0.21 &whole\\
345.7-0.2 & 345 & 6 & 1.8 & $\mathrm{<}0.4$  & $1.7\pm0.2$ & $4.8\pm1.0$ & $<$0.13 & 0.05 & 0.36 &  southern arc\\ 
348.5-0.0 & 348 & 10 & $3.6\times1.4$ & $1.3\pm0.2$ & $2.0\pm0.3$ & $13\pm3$ & 0.66 & 0.10 & 0.15 & central blob\\ 
348.5+0.1 & CTB37A & 15 & 8 & $27\pm6$ & $27\pm4$ & $336\pm86$ & 0.89 & 0.08 & 0.09  & eastern shell\\ 
348.7+0.3 & CTB37B & 17 & $8.2\times4$& $18\pm3$ & $32\pm4$ & $401\pm82$  & 0.58 & 0.05 & 0.08 & interface\\
349.7+0.2 & 349 & $2.5\times2$ & 2.1& $4.1\pm0.6$ & $32\pm4$ & $240\pm49$ & 0.13 & 0.02 & 0.13  & whole
\enddata
\tablenotetext{a}{\ The color ratios are obtained using the flux density value for the partial or whole region depending on the infrared morphology.  The radio sizes ($S_\mathrm{rad}$) are given in arc-minutes and were taken from Green's catalog. The photometry size ($S_\mathrm{phot}$) refers to the size used for obtaining the infrared fluxes. The integrated fluxes have not been color corrected since the mechanism for the infrared emission is uncertain. Also, no extinction correction has been applied. Furthermore, the 8 $\micron$ fluxes were multiplied by the extended source aperture correction factor as described in \cite{2005PASP..117..978R}.  The tabulated errors are the combination of the 1$\sigma$ errors plus the systematic calibration errors associated to each wavelength. Upper flux limits are given for remnants where point source contamination or apparent lack of emission (3$\sigma$ detection) is present.}
\tablenotetext{b}{\ Name or designation used to identify remnants in Figures \ref{colorplotflux} and \ref{colorplot}.}
\tablenotetext{c}{\ Locations and areas used for photometry of subregions of remnants as identified in the figures in the appendix.}
\end{deluxetable}

\begin{deluxetable}{lccc}
\tablewidth{0pt}
\tablecaption{color temperatures and dust masses based of the 24 and 70 $\micron$ fluxes for selected SNRs \label{mass} \tablenotemark{a}}
\tablehead{
\colhead{SNRs} & \colhead{T${_{\mathrm{24/70}}}$ \tablenotemark{b} } & \colhead{Distance \tablenotemark{c}} &
\colhead{M${_{\mathrm{dust}}}$ \tablenotemark{d}}}
\startdata
G11.2-0.3 & 63 & 4.4 & 0.034 \\
G15.9+0.2  & 60 & 8.5 & 0.081\\
Kes 73 & 63 & 8.5 & 0.11 \\
Kes 75 & 58 & $<$7.5 & $<$0.045 \\
3C 391 & 51 & 8.5 & 1.1\\
Kes 79 & 50 & 7.8 & 1.5 \\
W44 & 48 & 2.8 & 2.5 \\
3C 396 & 58 & 7.7$>$  & 0.046$>$\\
3C 397 & 65 & 7.5$>$  & 0.029$>$\\
W49B & 54 & 10 & 1.6 \\
RCW 103 & 55 & 3.1 &  0.18\\
G337.2-0.7 & 65 & 2.0-9.3 & $<$0.018 \\
G349.7+0.2 & 54 & 14.8  & 1.6  \\
CasA \tablenotemark{e} & 87 & 3.4 & 0.008 \\
IC443 \tablenotemark{e} & 48 & 0.7-2 & $<$0.27 
\enddata
\tablenotetext{a}{\ Color temperatures and dust masses are obtained using a dust emissivity index of $\beta=2$.}
\tablenotetext{b}{\ Approximate values for the dust temperature (K).}
\tablenotetext{c}{\ Distances (kpc) taken from Green's catalog.}
\tablenotetext{d}{\ Dust mass in solar masses (M$_{\sun}$).}
\tablenotetext{e}{\ Cas A and IC443 are included for comparison. Estimates are based on the flux densities obtained by \cite{2004ApJS..154..290H} and Noriega-Crespo et al. (\emph{in preparation}), respectively.}
\end{deluxetable}

\begin{deluxetable}{lrclr}
\tablewidth{0pt}
\tablecaption{F$_{\mathrm{MIR}}$-radio ratios for selected SNRs \label{table:qsnr}}         
\tablehead{
\colhead{Name} &  \colhead{$q{_{\mathrm{MIR}}}$} &
&
\colhead{Name} &  \colhead{$q{_{\mathrm{MIR}}}$}}
\startdata
G10.5-0.0&  2.06 && G35.6-0.4 & 1.37 \\
G11.2-0.3&  0.54 &&  3C396       &    0.06  \\
G14.3+0.1&  2.16 &&  3C397       &    0.10\\
G15.9+0.2&  0.86 && W49B        &    0.83    \\  
G18.6-0.2&  2.24 &&  Kes 17      &    0.69    \\
G20.4+0.1&  2.18 && G311.5-0.3  &    0.91    \\
G21.5-0.1&  2.60 &&   RCW103      &    1.07    \\
G23.6+0.3&  2.43 & &   G337.2-0.7  &    0.83    \\
Kes 73&  1.06 && G340.6+0.3  &    1.22    \\
Kes 75&  0.32 && G344.7-0.1  &    1.41   \\
3C391&  0.87  & & G349.7+0.2  &    0.76   \\
Kes 79 & 1.05 & & CasA    \tablenotemark{a}     &    -1.05 \\
W44 &    1.02 && IC443    \tablenotemark{b}    &    0.53 
\enddata                
\tablenotetext{a}{Values for infrared and radio emission from Cas A were taken from \cite{2004ApJS..154..290H}.} 
\tablenotetext{b}{Values for infrared emission from IC 443 were taken from Noriega-Crespo et al. (\emph{in preparation}).}
\end{deluxetable}

\begin{deluxetable}{lrrr}
\tablewidth{0pt}
\tablecaption{$q{_{\mathrm{IR_{\lambda}}}}$ values for different wavelengths \tablenotemark{a}\label{table:qlambda}}                    
\tablehead{
\colhead{$q{_{\mathrm{IR_{\lambda}}}}$} &  \colhead{All sample} & \colhead{Upper trend} & \colhead{Lower trend}}
\startdata      
$q{_{\mathrm{24}}}$ & $0.71\pm0.42$ & $0.39\pm0.11$ & $1.68\pm0.05$\\
$q{_{\mathrm{70}}}$ & $1.55\pm0.60$ &  $1.17\pm0.19$ & $2.70\pm0.04$\\
\hline
$q{_{\mathrm{MIR}}}$ & $1.19\pm0.53$ & $0.83\pm0.15$ & $2.28\pm0.04$ 
\enddata               
\tablenotetext{a} {\ $q{_{\mathrm{24}}}$ and $q{_{\mathrm{70}}}$ parameters are monochromatic. They are defined as $q{_\mathrm{IR_\lambda}}=\log(F{_\mathrm{IR_{\lambda}}}/F_\mathrm{21cm})$ where F${_\mathrm{IR_\lambda}}$ and F$_\mathrm{21cm}$ are the flux densities (in Jy) at specific mid-infrared wavelengths and in the radio.  $q{_{\mathrm{MIR}}}$ is a dimensionless parameter which represents the ratio of mid-infrared to radio (it is defined in Eq. \ref{qmir}). Individual $q{_{\mathrm{MIR}}}$ values for each remnant are presented in the Table~\ref{table:qsnr}.}
\end{deluxetable}

\begin{deluxetable}{lcccr}
\tablewidth{0pt}
\tablecaption{Infrared to X-ray ratios for selected SNRs\label{table:xray}}                    
\tablehead{
\colhead{Name} &  \colhead{Total MIR Flux} & \colhead{X-ray Flux$_{(0.3-10~\mathrm{KeV})}$ \tablenotemark{a}} & \colhead{IRX} & \colhead{Age\tablenotemark{b}}\\
& \colhead{(10$^{-9}$ erg cm$^{-2}$ s$^{-1}$)} & \colhead{(10$^{-9}$ erg cm$^{-2}$ s$^{-1}$)} & & }
\startdata     
G11.2-0.3   &    11 & 4.0 & 2.8 & 1600 (1)  \\ 
G15.9+0.2   &    5.3 & 0.96 & 5.6 & 500-2400 (2)\\
Kes73  &    9.9 & 1.8 & 5.6 & 500-1000 (3)\\ 
Kes75       &  2.7 & 0.21 & 13 & $884<$ /$>2400$ (4)\\ 
3C391    &   20 & 0.62 & 32 & ~4000 (5)\\ 
Kes79     &  16 & 0.17 & 93 & ~6000 (6) \\ 
W44\tablenotemark{c}  & 2.0 & 0.093 & 240 & ~20000 (7) \\ 
3C396       &   3.0 & 0.19 & 16 & 7100  (8) \\ 
3C397       &    8.1 & 1.3 & 6.2 &  ~5300 (9) \\ 
W49B    &  30 & 6.0 & 5.1 & ~1000-4000 (10) \\ 
RCW103  & 27 & 17 & 1.6  & ~2000 (11) \\ 
G337.2-0.7   &    1.6 & 0.93 & 1.7 & 750-3500 (12)\\ 
G349.7+0.2    &   14 & 0.33 & 42 & ~2800 (13) 
\enddata 
\tablenotetext{a}{\ X-ray fluxes were retrieved from the \emph{Chandra} Supernova Remnant Catalog.}
\tablenotetext{b}{\ Age is given in years and the values were taken from the following literature: (1) \cite{1977QB841.C58......}; (2) \cite{2006ApJ...652L..45R}; (3) \cite{2008ApJ...677..292T} ; (4) upper limit for the spin-down age of the associated pulsar \cite{2006ApJ...647.1286L}/ based on ionization time-scales \cite{2007ApJ...667..219M}; (5) \cite{2004ApJ...616..885C}; (6) \cite{2004ApJ...605..742S} ; (7) \cite{1991ApJ...372L..99W}; (8)\cite{1999ApJ...516..811H}; (9) \cite{2005ApJ...618..321S}; (10) e.g. \cite{2000ApJ...532..970H}; (11) \cite{1997PASP..109..990C}; (12) \cite{2006ApJ...646..982R} ; (13) \cite{2002ApJ...580..904S}}          
\tablenotetext{c}{\ Infrared and X-ray flux values are for a central rectangular region of W44.}  

\end{deluxetable}

%__________________________________________________________________

%:::::::::::PLOTS::::::::::::::::::::::::::::::::::

%*******************************
\appendix

\section{First Galactic quadrant: 10$^{\circ}$ $\leq$ l $\leq$ 60$^{\circ}$}

\subsection{G10.5-0.0}

\begin{figure}[h*]
\plotone{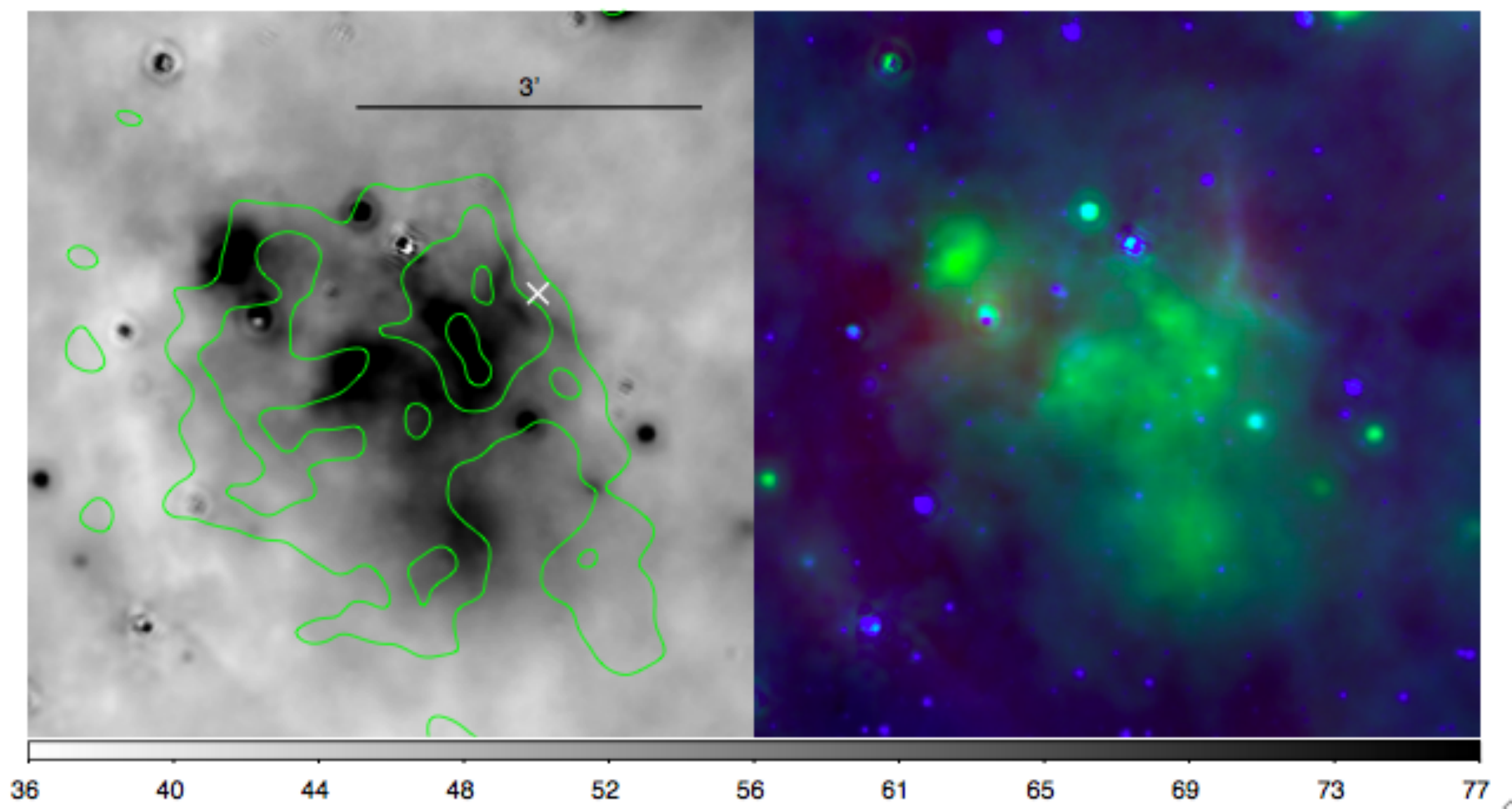}
\epsscale{.9}
\figcaption[f8.pdf]{SNR G10.5-0.0. Contours from VLA 20 cm observations: levels are 13, 17, 21 and 24 mJy/beam. The cross represents the location of the associated X-ray source (AX J180902-1948).\label{snrG10500}}
\end{figure}

First reported in \cite{2006ApJ...639L..25B} (hereafter, B06), this remnant was classified as a class II candidate (sources with the highest confidence index of detection are denominated as class I and the least as class III). The remnant has a possible ASCA (\emph{Advanced Satellite for Cosmology and Astrophysics}) X-ray counterpart \citep{2001ApJS..134...77S}. Figure \ref{snrG10500} shows that the 24 $\micron$ emission roughly matches the radio contours. At 70 $\micron$, it only emits in the northern part and at 8 $\micron$, the emission is confused with the background, although an arc of emission close to the X-ray source is well detected. This feature is also present at 24 and 70  $\micron$. There is a small \ion{H}{2} region inside the remnant which coincides with some of the 24  $\micron$ emission implying that the flux measurement is an upper bound.

\subsection{G11.1+0.1}

\begin{figure}[h*]
\plotone{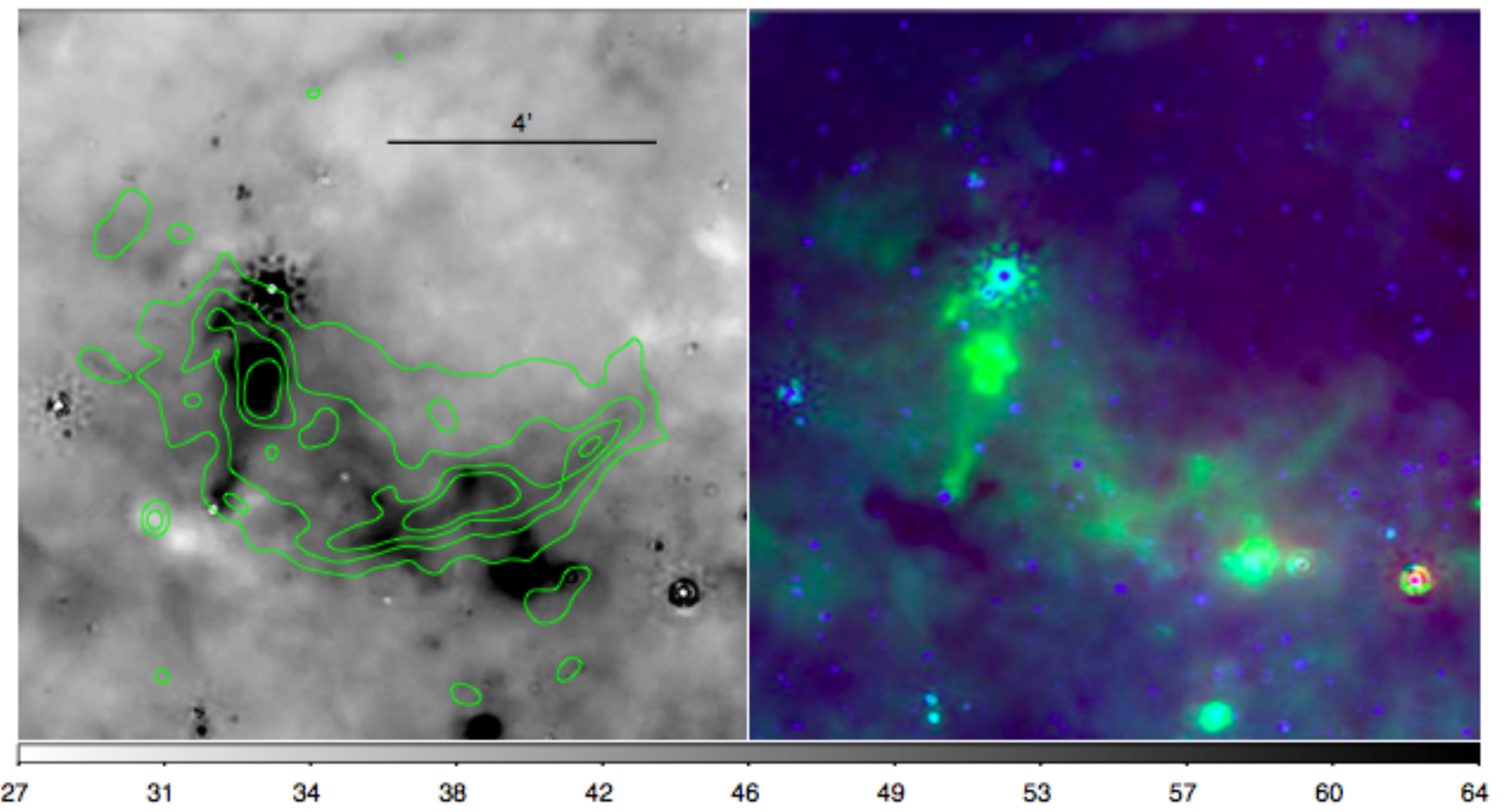}
\epsscale{0.9}
\figcaption[f9.pdf]{SNR G11.1+0.1. Contours from VLA 20 cm observations: levels are 14, 18, 22 and 26 mJy/beam. \label{snrG11101}}
\end{figure}

This is a class I object from B06 with a radio shell morphology. The infrared emission in all wavelengths is easily confused with the background on the southern part of the remnant. However, the brighter radio knots inside the radio contours show strong 24 $\micron$ emission (see Fig.~\ref{snrG11101}).

\subsection{G11.2-0.3}
\begin{figure}[h*]
\plotone{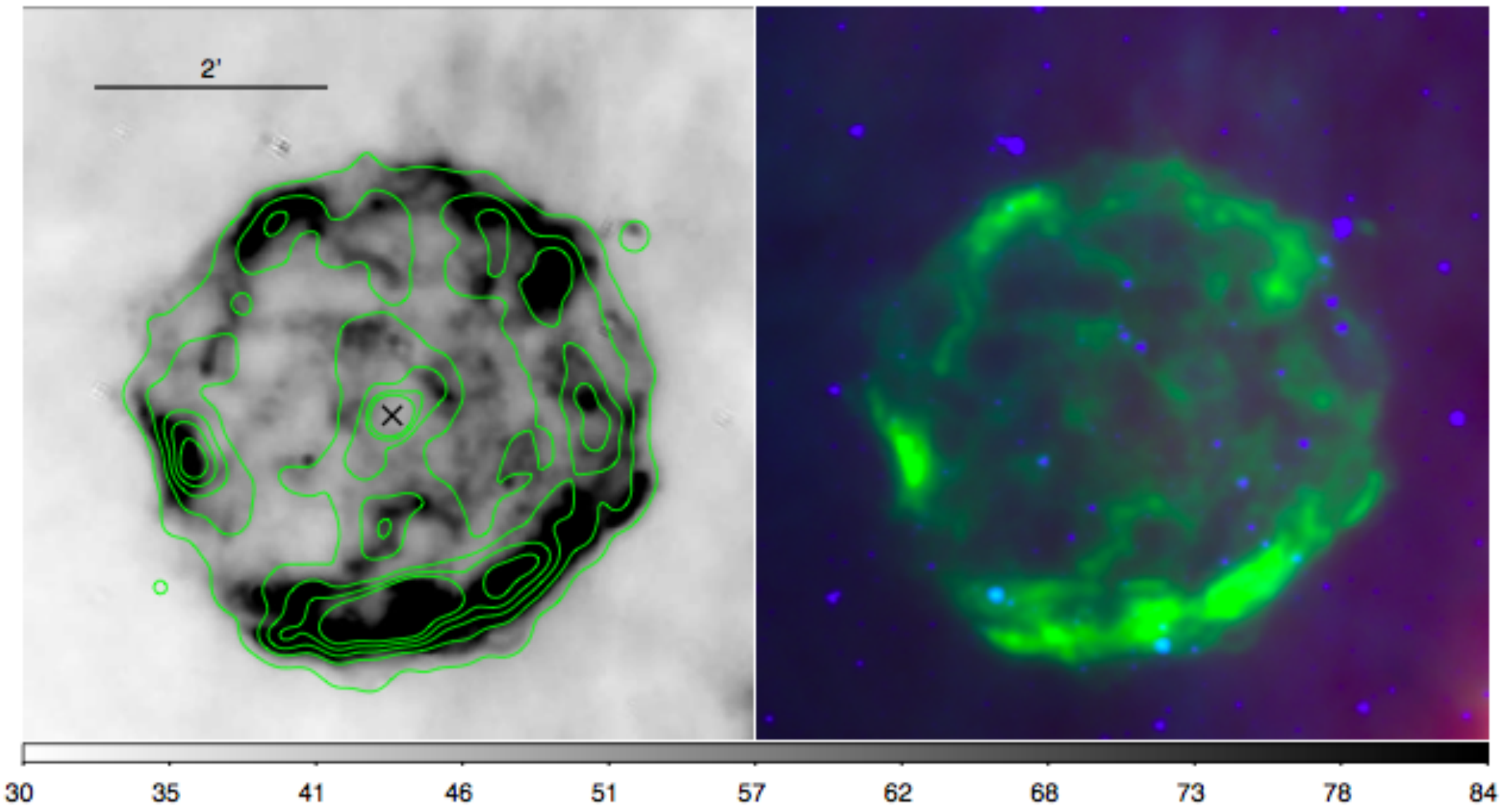}
\epsscale{0.9}
\figcaption[f10.pdf]{SNR G11.2-0.3. Contours from Chandra observations: levels are 2, 6, 9, 13, $16\times10^{-7}$ photons/cm${^2}$/sec/pixel. The cross marks the location of the associated pulsar (AX J1811.5-1926).\label{snrG11203}}  
\end{figure}

This fairly young remnant (1600 yr) contains a millisecond pulsar PSR J1811-1925 at the centre \citep{2001ApJ...560..371K}. It is clearly seen in the radio as a circular shell of approximately $4^{\prime}$ diameter (spectral index of 0.6) and it is located at a distance of 4.4 kpc (\citealp{2009BASI...37...45G}, and references therein). Using staring observations with \emph{Spitzer}, \citet{2009ASPC..414...22R} found Fe emission with temperature and density representative of ejecta material in regions with high Fe content. \cite{2007ApJ...657..308K} detected near-infrared [Fe II] and H{$_2$} filaments within and outside the radio borders of the remnant, respectively. While the H{$_2$} emission is most likely the result of some sort of interaction between the shockwave and the progenitor wind,  the [Fe II] emission is thought to be the combination of ejecta and surrounding ISM with a morphology characteristic of an `asymmetric explosion'. Recently, new near-infrared spectroscopic observations of dense Fe knots \citep{2009ApJ...703L..81M} strengthened the belief that the SN explosion happened along the north-south direction (in Galactic coordinates). The 24 $\micron$ image (Fig.~\ref{snrG11203}) reveals peaks of mid-infrared emission that coincide morphologically with X-ray emission (Fig.~\ref{snrG11203}). Previously detected in the IRAC wavelengths, this remnant shows 8 $\micron$ emission in the eastern part but since it is diffuse, a clear association is not possible. The same diffuse pattern is seen in the 70 $\micron$ image. Both filaments detected in the southern rim at 8 $\micron$ have correlated emission in the 24 and 70 $\micron$ images, although unresolved. The IRAC colors for these filaments suggest molecular emission from shocked gas (\citealp{2006AJ....131.1479R}, hereafter, R06). The blob in the most eastern part of the remnant is detected both at 24 and 70  $\micron$ and so it is most likely due to dust.

\subsection{G12.5+0.2}

\begin{figure}[h*]
\plotone{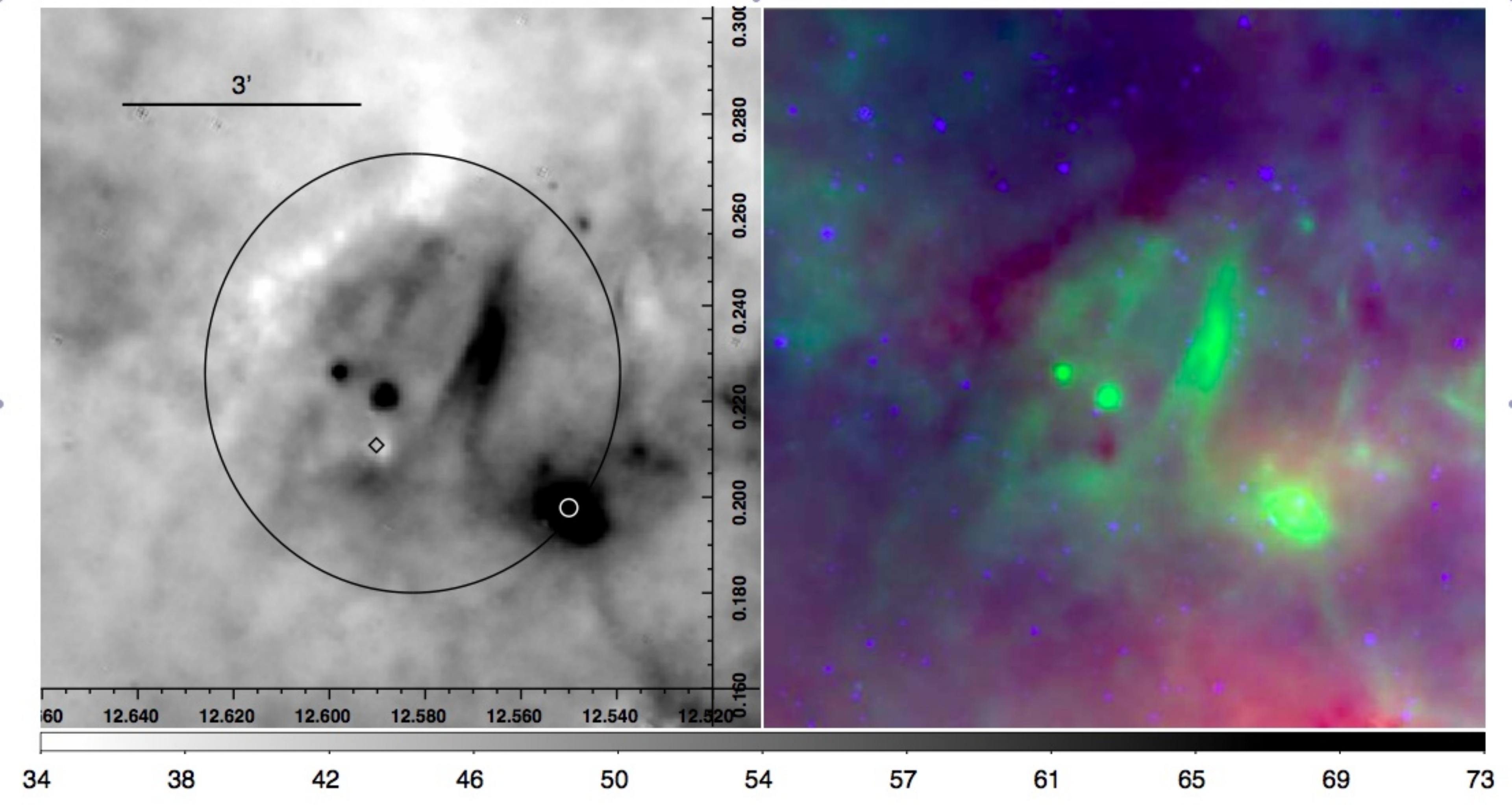}
\epsscale{.9}
\figcaption[f11.pdf]{SNR G12.5+0.2. Positions of a maser (diamond) and nearby infrared IRAS (circle) source are overlaid. Also shown is a circle showing the approximate size and location of the remnant in the radio as seen in Fig.2 from B06.\label{snrG12502}}
\end{figure}

This is a possible composite class II remnant from B06. Figure \ref{snrG12502} shows a striking ridge of 24 $\micron$ emission on the central-western part of the remnant. To the southwest it borders an IR source. Also, filamentary structure is seen towards the east at the boundary with a dark cloud. 

\subsection{G14.1+0.1}
\begin{figure}[h*]
\plotone{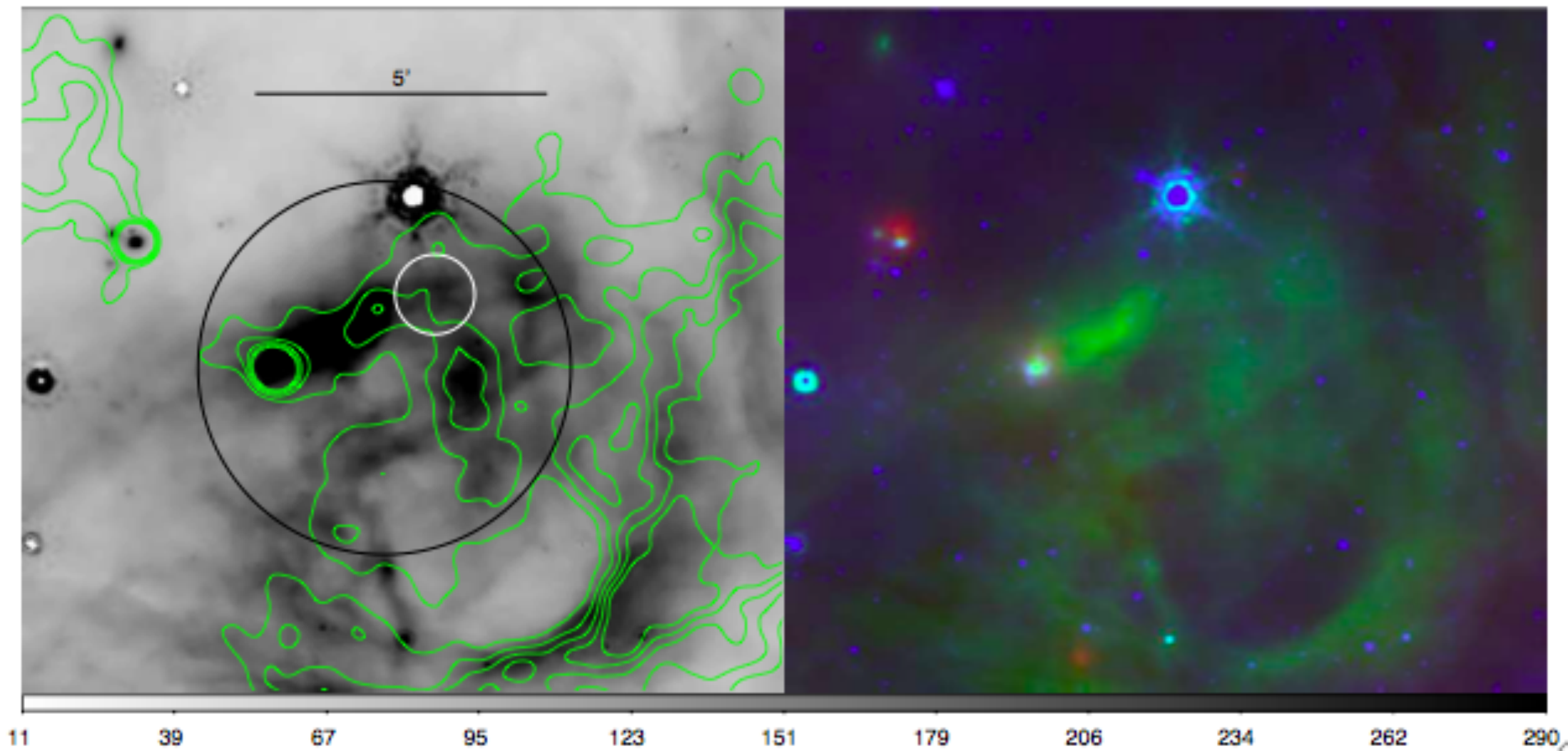}
\epsscale{.9}
\figcaption[f12.pdf]{SNR G14.1-0.1.Contours from VLA 20 cm observations: levels are 18.5 to 28, in steps of 2.4 mJy/beam. The bigger circle (in white) marks the approximate location of the remnant within the confused region (Fig.2 from B06); and the smaller circle (in black) indicates the area used to infer color ratios.\label{snrG14101}}
\end{figure}

This is a class II shell remnant from B06. The 24 $\micron$ emission has a horseshoe shape (Fig.~\ref{snrG14101}) which roughly matches the radio morphology. At 8 and 70 $\micron$ only diffuse emission is seen. The remnant is surrounded by all kinds of infrared filaments which without the radio contours would be hard to distinguish from the overall structure.

\subsection{G14.3+0.1}

\begin{figure}[h*]
\plotone{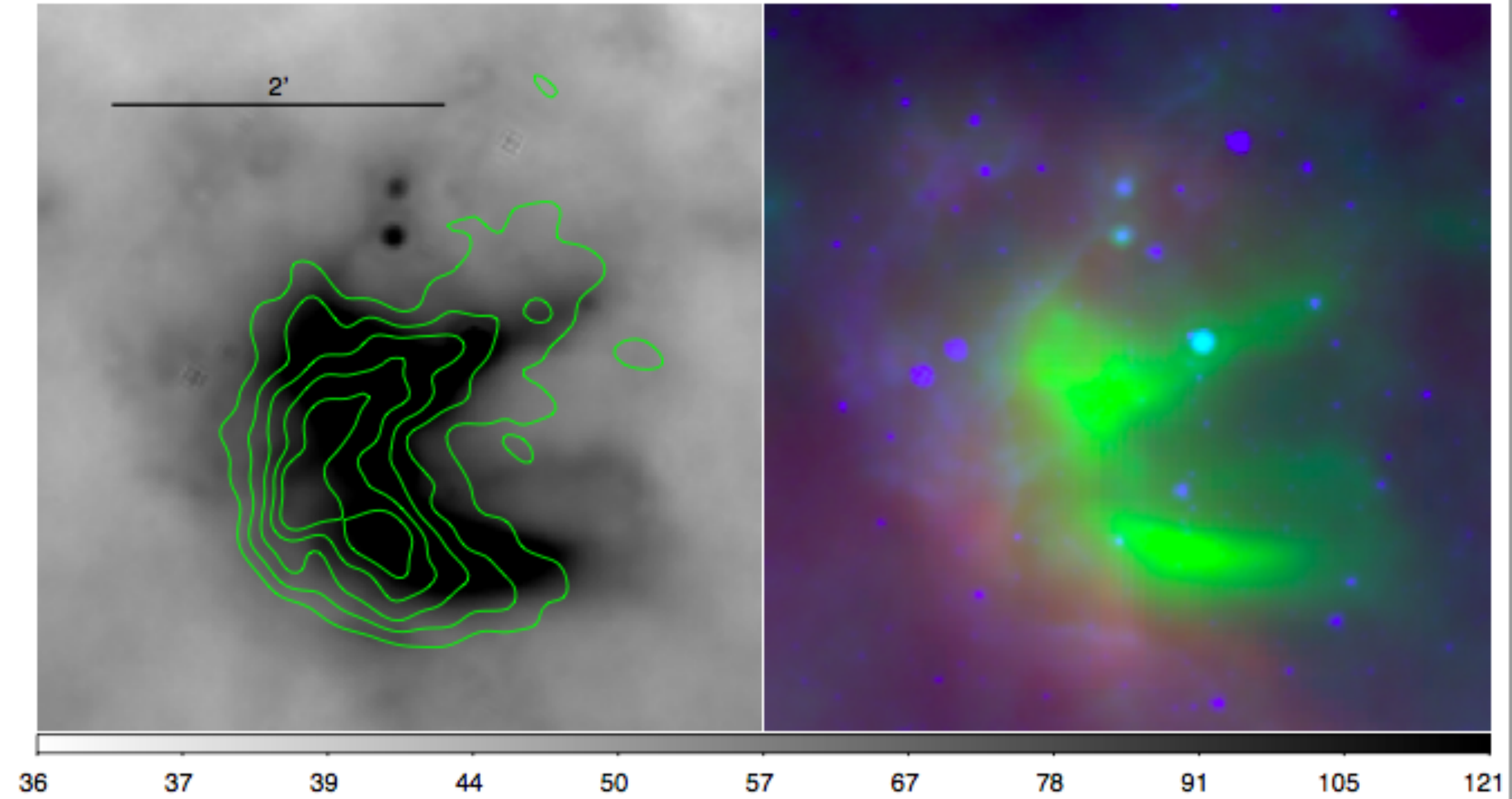}
\epsscale{.9}
\figcaption[f13.pdf]{SNR G14.3+0.1. Contours from VLA 20 cm observations: levels are 16, 20, 24 and 28 mJy/beam. \label{snrG14301}}  

\end{figure}

This is another class II remnant from B06. While the 8 $\micron$ IRAC image shows emission localized in the eastern part, the 70 $\micron$ emission fills the radio contours uniformly. The 24 $\micron$ emission also fills the contours with two peaks located north and south of the remnant's centre (see Fig.~\ref{snrG14301}).

\subsection{G15.9+0.2}
\begin{figure}[h*]
\plotone{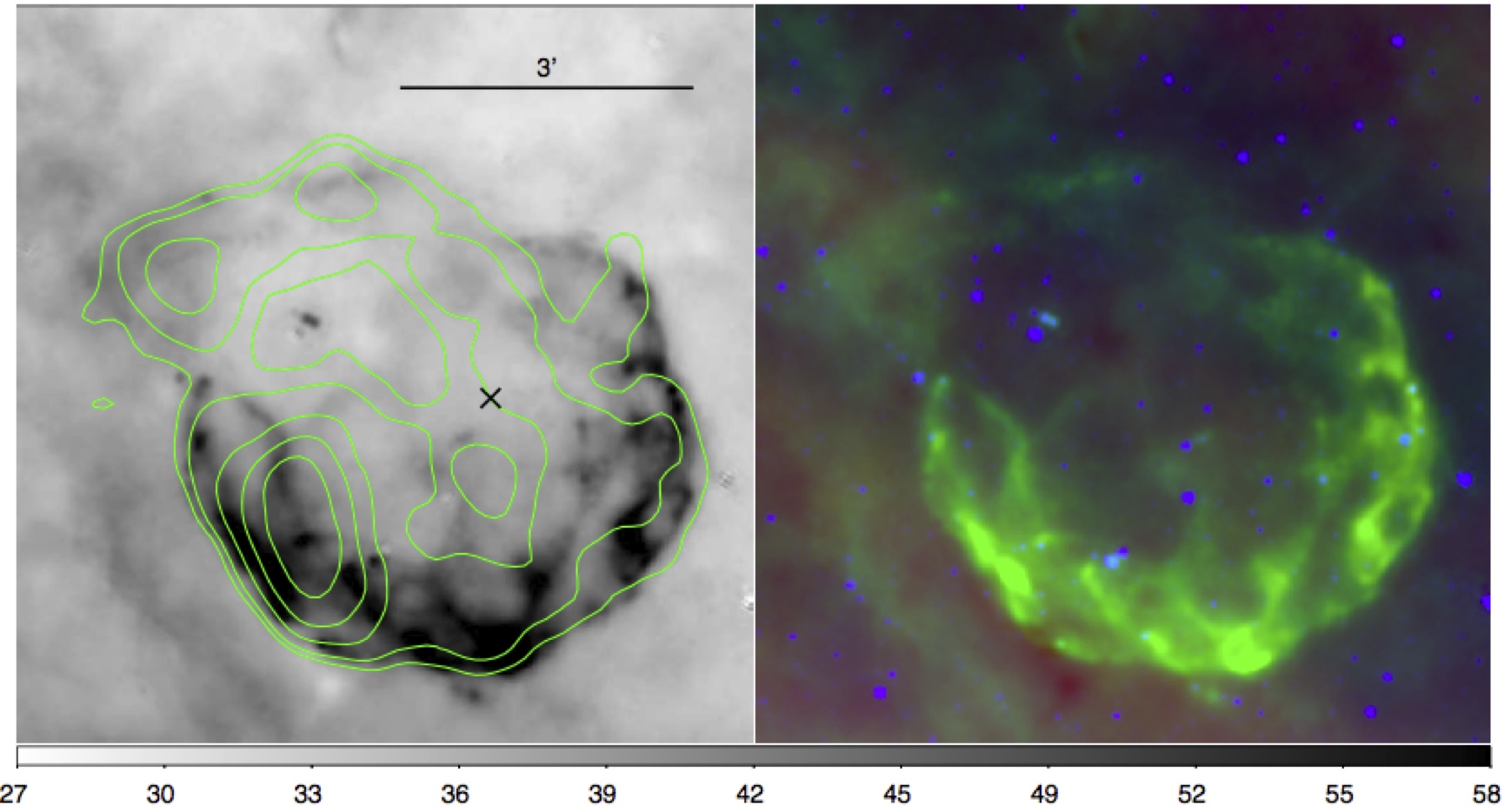}
\epsscale{.9}
\figcaption[f14.jpeg]{SNR G15.9+0.2. Contours from VLA 20 cm observations: levels are 0.05, 0.12, 0.34, 0.7 and 1.2 Jy/beam. The cross marks the location of an X-ray source (CXOU J181852.0-150213)\label{snrG15902}}  
\end{figure}

Chandra observations showed that this remnant is young ($\leq2400$ yr) with an X-ray point source \citep{2006ApJ...652L..45R}. The obtained X-ray spectrum implied high element abundances characteristic of ejected material. The infrared detections are presented in Figure \ref{snrG15902}. This remnant was not detected in either the GLIMPSE or IRAS surveys. The 24 $\micron$ emission is bright in the southern-most part. There are also traces of emission in the north rim. At 8 $\micron$, the emission is very faint and mainly seen around the south-eastern radio contours. The 70 $\micron$ image shows diffuse emission extending along the south-eastern part; nevertheless, it is unclear if this is associated with the remnant. Also, in the same area, the remnant seems to be encountering interstellar material, although no maser emission has been reported \citep{1997AJ....114.2058G}.

\subsection{G16.4-0.5}

\begin{figure}[h*]
\plotone{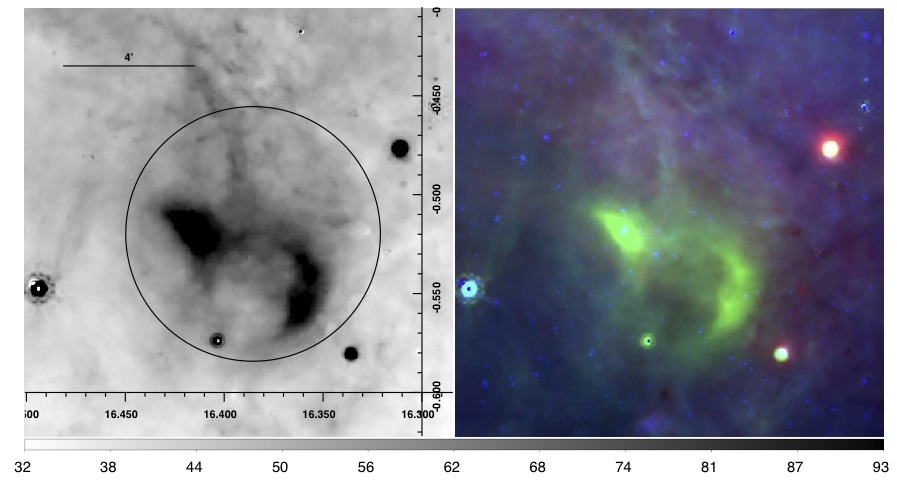}
\epsscale{.9}
\figcaption[f15.jpeg]{SNR G16.4-0.5. Also shown is a circle (in black) which marks the approximate location of the remnant in the radio as seen in Fig.2 from B06.\label{snrG16405}}
\end{figure}

B06 classified this as a class II partial shell. The 24 $\micron$ emission has a shape similar to a fork with the strongest emission coming from its prongs. The western arc feature does not seem to have a strong counterpart at the other infrared wavelengths, unlike the rest of the structure which shows up diffusely (Fig.~\ref{snrG16405}).

\subsection{G18.6-0.2}

\begin{figure}[h*]
\plotone{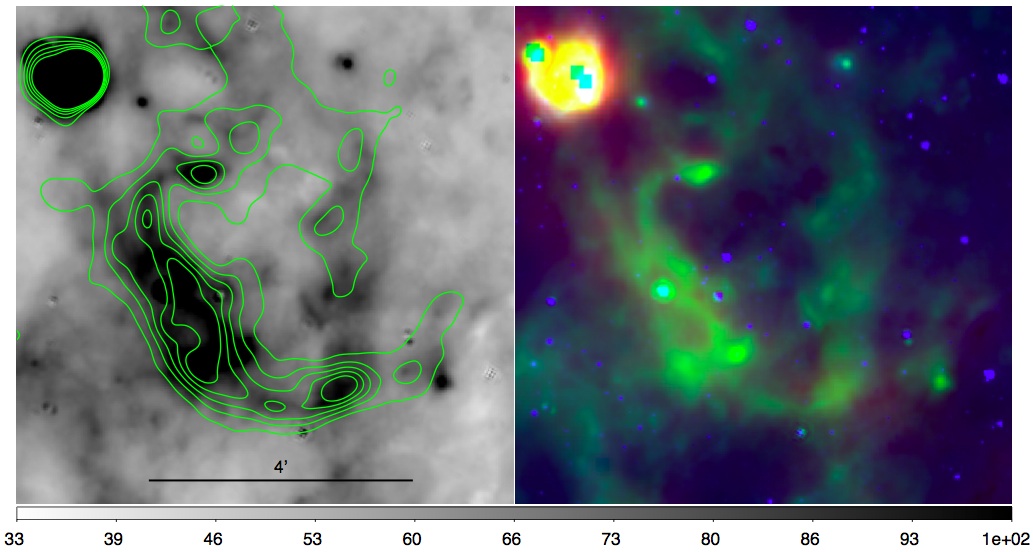}
\epsscale{.9}
\figcaption[f16.jpeg]{SNR G18.6-0.2. Contours from VLA 20 cm observations: levels are from 18 to 30, in steps of 4 mJy/beam.\label{snrG18602}}  
\end{figure}

Another partial shell class II remnant from BG06 is shown in Figure \ref{snrG18602}. There is a good correspondence between the radio contours and the infrared emission, in particular, for 24 $\micron$. At 70 $\micron$ the strongest infrared emission matches the brightest radio peaks and, at 8 $\micron$, there is diffuse emission which seems to occupy those regions as well, but no clear association can be established based on this filter alone.

\subsection{G20.4+0.1}

\begin{figure}[h*]
\plotone{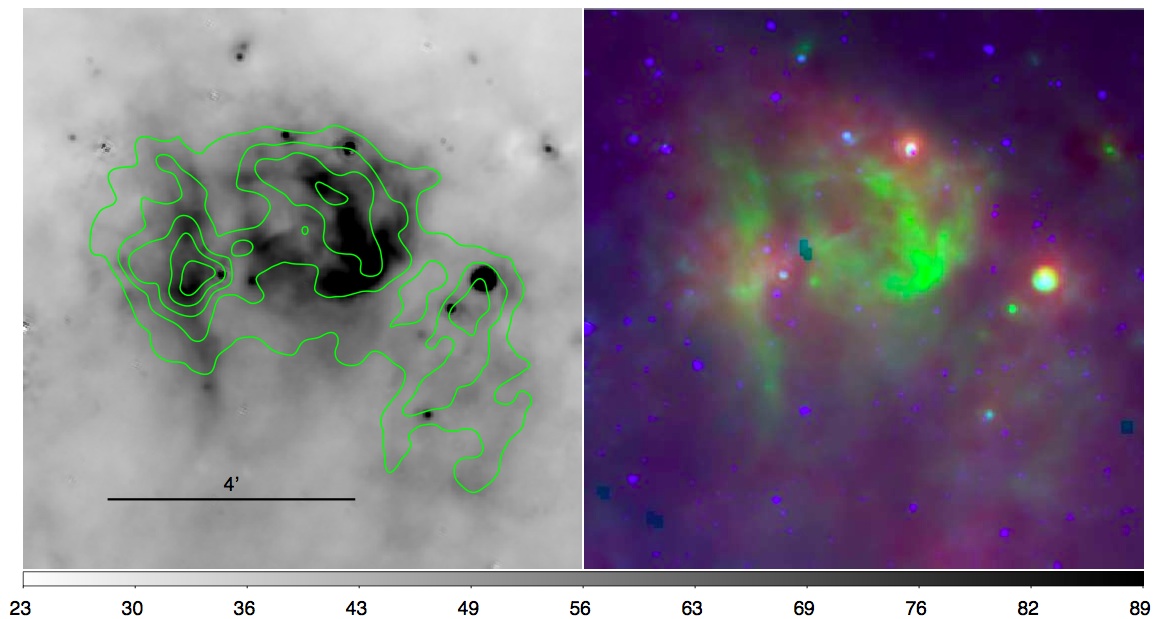}
\epsscale{.9}
\figcaption[f17.jpeg]{SNR G20.4+0.1. Countours from VLA 20 cm observations: levels are from 10 to 20, in steps of 2.5 mJy/beam.\label{snrG20401}} 
\end{figure}

Figure \ref{snrG20401} shows a shell class I remnant from B06. The 24 and 70 $\micron$ emission fill the radio contours. The northwestern radio shell is well matched by the 24 and 70 $\micron$ emission and it is surrounded by 8 $\micron$ filaments.

\subsection{G21.5-0.9}

\begin{figure}[h*]
\plotone{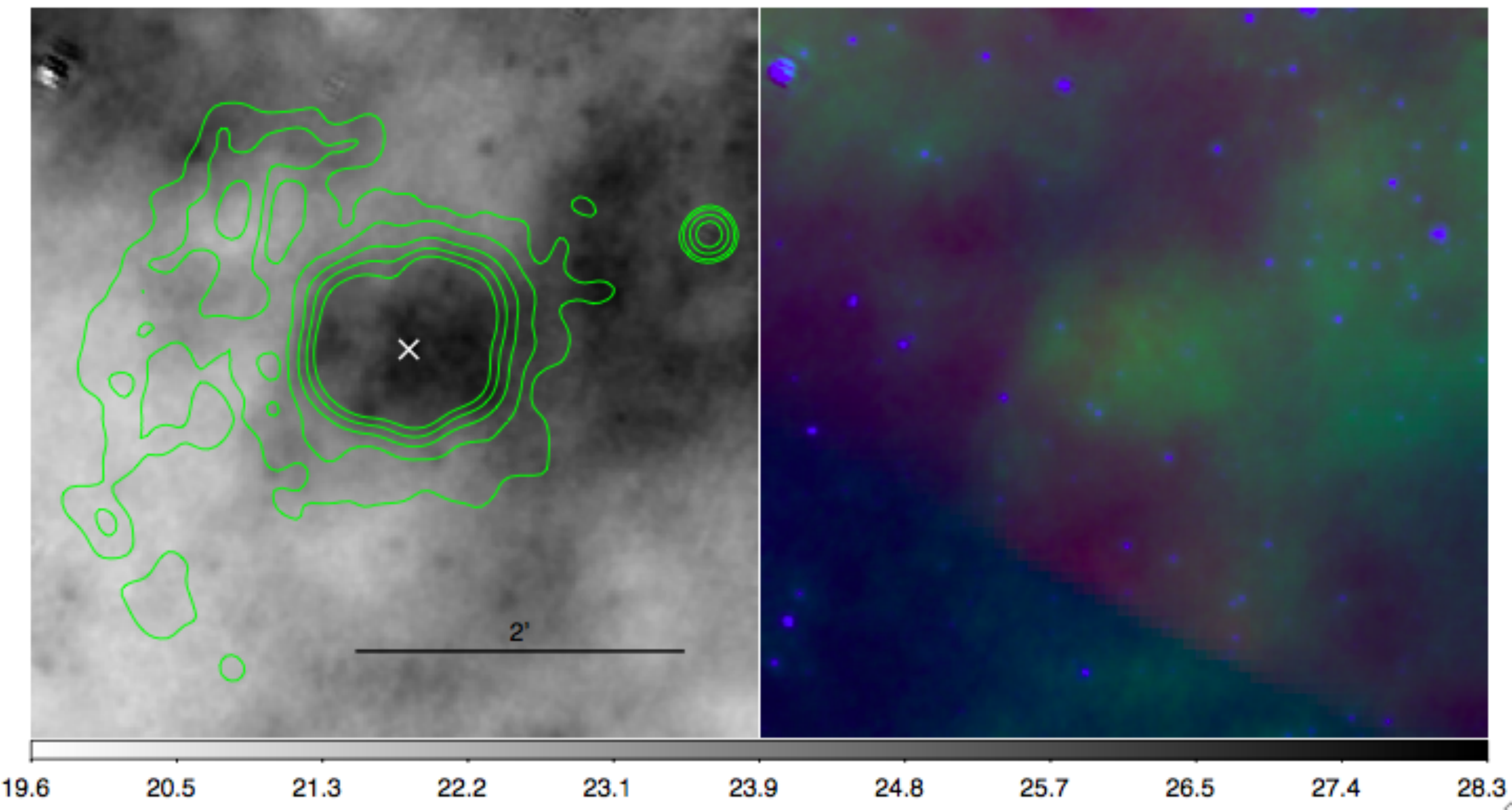}
\epsscale{.9}
\figcaption[f18.pdf]{SNR G21.5-0.9. Contours from Chandra observations: levels are 2, 3, 6.5, 12 and $20\times10^{-7}$ photons/cm${^2}$/sec/pixel. The cross locates PSR J1833-1034.\label{snrG21509}}
\end{figure}

\cite{2008MNRAS.391L..54T} suggested that this remnant is at 4.8 Kpc based on HI and CO observations and it has an associated pulsar (PSR J1833-1034) at its centre (\citealp{2006ApJ...637..456C}).
Figure \ref{snrG21509} shows the infrared emission at 24 and 70 $\micron$ associated with the central X-ray contours. There is no distinctive association between the 8 $\micron$ diffuse emission and the Chandra image.

\subsection{G21.5-0.1}

\begin{figure}[h*]
\plotone{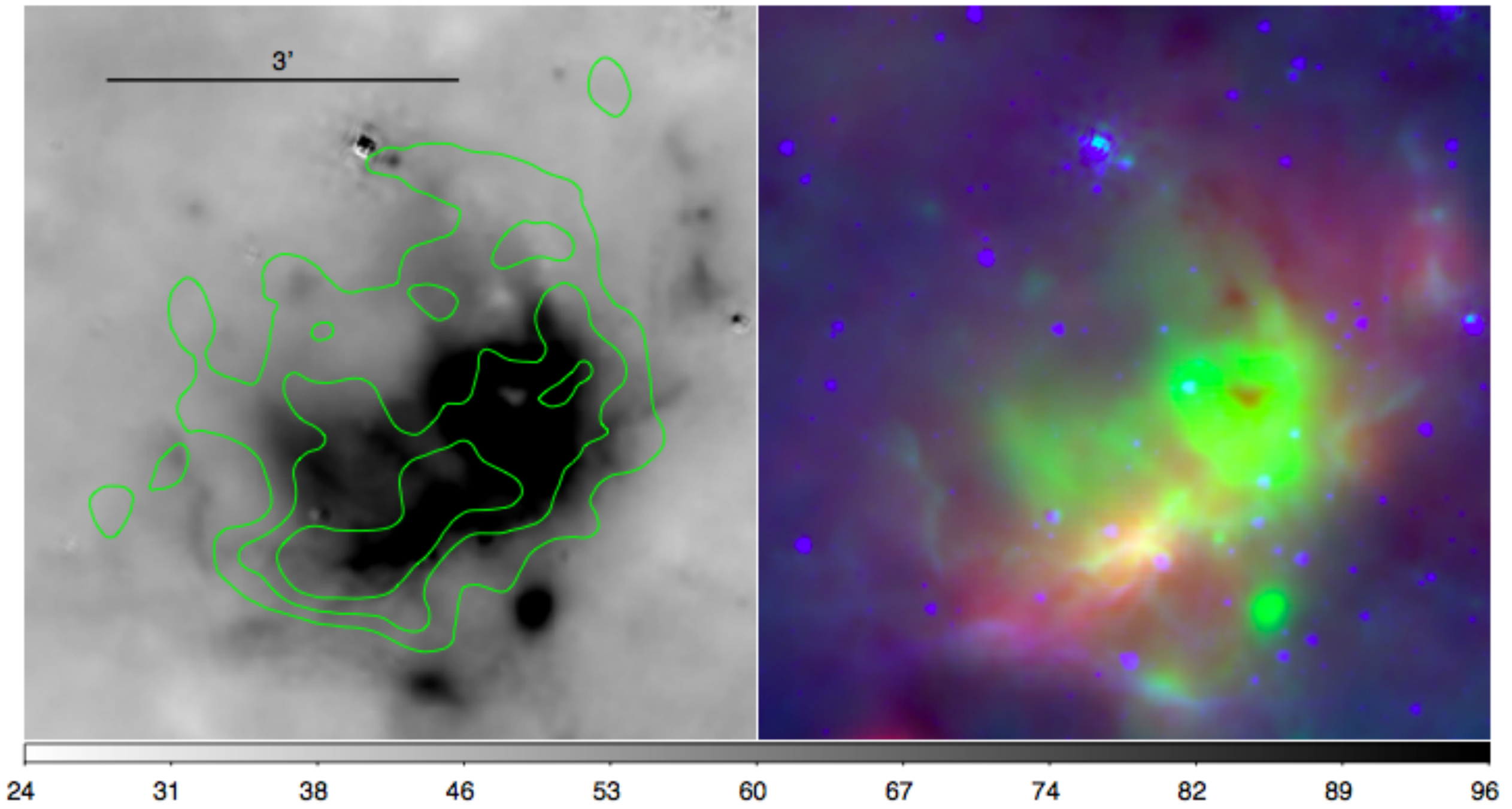}
\epsscale{.9}
\figcaption[f19.pdf]{SNR G21.5-0.1. Contours from VLA 20 cm observations: levels are from 7 to 10, in steps of 1 mJy/beam.\label{snrG21501}}  
\end{figure}

This is a partial shell remnant identified in BG06, with a possible ASCA X-ray counterpart \citep{2001ApJS..134...77S}. Centred at coordinates $21.6$ and $-0.1$, a ring with $0.5^{\prime}$ radius accounts for most of the 24 $\micron$ emission (Fig.~\ref{snrG21501}). The southern filament which coincides with a bright radio ridge is seen in all of the infrared wavelengths. Below this, there is a filament of 8 $\micron$ immediately outside the radio contours.

\subsection{G21.8-0.6 or Kes 69}

\begin{figure}[h*]
\plotone{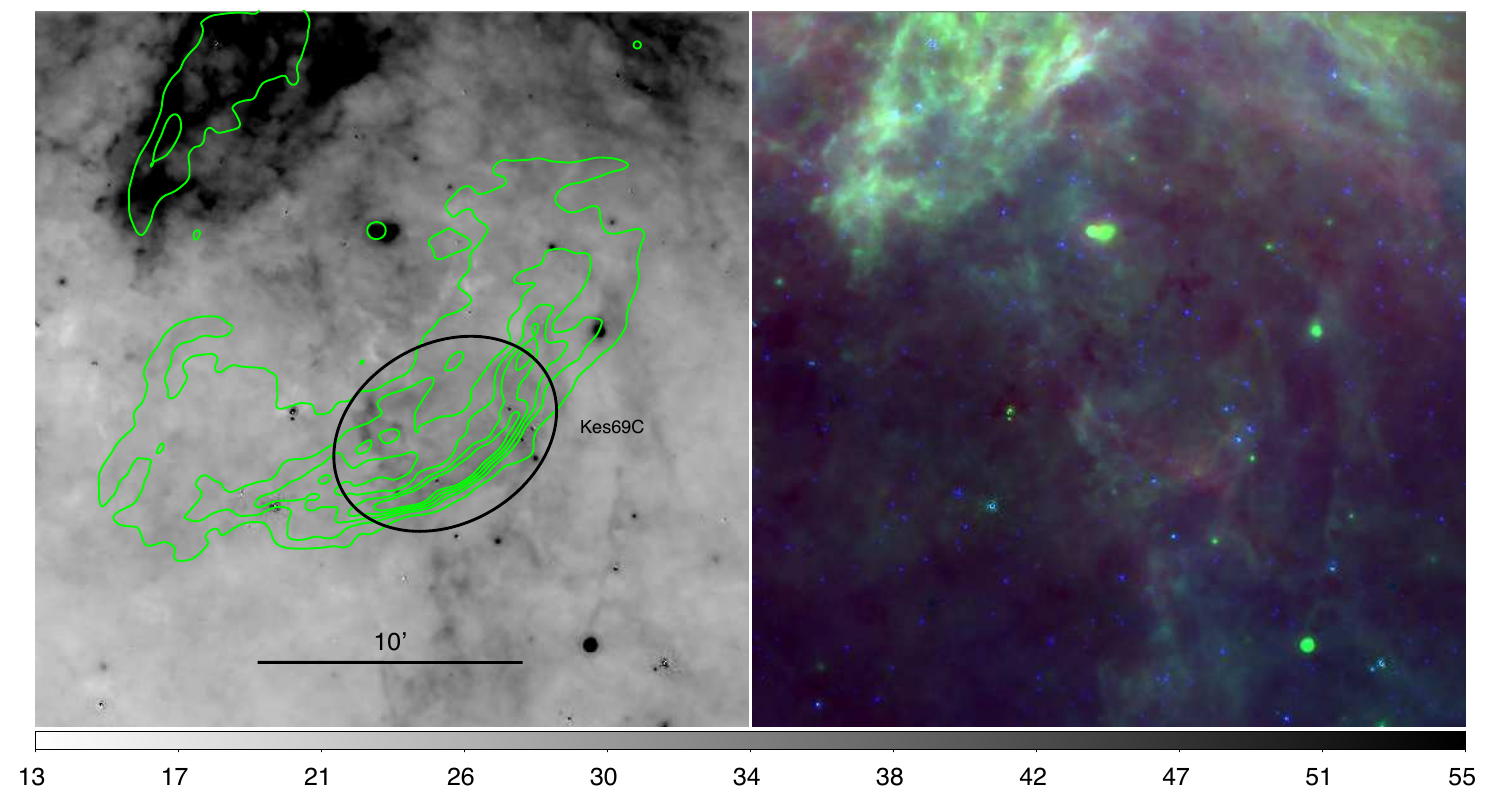}
\epsscale{.9}
\figcaption[f20.pdf]{SNR G21.8-0.6. Contours from VLA 20 cm observations: levels are from 20 to 100, in steps of 20 mJy/beam. Region used for partial photometry is also indicated in the figure (in black)\label{snrG218-06}}  
\end{figure}

At a distance of 5.5 Kpc (\citealp{2008MNRAS.391L..54T}), Kes 69 was detected in the GLIMPSE survey and shocked lines are most likely responsible for the majority of the emission seen in the south part of the radio contours (R06). The 8 and 24 $\micron$ images (Fig.~\ref{snrG218-06}) show filaments that might be associated with the remnant centre (\emph{Kes69C}), though there are similar faint structures to the south beyond the radio contours. There is prominent mid-infrared emission north of the remnant due to a bright \ion{H}{2} region.

\subsection{G23.6+0.3}
\begin{figure}[h*]
\plotone{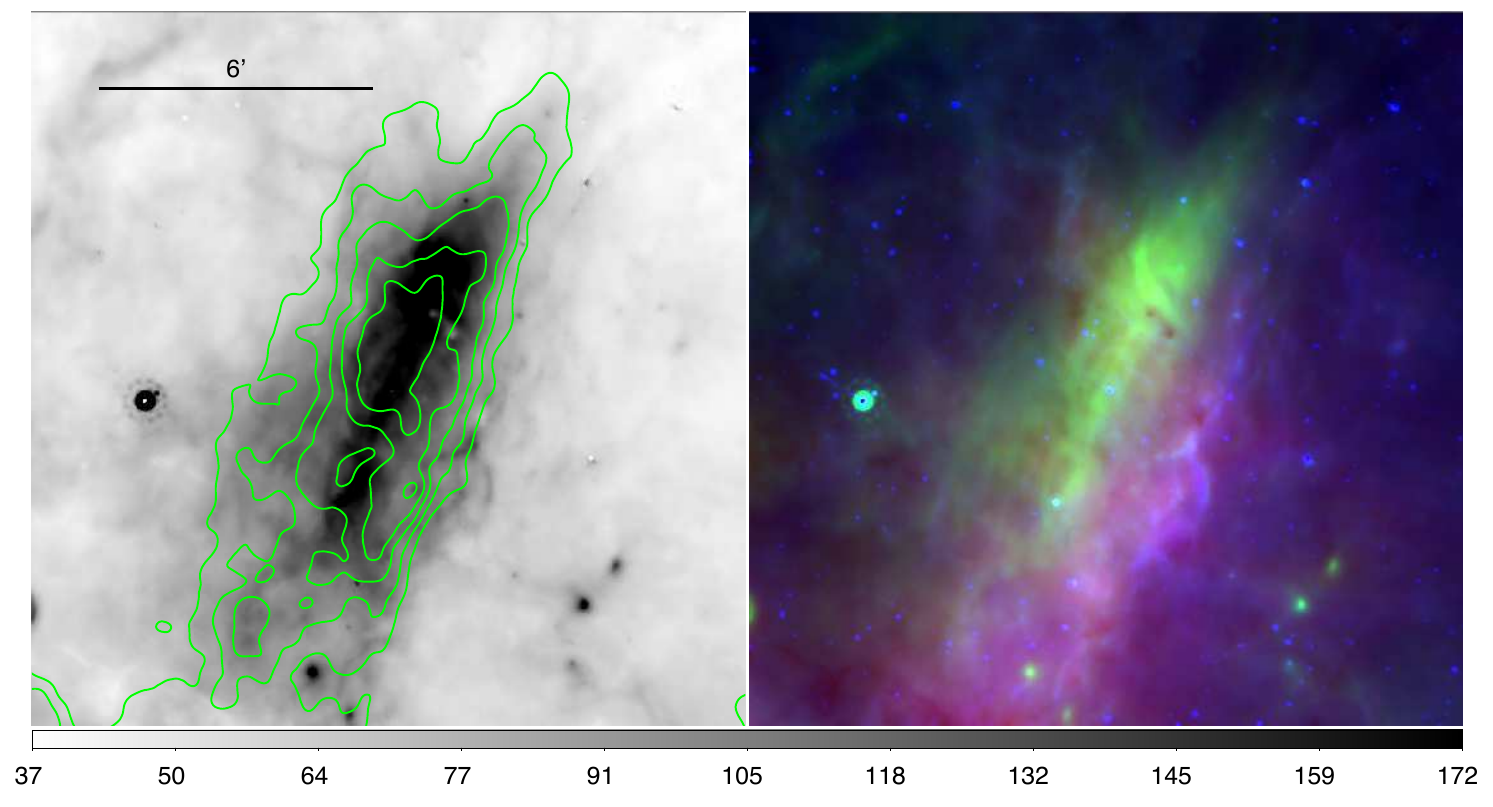}
\epsscale{.95}
\figcaption[f21.pdf]{SNR G23.6+0.3. Contours from VLA 20 cm observations: levels are from 10 to 22, in steps of 3 mJy/beam.\label{snrG236-03}}  
\end{figure}

This remnant has a radio spectral index of -0.34 (\citealp{1970AuJPA..14..133S}). Its odd elongated radio shape is seen to spatially coincide with the 24 $\micron$ emission (Fig.~\ref{snrG236-03}). The 8 and 70 $\micron$ emissions are more prominently displaced towards the south and west of the radio boundaries. No maser association has been reported.

\subsection{G27.4+0.0 or Kes 73}

\begin{figure}[h]
\plotone{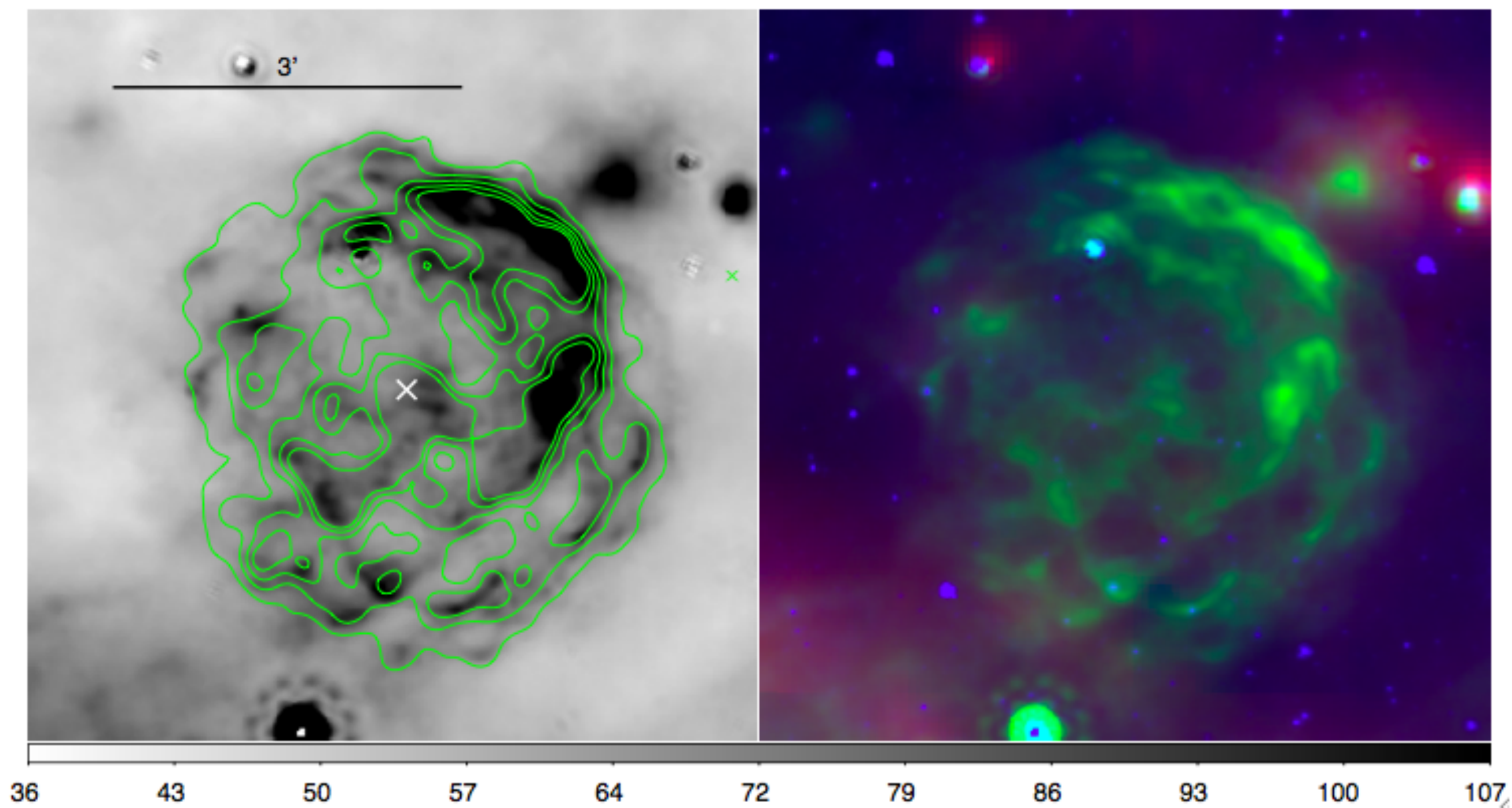}
\epsscale{.9}
\figcaption[f22.pdf]{SNR G27.4+0.0. Contours from Chandra observations: levels are from 2 to $10\times10^{-7}$, in steps of $2\times10^{-7}$ photons/cm$^2$/sec/pixel. The cross represents the location of the central X-ray source AX J1841.3-0455.\label{snrG274-00}}  
\end{figure}

This remnant has a $4^{\prime}$ diameter shell-like structure in the radio. \citet{2008ApJ...677..292T} reported that this SNR can be as far as 9.8 kpc and be 500-1000 yr old.  \cite{1997ApJ...486L.129V} discovered a low period (12s) X-ray signal from the central compact source (1E 1841-045). The SNR is not detected at the IRAC wavelengths but it is spectacularly seen at 24 $\micron$  (Fig.~\ref{snrG274-00}).  This mid-infrared emission traces the X-ray contours extremely well, with both peaking in the northwestern ridge. 

\subsection{G28.6-0.1}
\begin{figure}[h*]
\plotone{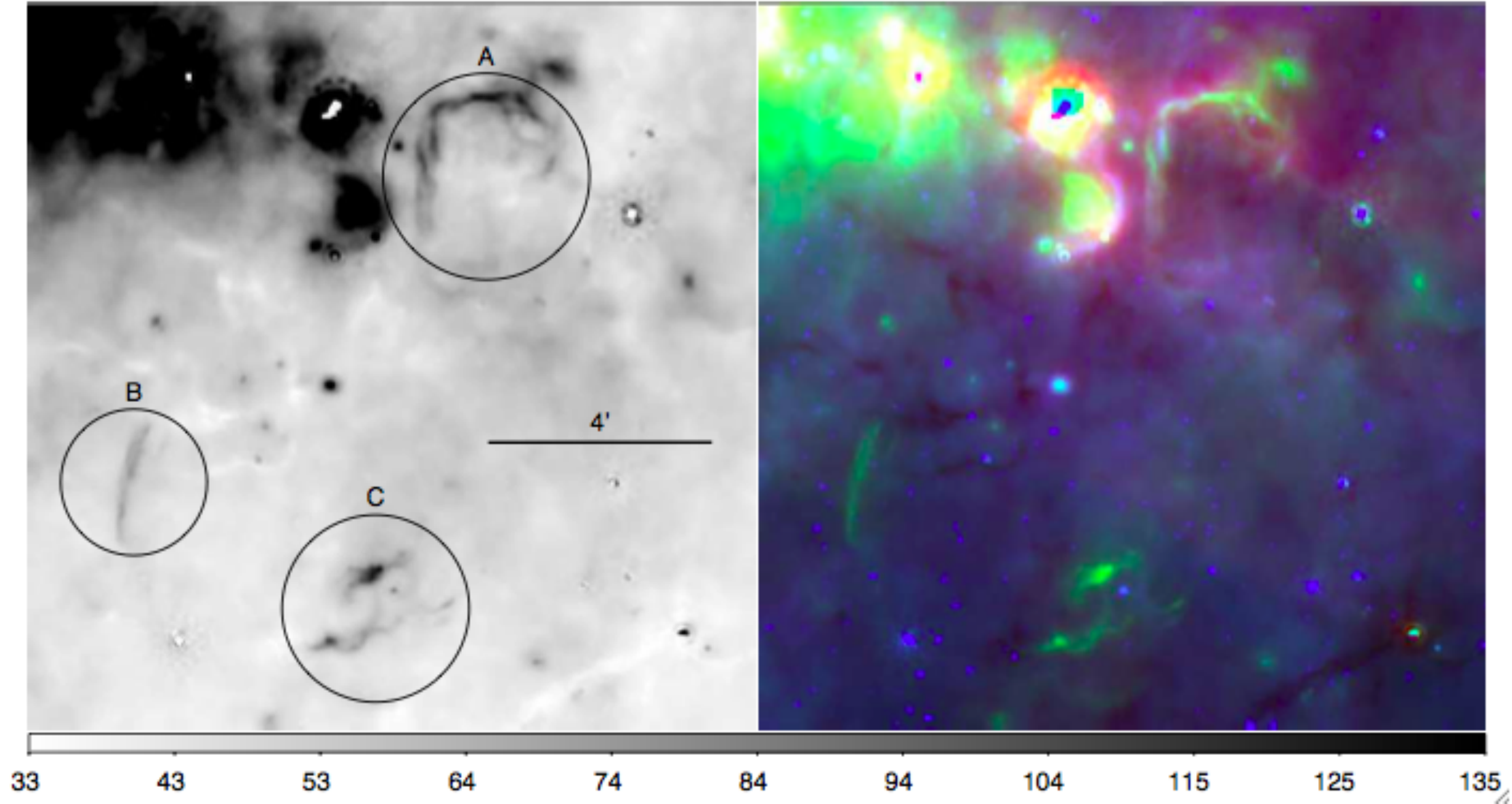}
\epsscale{.9}
\figcaption[f23.pdf]{Vicinity of SNR G28.6+0.1. Regions A, B and C are known to have non-thermal radio emission (\citealp{1989ApJ...341..151H}) and represent the regions used to estimate color ratios.\label{snrG28601}}  
\end{figure}

This region is highly contaminated by thermal emission and therefore extremely bright in infrared wavelengths. The three areas chosen to calculate infrared fluxes are identified in Figure~\ref{snrG28601} by A, B and C. These sources showed a non-thermal spectra in the radio \citep{1989ApJ...341..151H} and 20 cm flux densities of 0.73, 1.05 and 1.05 Jy, respectively. The radio morphology of all sources trace closely the 24 $\micron$ shape. Sources B and C do not show any 8 and 70 $\micron$ emission and are only well detected at 24 $\micron$.

\subsection{G29.7-0.3 or Kes 75}

\begin{figure}[h*]
\plotone{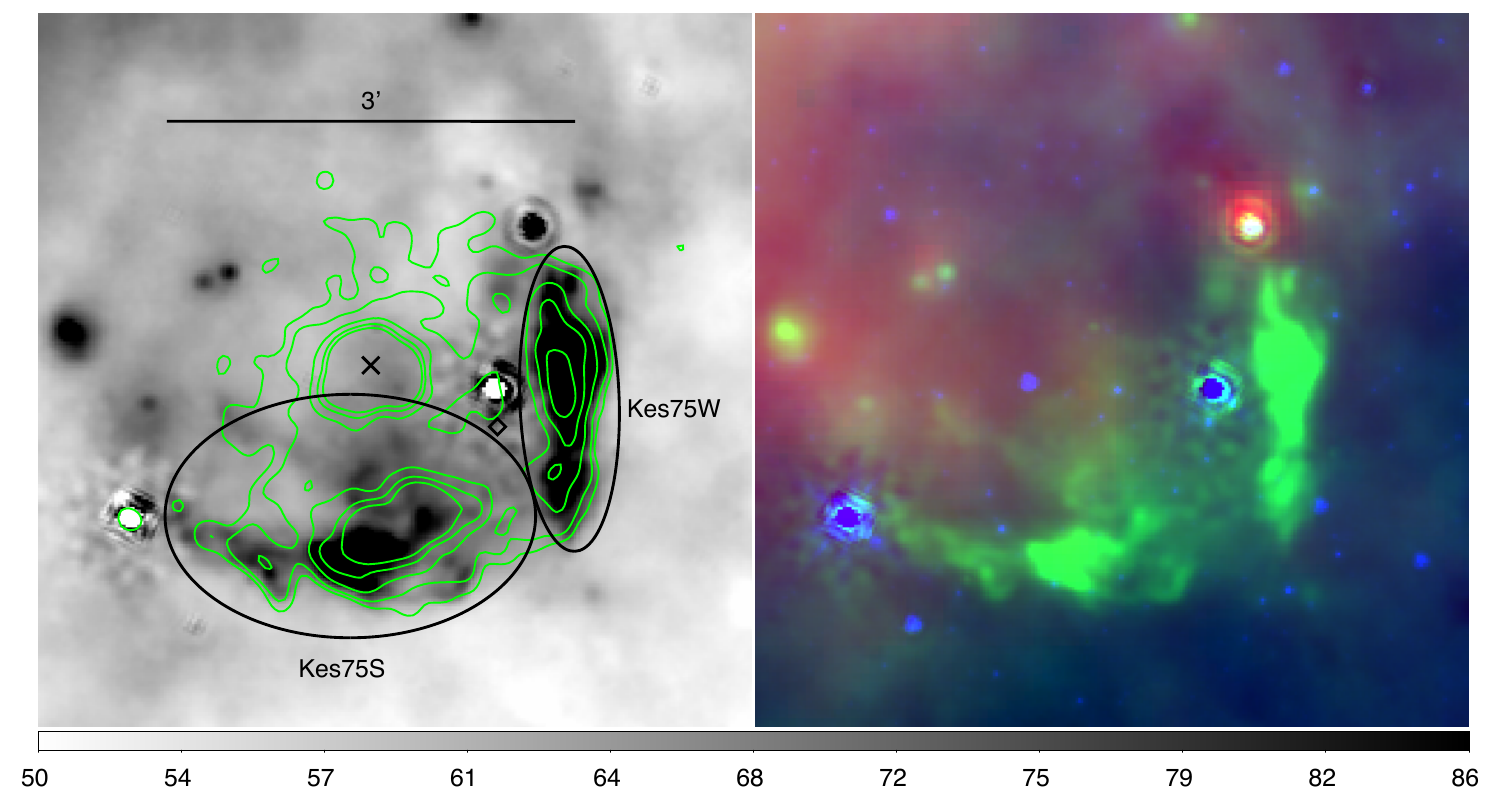}
\epsscale{.9}
\figcaption[f24.pdf]{SNR G29.7-0.3. Contours from Chandra observations: levels are 1.3, 1.8, 3.2, 5.5 and $8.7\times10^{-7}$ photons/cm$^2$/sec/pixel. The cross marks the location of the pulsar AX J1846.4-0258. Regions used for partial photometry are also indicated in the figure (in black). \label{snrG29703}}  
\end{figure}

This composite remnant has an associated pulsar (PSR J1846-0258; \citealp{2000ApJ...542L..37G}) and is thought to have an upper distance limit of 7.5 kpc (\citealp{2008A&A...480L..25L}). This remnant was not detected in GLIMPSE or IRAS surveys but was well identified using targeted MIPS observations by \cite{2007ApJ...667..219M} where most of the emission at 24 $\micron$ originates from two bright shells located south and west of the remnant's centre. Figure~\ref{snrG29703} shows a clear detection at 24 $\micron$ which traces accurately the brighter X-ray regions with the central X-ray peak being the only exception; there is no 8 $\micron$ emission there and the 70 $\micron$ emission appears to be non-existent too.  There is also no significant infrared emission at 8 or 70 $\micron$ associated with the bright 24 $\micron$ features. Unrelated 70 $\micron$ emission shows up diffusely in the eastern part of the remnant which overestimates the 70/24 color ratio measured for this region (\emph{Kes75S}).

\subsection{G31.9+0.0 or 3C 391}

Green's SNR catalog describes this object as a shell-type remnant with a spectral index of about 0.49. Previous infrared spectral observations with ISO revealed the existence of ionic lines \citep{2000ApJ...544..843R}, in particular [Fe II] at 26 $\micron$ and [S II] at 35 $\micron$ towards the direction of the BML (Broad Molecular Line) region at Galactic coordinates 31.84, 0.03 (Fig.~\ref{snrG31900}). Besides the previous elements, [O IV] at 26 $\micron$ was also detected but found to be 3 times brighter in the radio peak (bar at Galactic coordinates 31.87, 0.06) than in the BML region. Right above this radio peak, the remnant is encountering a nearby molecular cloud (\citealp{1998AJ....115..247W}). 
All of the above line emitters contaminate the MIPS channels and therefore contribute to fluxes reported in Table \ref{table:1}. Furthermore, \cite{2002ApJ...564..302R} also found clumps of H${_2}$ emission in the BML which was again confirmed with IRAC color ratios by R06.
Two 1720 MHz OH masers were detected within the remnant \citep{1996AJ....111.1651F}. IRAC images show emission associated with the western maser where the previously mentioned BML is located. Mid-infrared 24 and 70 $\micron$ emission from MIPSGAL is clearly seen in this region as well (Fig.~\ref{snrG31900}). Previous IRAC observations revealed a rim of [Fe II] emission which matches the brightest radio emission in the north part of the remnant. The same feature is seen in the 24 $\micron$ image which suggests that the bulk of emission at this location can be due to [Fe II] at 26 $\micron$.  R06 also found that the infrared emission in the middle knots (east and west at the `waist'; the one in the west has the same location as the BML) is produced by shocked molecular gas. These knots are well detected at 24 $\micron$ (with some likely contribution from H$_{2}$ S(0) at 28 $\micron$) and are especially strong at 70 $\micron$.  The southeastern rim structure (at 24 $\micron$) coincides with an enhancement in the radio and the second maser.

Remnant 3C391 is characterized by a faint outer rim that can be seen both in the radio at 20 cm and in the infrared at 24 $\micron$. This is the collisionless shock where the blast wave encounters the ISM. Although the radio structure is spatially correlated with the faint mid-infrared, a spectral index analysis shows that this 24 $\micron$ emission is too strong to have only a synchrotron origin. 

\subsection{G33.6+0.1 or Kes 79}
This a `mixed-morphology' remnant with a centrally compact X-ray source \citep{2003ApJ...584..414S}. It has an incomplete radio shell of $10^{\prime}$  in size and it is X-ray centre filled. The southern filaments seen in the Chandra image are well detected and have a similar morphology in the 24 $\micron$ image (Fig.~\ref{snrG33601}). Those are located in the same region as a dark cloud, where previous CO and HCO$^{+}$ observations suggested an interplay with a nearby molecular cloud \citep{1992MNRAS.254..686G}. These authors also noted that this must be an evolved remnant given its large linear size.
Furthermore, using Chandra observations \cite{2004ApJ...605..742S} did not encounter evidence of ejecta material, reinforcing the conclusion that this must be an older remnant (around 6000 yr old). The 24 $\micron$ emission fills and traces the brightest X-ray contours (so called inner-shell) in the central part of the remnant while 8 $\micron$ diffuse emission is present in the northern part of the remnant, but not likely to be associated. Likewise, at 70 $\micron$ it appears that the main contribution to the flux is probably the IRAS source 18501+0038 (represented by a white circle in Fig.~\ref{snrG33601}).

\subsection{G34.7-0.4 or W44}

This evolved object is classified as a `mixed-morphology' remnant \citep{1998ApJ...503L.167R} due to its radio shell with inner X-ray filled emission. It is mainly detected at 70 $\micron$ possibly indicating that it has had time to evolve (see Fig.~\ref{snrG34704}). This is in agreement with the inferred age of about 20000 yr from the characteristic age of the associated pulsar \citep{1991ApJ...372L..99W}. In the 70 $\micron$ image, the western filaments trace well the radio contours. In all the infrared wavelengths, there is emission associated to the northern region. This is one of the most stunning detections in GLIMPSE data where R06 found that emission at 4.5 $\micron$ (which is mostly from shocked molecular gas) matches well the radio contours. Previously, \cite{2005ApJ...618..297R} also detected near-infrared shocked H${_2}$ and millimeter wave molecular lines, which revealed the interaction of the remnant with a GMC. Moreover, the existence of extended maser emission has also been reported \citep{2008ApJ...683..189H} inside of the remnant. Recent \emph{Spitzer} spectral line analysis \citep{2007ApJ...664..890N} revealed the emission of [Fe II] around 26 $\micron$ and some H${_2}$ at 28 $\micron$ which can be contributors to the faint emission seen in the MIPSGAL 24 $\micron$ image. At 8 $\micron$ the object is easily confused with the diffuse background.

\subsection{G35.6-0.4}
This object was recently re-classified as a SNR \citep{2009MNRAS.399..177G}. It has a spectral index of 0.47. The radio emission is peaked, with size of $15^{\prime}\times11^{\prime}$. IRAS emission has not been previously reported. Its emission at 8 and 70 $\micron$ can be more easily confused with the background material and there seems to be no convincing association with the radio contours (Fig.~\ref{snrG35604}). At all wavelengths, there is emission associated with a gamma-ray source HESS 1858+020 and a narrow ridge (jets?) can be seen at 24 $\micron$.

\subsection{G39.2-0.3 or 3C396}
Several filaments are observed at 24 $\micron$ (Fig.~\ref{snrG39203}). Specifically, the top interior part of the remnant is only observed in this wavelength. Also, recent near-infrared observations \citep{2009ApJ...691.1042L} revealed the presence of [Fe II] and H${_2}$ in the northwestern part of the shell matching the radio emission. The 24 $\micron$ image shows the exact same features which suggests that some emission is probably due to lines in the MIPSGAL band. Other filaments near the centrally projected hot point source (IRAS 19017+0522) are seen at 8, 24 and 70 $\micron$. Previously, R06 had argued that those filaments had colors consistent with PDRs . In that case, the suite of \emph{Spitzer} passbands is probing the dust emission from PAHs to big grains. The tail of radio emission mentioned in \cite{1990A&A...232..467P} is partially traced by 24 and 70 $\micron$ with some spatial mismatch. There is a bright filament in the south part of this tail, which is detected in all of the infrared wavelengths. Moreover, in the eastern part of the tail, a filament of infrared emission also seems to match exactly the radio contour. The infrared emission that is in between the tail and the remnant is most likely to be caused by thermal dust grains.  

\subsection{G41.1-0.3 or 3C 397}

This is another shell-type morphology remnant with spectral index of 0.48 \citep{2009BASI...37...45G}.  Its uncommon rectangular shape on the upper edge suggests interaction with the ISM, specifically a dense molecular cloud inferred through CO observations \citep{2005ApJ...618..321S}. Based on Chandra data, the same authors propose that the remnant is about 5300 yr old and is now starting the radiative stage.  The 8 and 70 $\micron$ images show some emission in the north part of the remnant (Fig.~\ref{snrG41103}). The 24 $\micron$ image reveals rich filamentary structure that nicely resembles the entire X-ray counterpart. The GLIMPSE survey reported a faint detection in regions where IRAC colors suggested shocked emission from ionic lines (R06).

\subsection{G43.3-0.2 or W49B}

W49B is characterized by an odd `barrel'-shaped morphology in the radio (\citealp{1994ApJ...437..705M}). 
R06 reported that most of the emission at the IRAC wavelengths is consistent with shocked ionic and molecular emission lines. Later on, near-infrared observations revealed that the [Fe II] morphology, at 1.64  $\micron$, is similar to the radio structure and the 2.12 $\micron$ H$_{2}$ can be found primarily in the south (right below the [Fe II] ridge) and in the northwestern part of the remnant \citep{2007ApJ...654..938K}. With the help of Chandra data, these authors also reported an overabundance of heavy elements thus suggesting that the X-ray emission is dominated by the ejected material as found previously by \cite{2000ApJ...532..970H} using ASCA data. Both MIPS channels (see Fig.~\ref{snrG43302}) show interesting filamentary structure and are brighter in the same locations, thus indicating that there is a strong dust continuum contribution as the primary emission mechanism in the far-infrared.  However, note that the 24 $\micron$ image seems to trace well the ionic shocked component within the X-ray contours.

\subsection{G54.4-0.3}
GLIMPSE data (R06) revealed two regions with the same color ratios as photodissociation regions whose infrared filaments spatially match the radio profile. Those are also seen as filaments in the 24  $\micron$ and diffuse emission at 70  $\micron$ (see Fig.~\ref{snrG54403}).  Fluxes for those two regions are reported in Table \ref{table:1}. 

\section{Fourth Galactic quadrant: 295$^{\circ}$$\leq$l$\leq$350$^{\circ}$}

\subsection{G296.8-0.3}
This remnant with an elongated radio shell is at 9.6 kpc and has an age between $(2-10)\times10^3$ yr \citep{1998MNRAS.296..813G}. These authors also concluded that its radio morphology might have been shaped either due to mass loss from winds or interaction with the ISM. There is an arc of bright 24 $\micron$ emission which matches the strongest radio synchrotron emission in the northwest region. No IRAS or GLIMPSE detection has been reported. As seen in Fig.~\ref{snrG296803}, there is some diffuse 70 $\micron$ emission around it that is unlikely to be associated with the remnant. 

\subsection{G304.6+0.1 or Kes 17}

Infrared emission from the northwestern part was detected in the GLIMPSE data and attributed to shocked molecular gas (R06). Those filaments are also present in the 24  $\micron$ image (Fig.~\ref{snrG304601}). Moreover, at that same wavelength, there is emission along the south which spatially coincides with the radio ridge. At 70 $\micron$, there is just filled emission around the northwestern shell. 

\subsection{G310.8-0.4 or Kes 20A}

MOST observations show a well-defined eastern radio shell with a weak western counterpart \citep{1996A&AS..118..329W}. Diffuse thermal emission in the western counterpart of the remnant has led to previous confusion regarding the thermal or non-thermal origin of this SNR. Following the radio contours to the south west of the bright compact eastern radio peak is an arc of 24 $\micron$ emission (Fig.~\ref{snrG310}). This feature is also present at 70 $\micron$ but it is not clearly distinguishable at 8 $\micron$. There are also filaments of diffuse infrared emission that follow the north-south eastern radio shell.

\subsection{G311.5-0.3}

\cite{1996A&AS..118..329W} first identified this as a shell type SNR.  With GLIMPSE data, R06 detected an incomplete shell (brightest in the western part). Color-color ratios led to the conclusion that infrared emission was due to shocked molecular gas. Both MIPS images show the same approximate morphology following a circular shell (Fig.~\ref{snrG311503}).

\subsection{G332.4-0.4 or RCW 103}

This young remnant (age about 2000 yr) has a shell-like radio morphology and is at a distance of 3.1 kpc (\citealp{2009BASI...37...45G} and references herein). It has a soft X-ray point source near the centre of the remnant \citep{1980ApJ...239L.107T}.  The 24 $\micron$ image (Fig.~\ref{snrG332404}) shows extensive diffuse and filamentary emission related to the remnant, spatially correlated with the radio. Numerous small filaments trace the outer borders of the synchrotron emission. There is a dark arc in the centre left of the remnant where the IR and X-ray shells seem to be incomplete. For the southern bright rim, R06 found colors indicating shocked molecular gas along with some ionic emission, which corroborated the reported detection of ionized species by \cite{1999A&A...343..943O}. Around 24 $\micron$, the ISO spectrum showed strong emission mainly from [Fe II] as well as [O IV]. These authors also reported that heavy elements are in the gas phase given their solar-like abundances and therefore most of the infrared emission in that rim must be the result of fine-structure lines. The 24 $\micron$ morphology is very similar to the X-ray emission, as seen in other young remnants.

\subsection{G336.7+0.5}

There is an arc in the 24 $\micron$ image which matches roughly a radio counterpart emission peak (see Fig.~\ref{snrG336705}). At 8 and 70 $\micron$, there is some diffuse emission in the south which coincides with a brighter radio region but it seems unlikely that they are associated. The flattened indented outer radio contours to the south west coincide with with a dark cloud, possibly the locus of interaction of the SNR with the ISM.

\subsection{G337.2-0.7}

The radio contours from MOST observations indicate that this is a compact circular shell remnant (\citealp{1996A&AS..118..329W}). It has not been detected in the IRAC bands or IRAS surveys. The remnant is more complex in the X-ray and mid-infrared (Fig.~\ref{snrG337202}), with the 24 $\micron$ emission composed of several filaments forming two distorted apparent shells. The interior one is almost complete and is located in the western part. The other is visible as two long filaments mainly in the north-south direction in the eastern part of the remnant. Both relate closely to the contours of the Chandra X-ray emission. There is some 70 $\micron$ diffuse emission in the north-eastern part but unlikely to be related with the remnant.

\subsection{G340.6+0.3}

This is a circular shell remnant in the radio located at 15 kpc (\citealp{2007A&A...468..993K}). Filaments of 24 $\micron$ emission follow the radio contours especially in the south west (see Fig.~\ref{snrG340603}). Also, emission from two projected IRAS point sources, 16441-4427 and 16441-4429, situated in the north region of the SNR at opposing sides, enhance the flux considerably at the MIPSGAL wavelengths; note that in the \emph{Spitzer} bands there is considerable extended emission at these peaks.

\subsection{G344.7-0.1}

\cite{1993AJ....105.2251D} reported an angular size of 33 pc in the radio which agrees with the estimate of a middle-aged remant as based on the X-ray plasma temperature found using ASCA data \cite{2005PASJ...57..459Y}. Figure~\ref{snrG344701} shows that the emission at all three infrared wavelengths is brightest at the central radio peak. Previous GLIMPSE analysis showed that this structure had colors consistent with shocked ionized gas (R06). Although now clearly diffuse emission, this peak was catalogued as IRAS 17003-4136. The 24 $\micron$ image also shows a few striking filaments associated with the outer radio boundaries, especially where compressed at the northwest.

\subsection{G345.7-0.2}

A rim of emission at 24 and 70 $\micron$ is detected in the south-eastern part of the remnant (Fig.~\ref{snrG345702}), possibly shocked dust from interaction with the adjacent dark cloud. No maser association is reported (\citealp{1997AJ....114.2058G}).

\subsection{G348.5-0.0}

This SNR is independent from G348.5+0.1 (see below), although its radio morphology looks like a jet of material emanating from CTB 37A \citep{1991ApJ...374..212K}.  GLIMPSE data revealed that the arc of associated infrared emission is mainly dominated by ionic shocks (R06). Figure~\ref{snrG348db} shows the same structure at 24 and 70 $\micron$, the 24 $\micron$ emission being the strongest of all.

\subsection{G348.5+0.1 or CTB 37A}\label{snrG31900}

GLIMPSE analysis supports the idea that this remnant is interacting with dense interstellar gas, first indicated by the presence of masers \citep{1996AJ....111.1651F} within the remnant. \cite{2000ApJ...545..874R}  found CO clouds at approximately the same velocities and locations of the OH 1720 MHz masers. Figure~\ref{snrG348db} shows that the MIPSGAL images have the same diffuse emission seen at the IRAC wavelengths including the eastern patch of infrared emission which is thought to be physically associated with a maser (and whose IRAC colors suggest emission from shocked molecular gas as noted by R06). At 24 $\micron$ there are also southern filaments which closely trace the radio boundaries. Bright infrared emission on the northwest fills the faint radio emission part of the remnant and its IRAC colors are similar to photodissociation regions (R06).

\subsection{G348.7+0.3 or CTB 37B}

In this very confused field, Figure~\ref{snr3487} shows that the radio peak from MOST contours has an associated mid-infrared peak at 24 $\micron$. In addition there are some filaments within the remnant that seem to be the result of interaction with the ISM. Chandra follow-up observations of a Gamma-ray source (HESS J1713-381) thought to be associated with CTB 37B (\citealp{2008A&A...486..829A} and references therein) also detected X-ray thermal emission close to the remnant centre whose temperature suggested a middle-aged SNR (about 5000 yr). 

\subsection{G349.7+0.2}

OH maser emission was found in the remnant's interior by \cite{1996AJ....111.1651F} and evidence of interaction with a molecular cloud was later strengthened by \cite{2001AJ....121..347R} with CO observations. With extremely similar X-ray and radio morphology, its corresponding brightness in both regimes is also high (\citealp{2005ApJ...618..733L}). The infrared emission is present with a similar morphology at all three wavelengths (Fig.~\ref{snrG349702}). The infrared is strongly peaked within the X-ray contours, but not closely related, except for the 24 $\micron$ emission. Previous GLIMPSE analysis showed infrared emission in all of the IRAC bands (R06).

%_______________________________________________________________

\section{Individual Images}

Images of the infrared emission for each detected remnant follow. The order is the same for all the figures and consists of the black and white 24 $\micron$ image (colorbar in units of MJy/sr) on the left and the 3-color image on the right, with 8, 24 and 70 $\micron$ being blue, green and red, respectively. Overlaid in each image are contours either from radio or X-ray observations when available (see \S\ref{otherdata}). Images are all in Galactic coordinates, and our remarks on north (up), east (left), etc., are with respect to this system. Also indicated is the angular scale which can be compared with the few arcminute resolution of \emph{IRAS}. 

\clearpage

\begin{figure}[h*]
\plotone{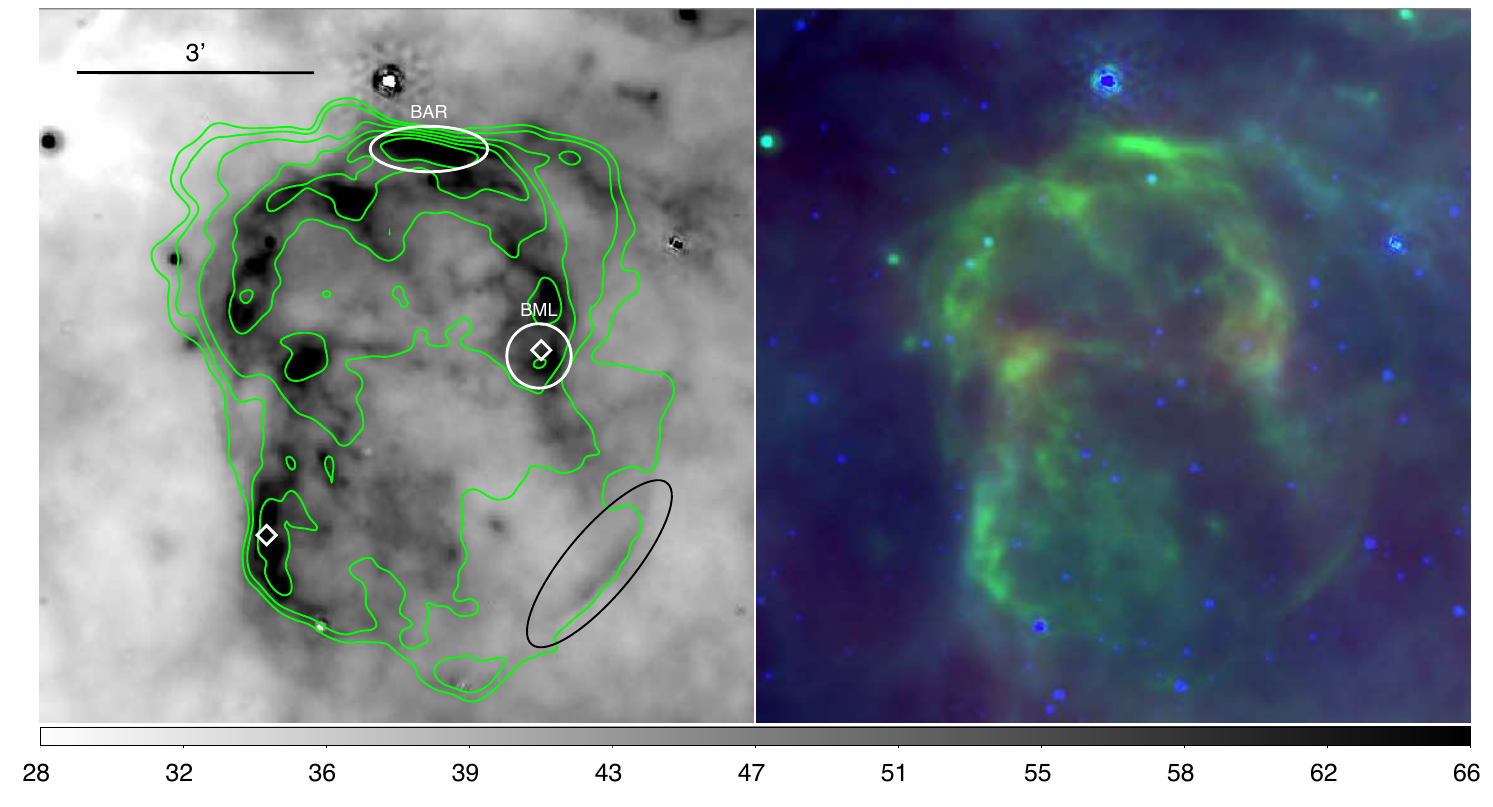}
\epsscale{.9}
\figcaption[f25.pdf]{SNR G31.9+0.0. Contours from VLA 20 cm observations: levels are 15, 28, 69, 140, 230 and 350 mJy/beam. The two maser locations are represented by diamonds. The ellipse (in black) in the southwestern part of the remnant locates a portion of the possible collisionless shock. Regions used for partial photometry are also indicated in the figure (in white).\label{snrG31900}}  
\end{figure}

\begin{figure}[h*]
\plotone{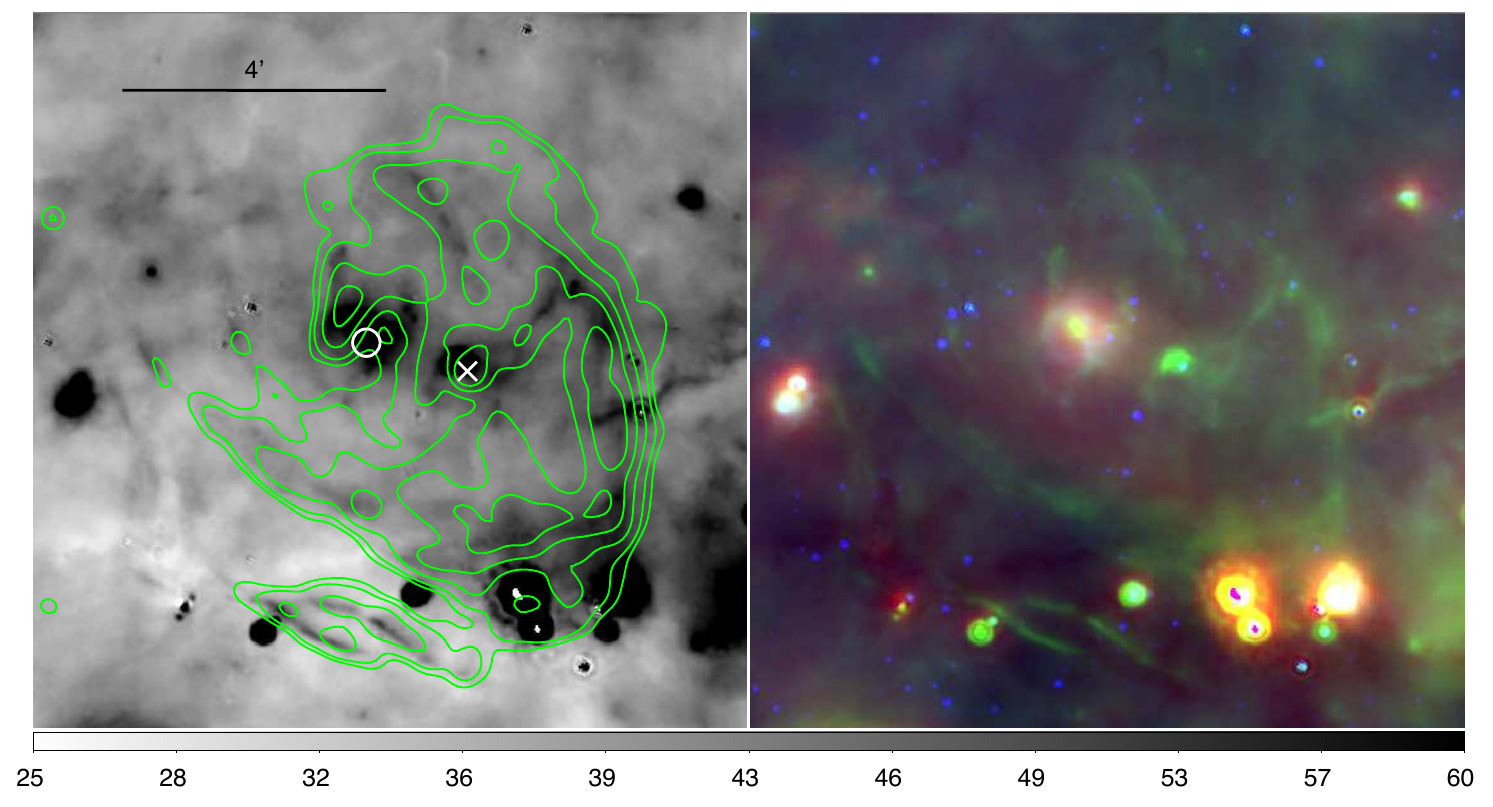}
\epsscale{.9}
\figcaption[f26.pdf]{SNR G33.6+0.1. Contours from Chandra observations: levels are 1.5, 2, 3.6, 6.3 and $10\times10^{-7}$ photons/cm$^2$/sec/pixel. The cross marks the location of PSR J1852+0040 and the circle represents an IRAS point source.\label{snrG33601}}  
\end{figure}

\begin{figure}[h*]
\plotone{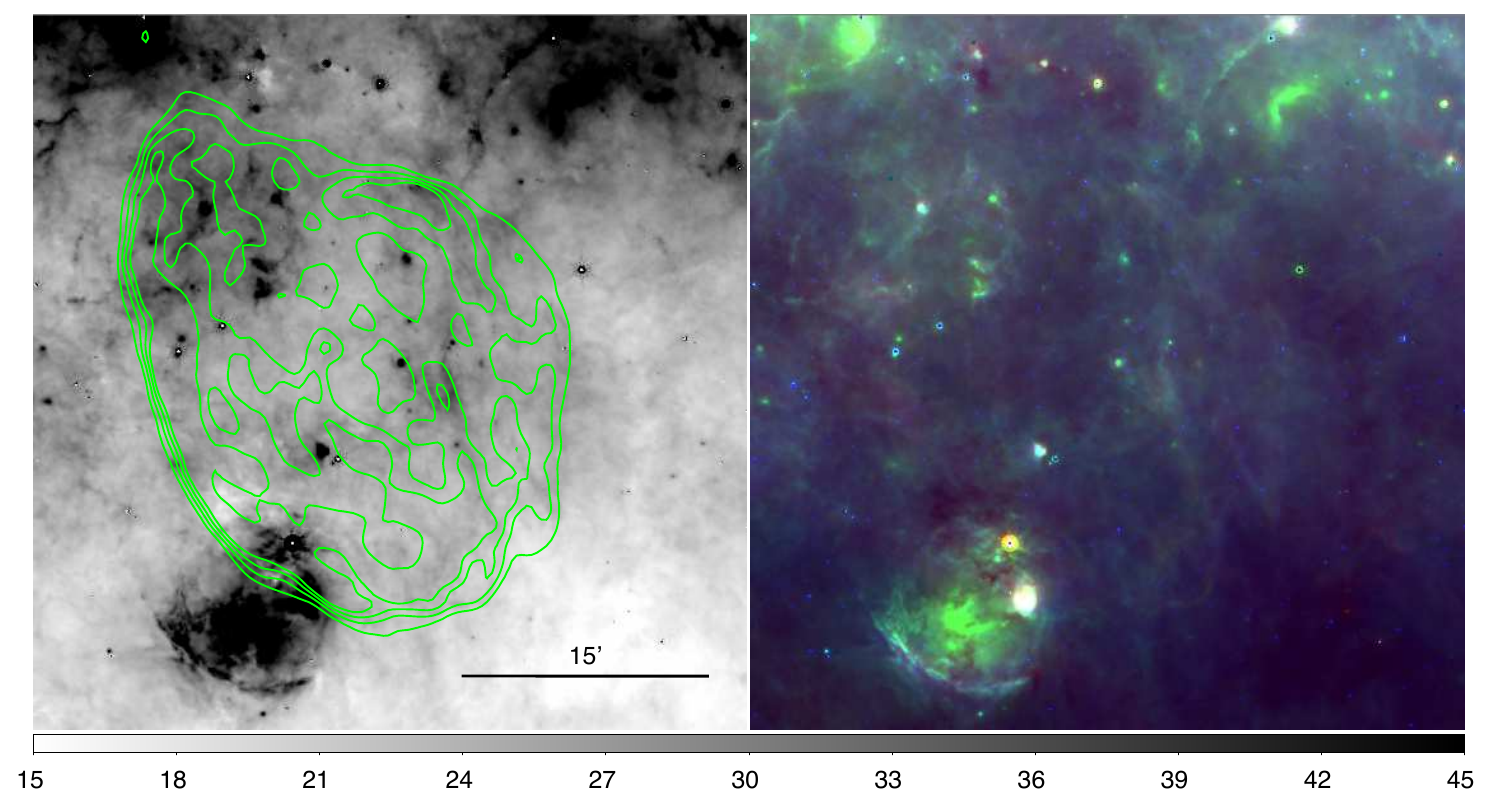}
\epsscale{.9}
\figcaption[f27.pdf]{SNR G34.7-0.4. Contours from VGPS 21 cm observations: levels are 35, 52, 68 and 85 Jy/beam.\label{snrG34704}}  
\end{figure}

\begin{figure}[h*]
\plotone{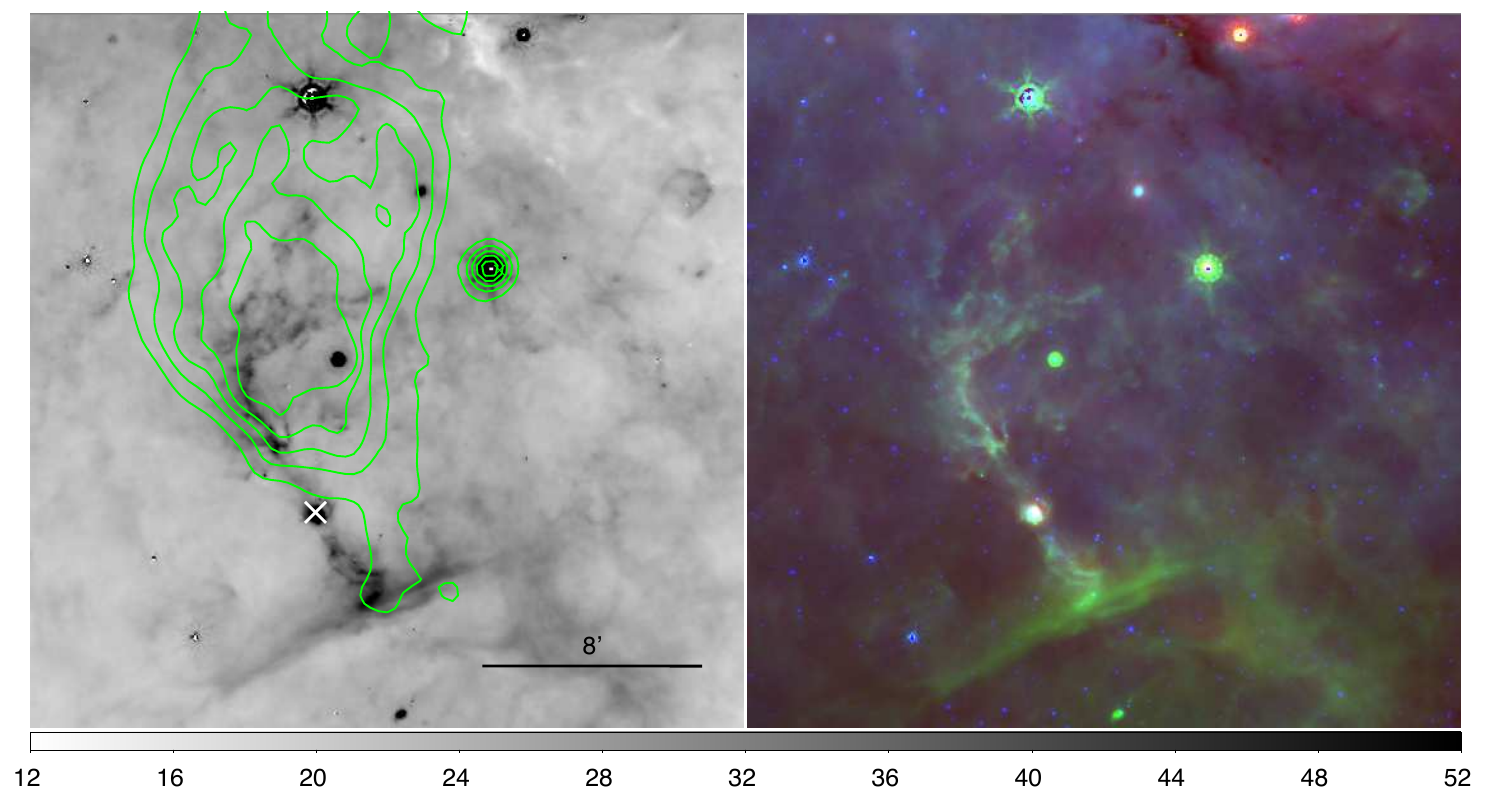}
\epsscale{.9}
\figcaption[f28.pdf]{SNR G35.6-0.4. Contours from VGPS 21 cm observations: levels are from 18 to 30, in steps of 3 Jy/beam. The cross marks the location of $\gamma$-ray source J1858+020.\label{snrG35604}}  
\end{figure}

\begin{figure}[h*]
\plotone{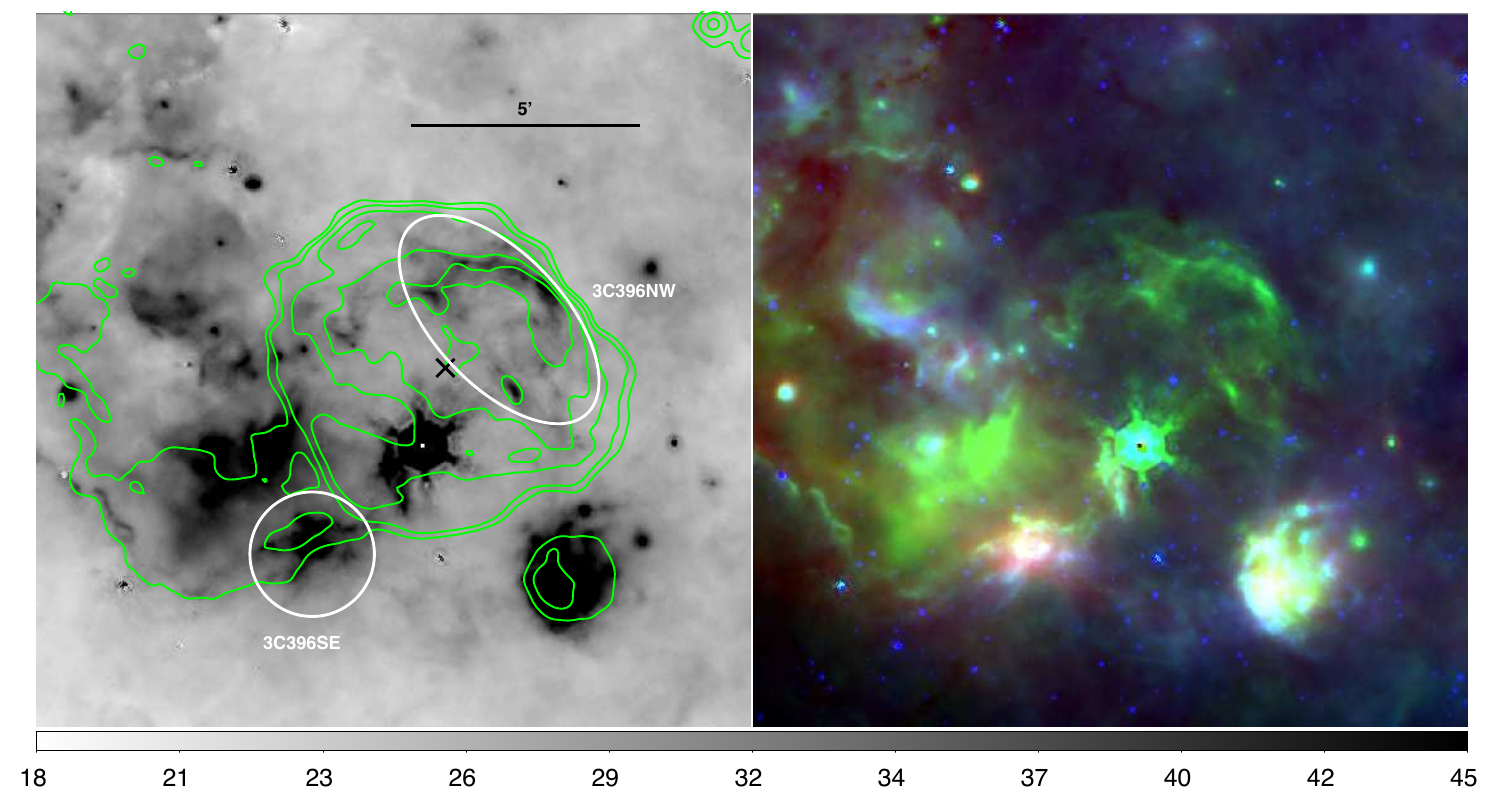}
\epsscale{.9}
\figcaption[f29.pdf]{SNR G39.2-0.3. Contours from 20cm VLA observations: levels are 1, 4, 12, 26 and 45 mJy/beam. The cross represents the location of the X-ray source J190406.5+052646. Regions used for partial photometry are also indicated in the figure (in white). \label{snrG39203}}  
\end{figure}

\begin{figure}[h*]
\plotone{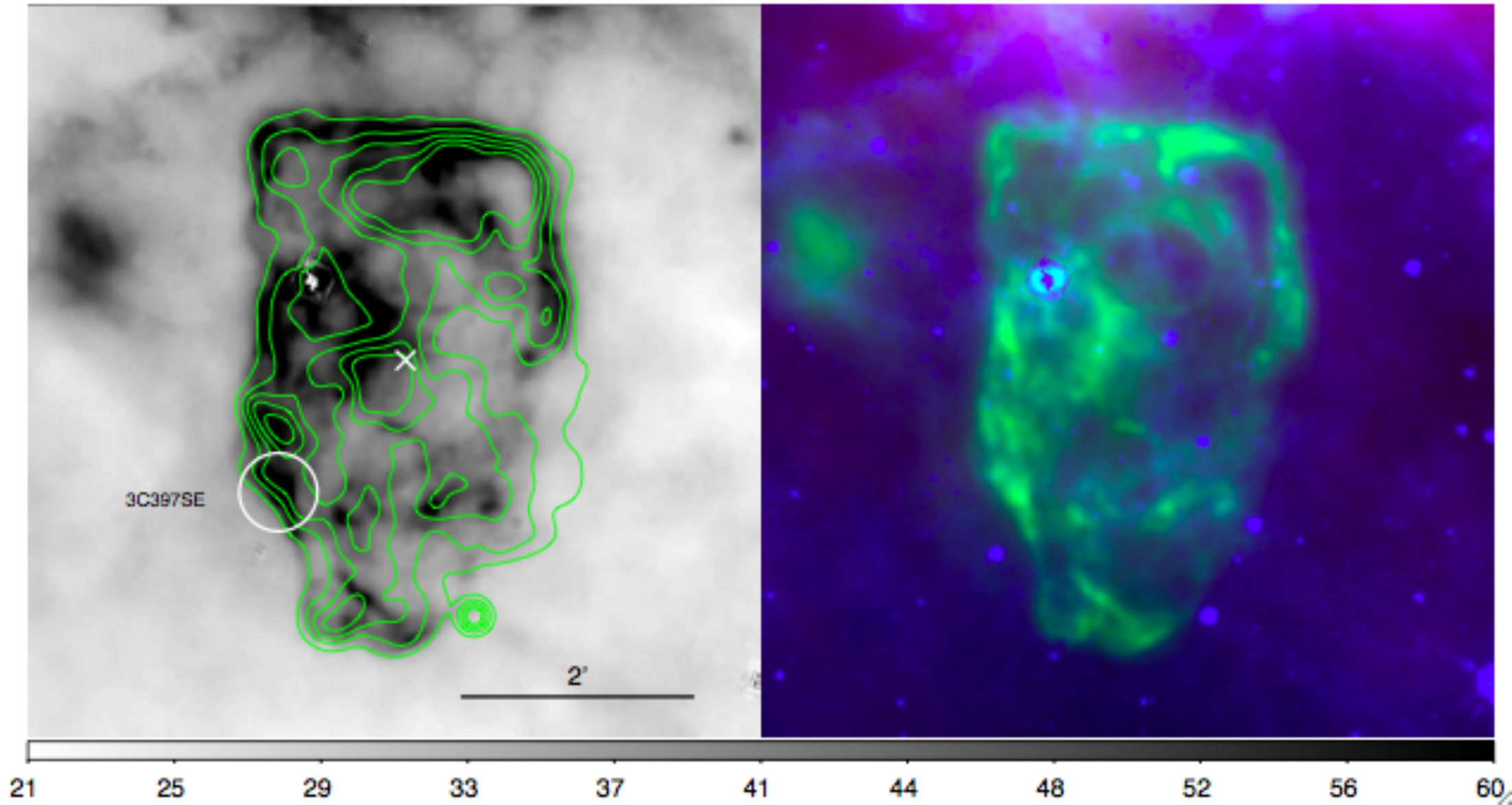}
\epsscale{.9}
\figcaption[f30.pdf]{SNR G41.1-0.3. Contours from Chandra observations: levels are 2, 4.3, 6.5, 8.8 and $11\times10^{-7}$ photons/cm$^2$/sec/pixel. The cross represents the location of the X-ray source J1907.5+0708. Region used for partial photometry is also indicated in the figure (in white).\label{snrG41103}}  
\end{figure}

\begin{figure}[h*]
\plotone{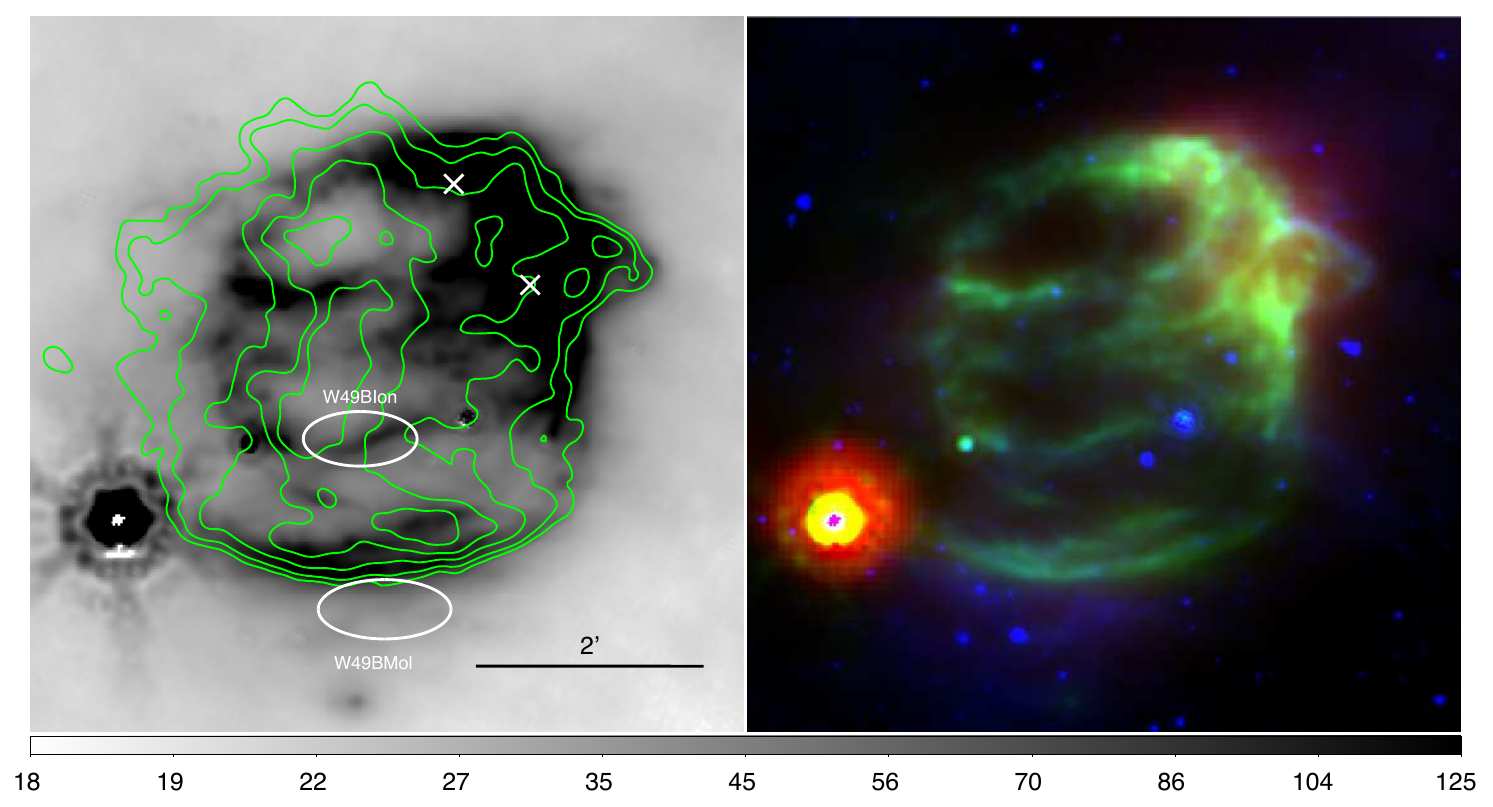}
\epsscale{.9}
\figcaption[f31.pdf]{SNR G43.3-0.2. Contours from Chandra observations: levels are 2, 3.1, 6.5, 12 and $20\times10^{-7}$ photons/cm$^2$/sec/pixel. The crosses represent the location of the X-ray source J1911.0+0906 and $\gamma$-ray source J1911.0+0905. Regions used for partial photometry are also indicated in the figure (in white).\label{snrG43302}}  
\end{figure}

\begin{figure}[h*]
\plotone{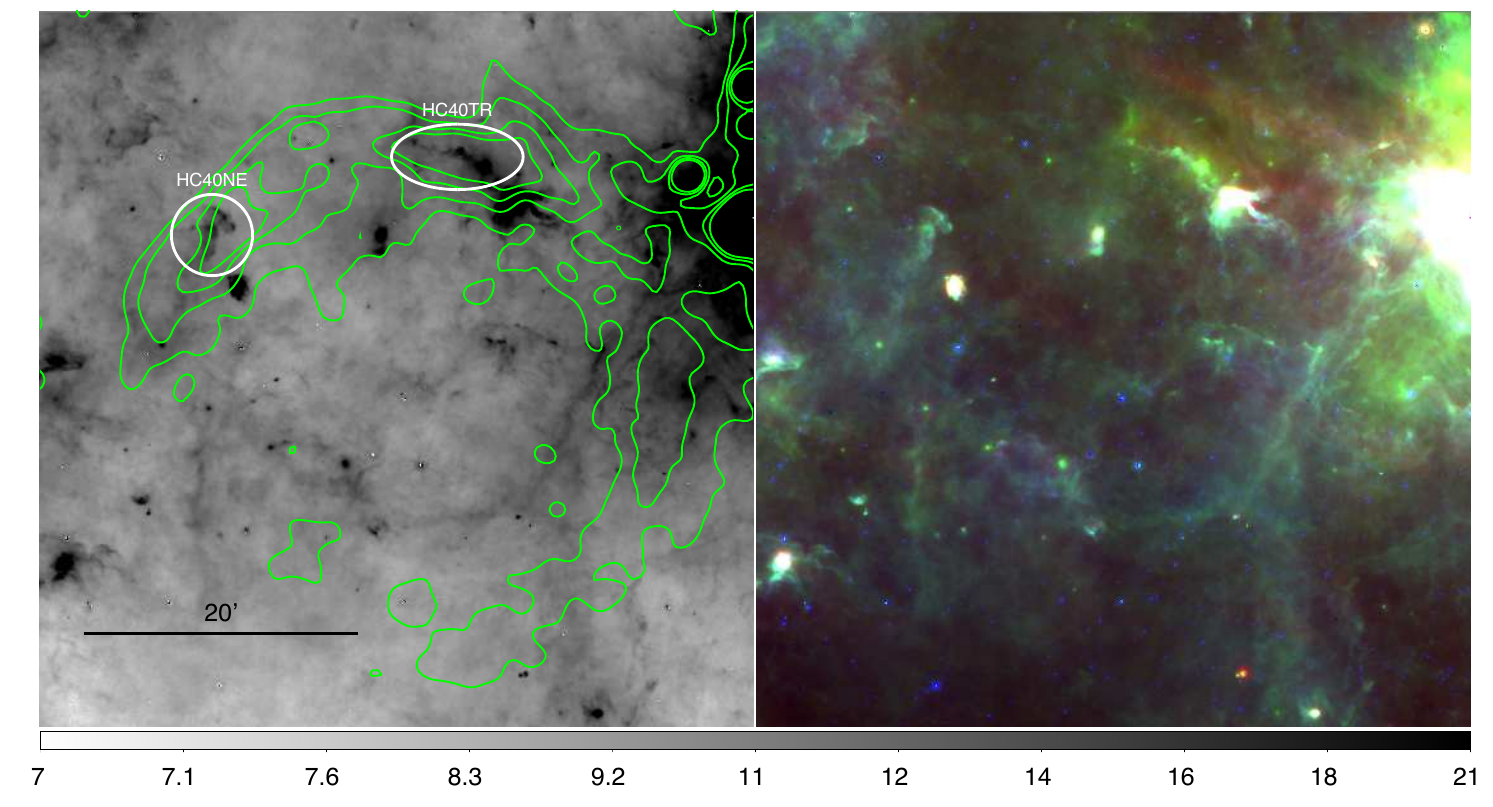}
\epsscale{.9}
\figcaption[f32.pdf]{SNR G54.4-0.3. Contours from VGPS 21 cm observations: levels are from 9.5 to 14, in steps of 1.5 Jy/beam. Regions used for partial photometry are also indicated in the figure (in white).\label{snrG54403}}  
\end{figure}

\begin{figure}[h*]
\plotone{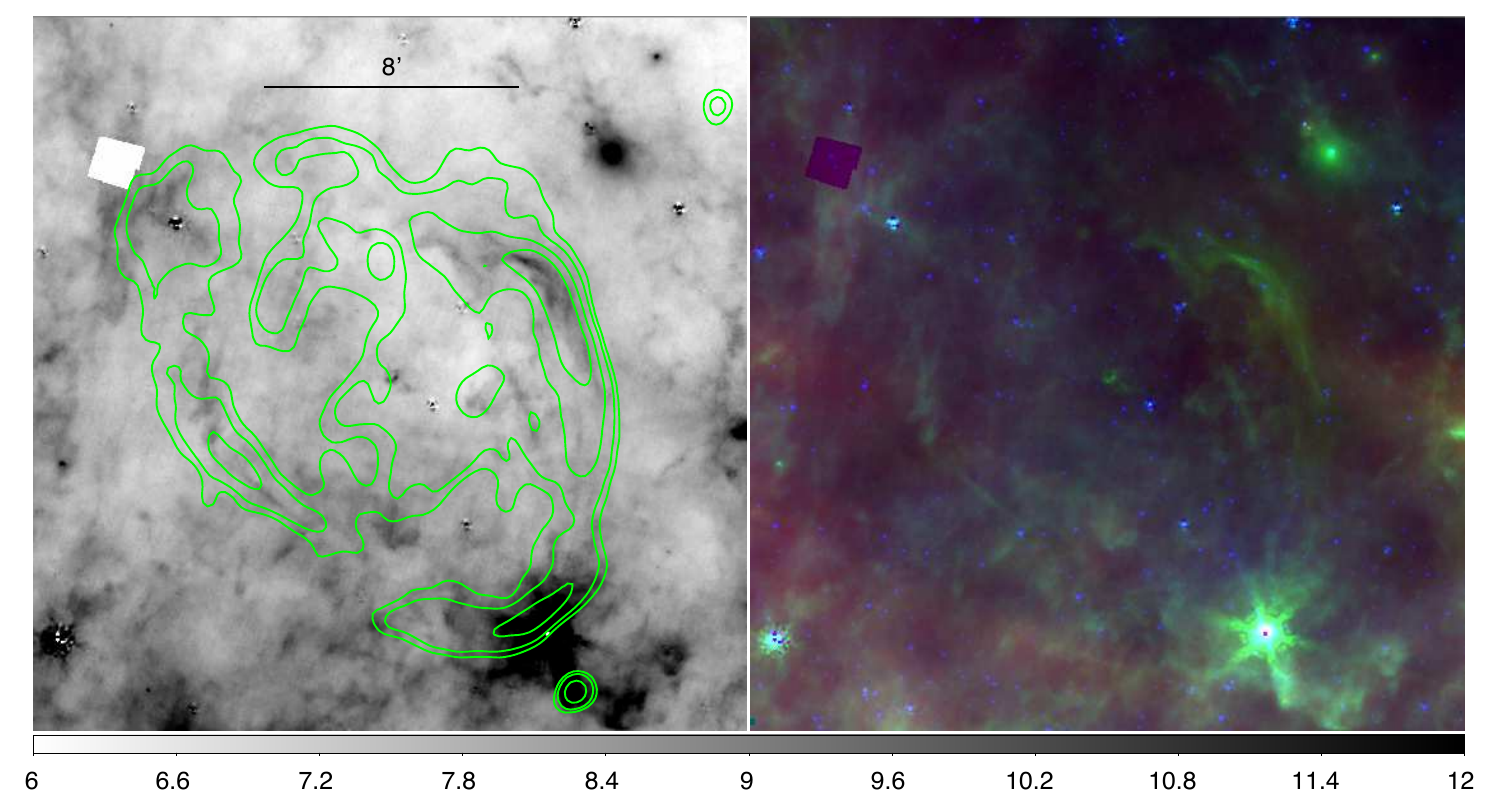}
\epsscale{.9}
\figcaption[f33.pdf]{SNR G296.8-0.3. Contours from MOST observations: levels are 10, 22, 58, 120 and 200 mJy/beam.\label{snrG296803}}  
\end{figure}

\clearpage

\begin{figure}[h*]
\plotone{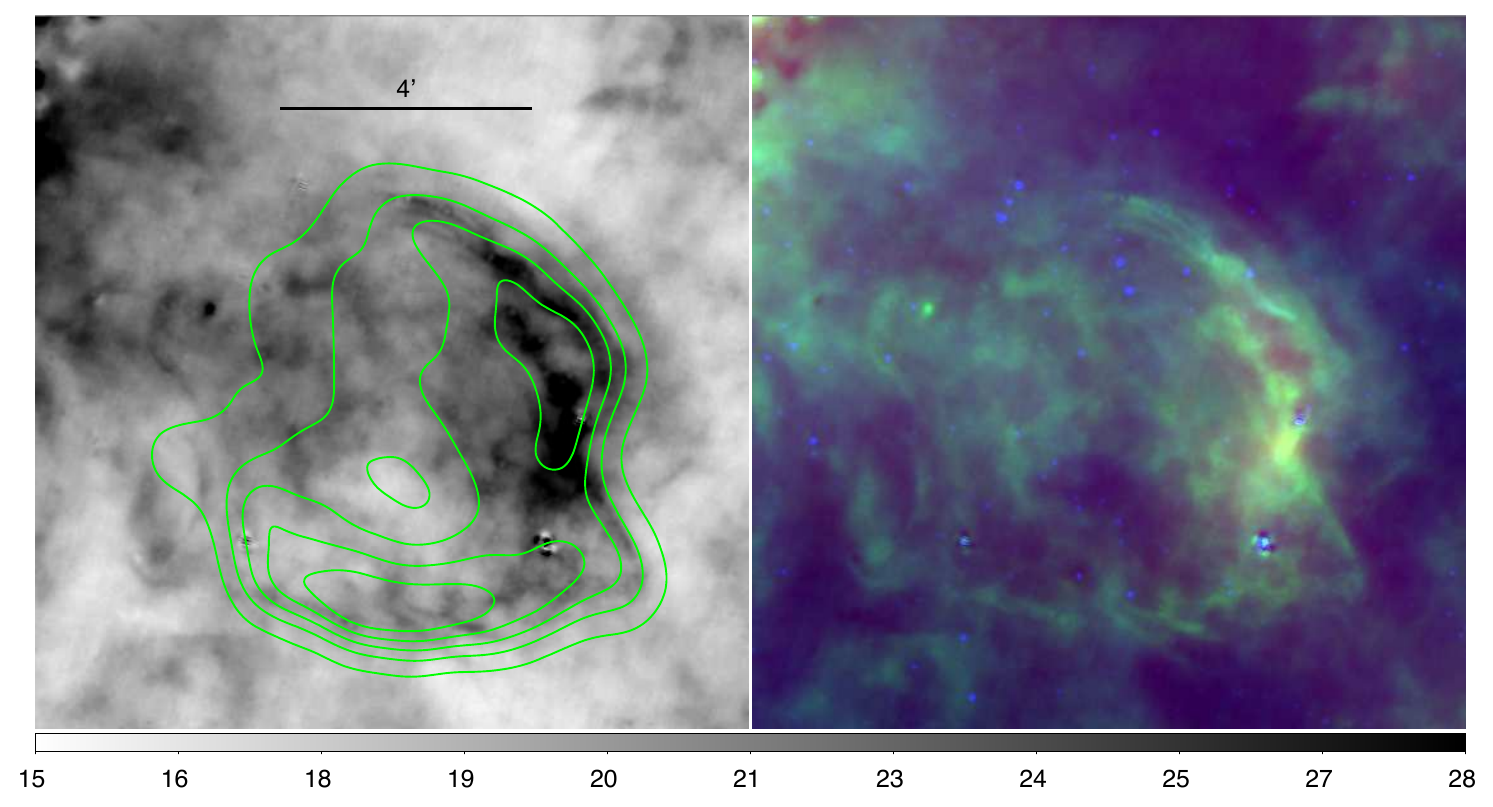}
\epsscale{.9}
\figcaption[f34.pdf]{SNR G304.6+0.1. Contours from MOST observations: levels are 0.06 to 0.5, in steps of 0.11 Jy/beam.\label{snrG304601}}  
\end{figure}

\begin{figure}[h*]
\plotone{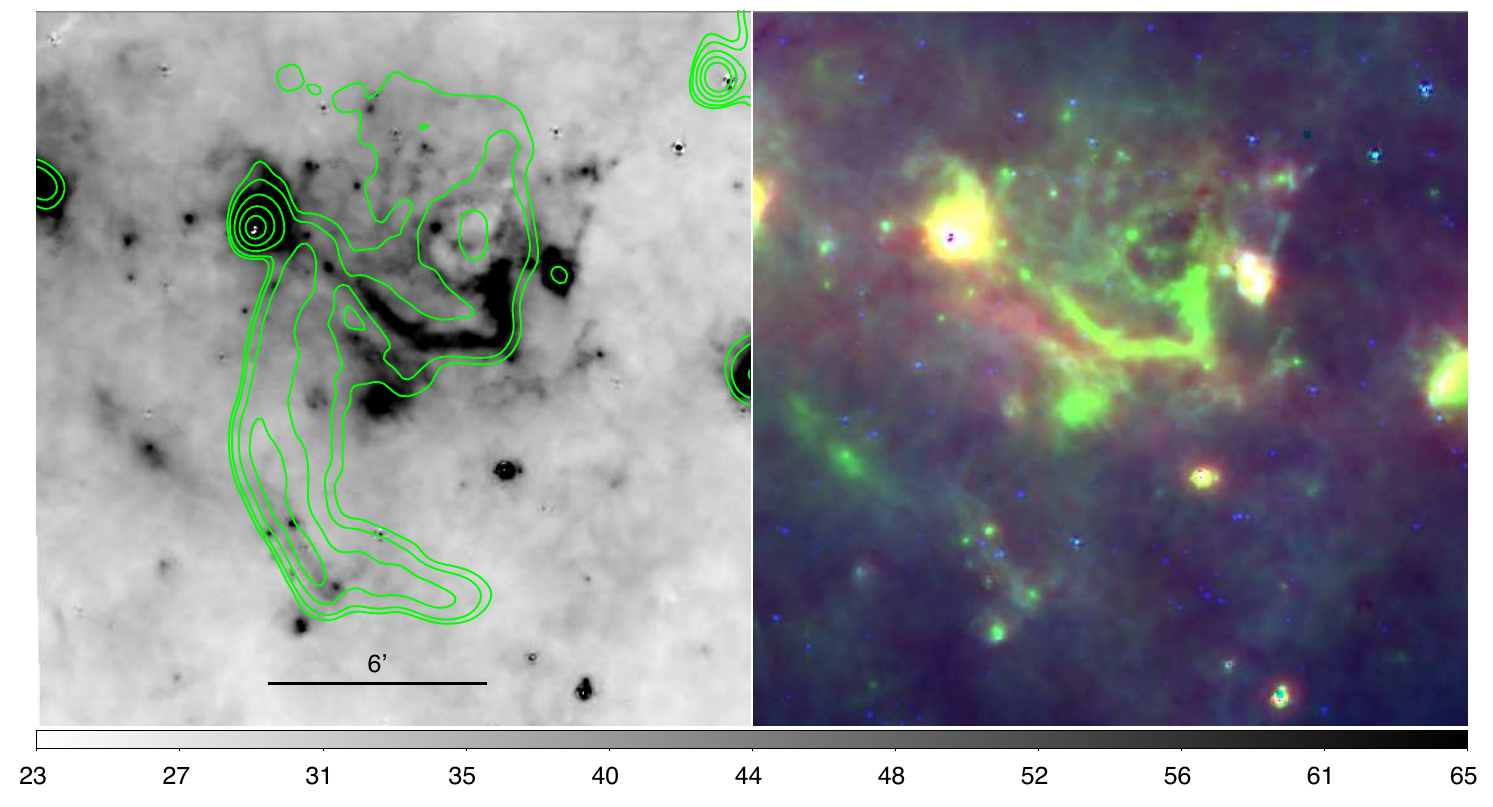}
\epsscale{.9}
\figcaption[f35.pdf]{SNR G310.8-0.4. Contours from MOST observations: levels are 35, 58, 126, 240 and 400 mJy/beam.\label{snrG310}}  
\end{figure}

\begin{figure}[h*]
\plotone{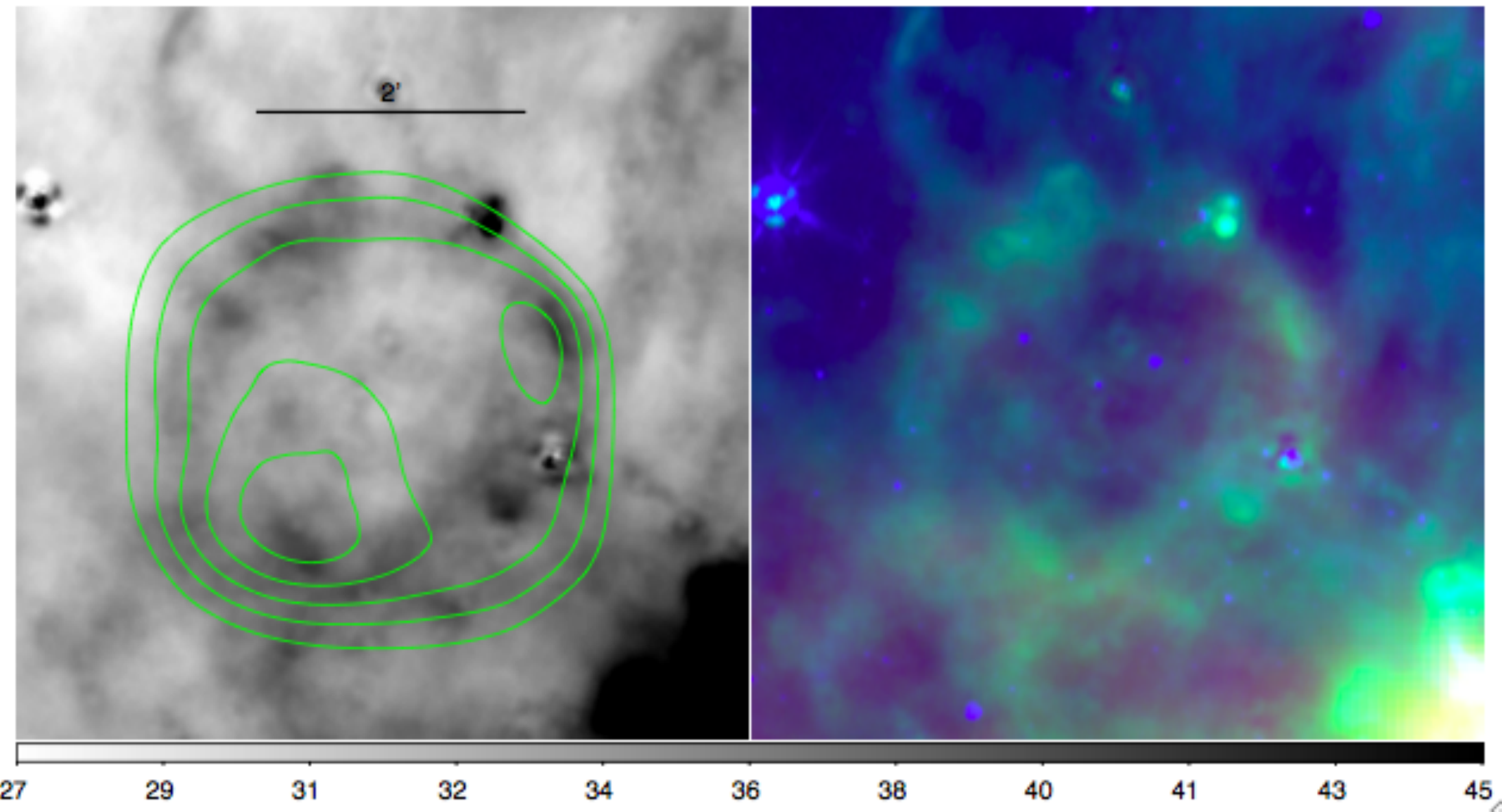}
\epsscale{.9}
\figcaption[f36.pdf]{SNR G311.5-0.3. Contours from MOST observations: levels are 0.1, 0.14, 0.18, 0.24 and 0.27 Jy/beam.\label{snrG311503}}  
\end{figure}

\begin{figure}[h*]
\plotone{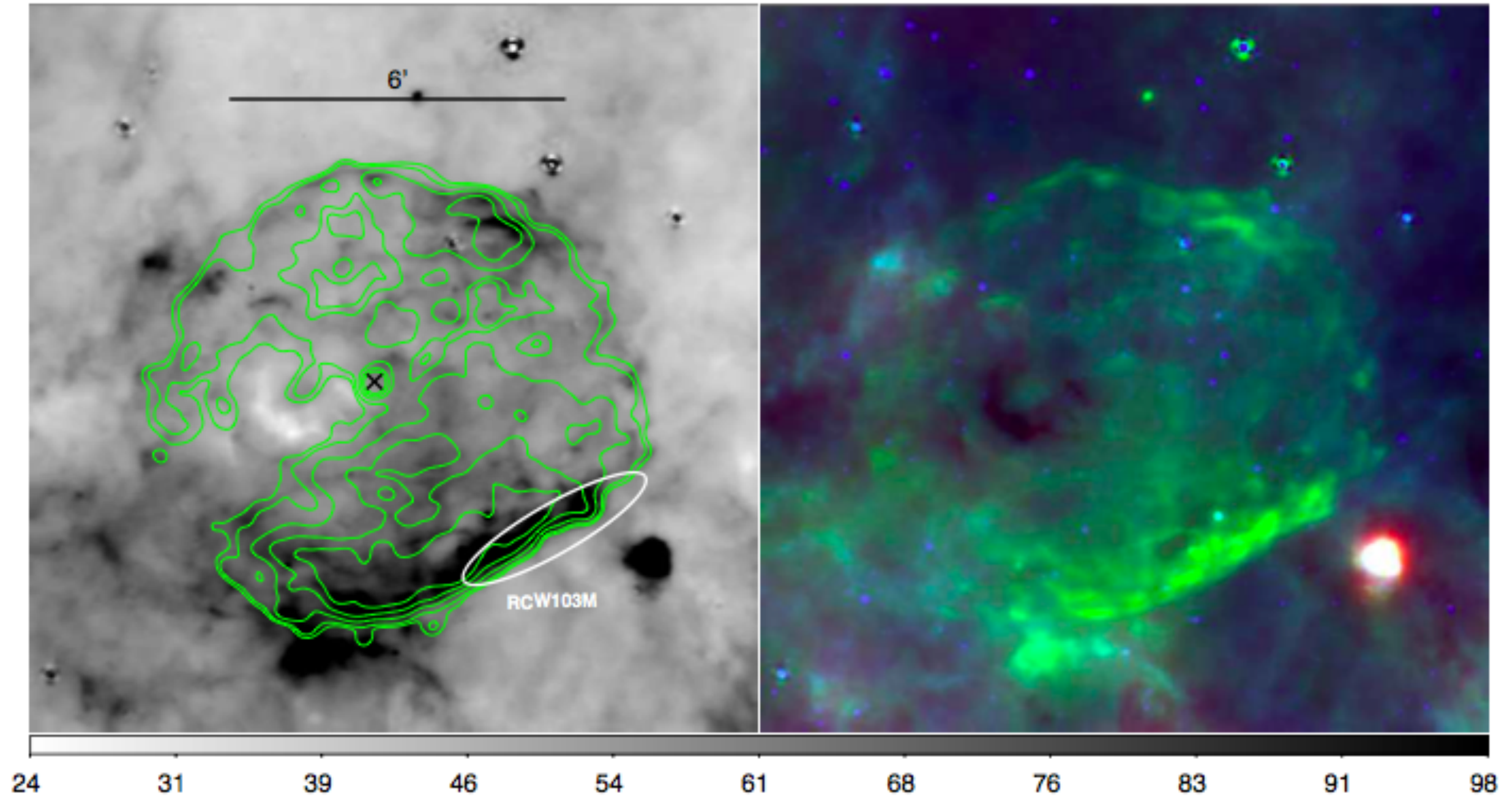}
\epsscale{.9}
\figcaption[f37.pdf]{SNR G332.4-0.4. Contours from Chandra observations: levels are 0.8, 1.1, 2, 3.4 and $5.5\times10^{-6}$ photons/cm$^2$/sec/pixel. The cross shows the location of the X-ray source 2E 1613.5-5053. Region used for partial photometry is also indicated in the figure (in white). \label{snrG332404}}  
\end{figure}

\begin{figure}[h*]
\plotone{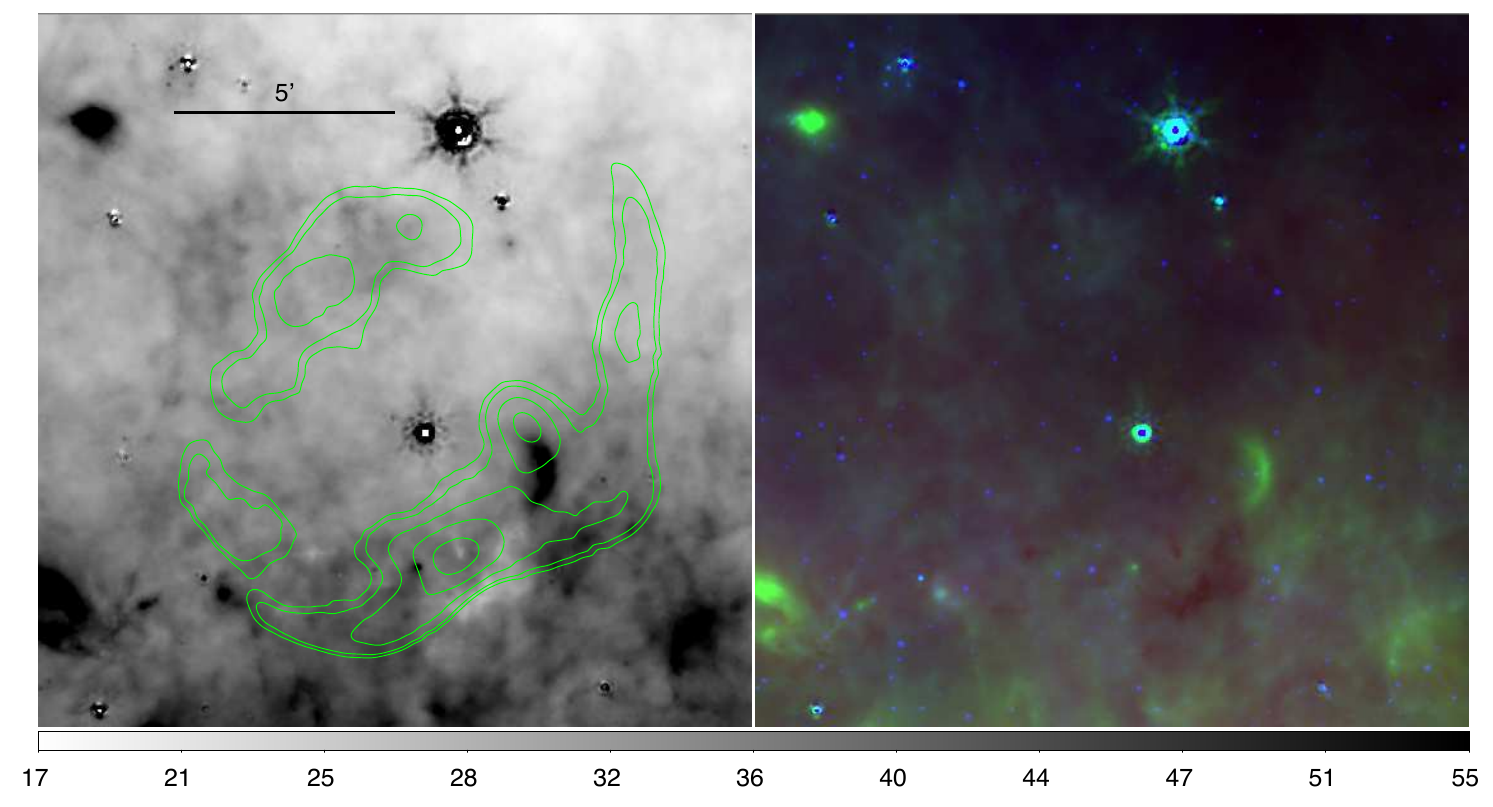}
\epsscale{.9}
\figcaption[f38.pdf]{SNR G336.7+0.5. Contours from MOST observations: levels are 35, 45, 76, 130 and 200 mJy/beam.\label{snrG336705}}  
\end{figure}

\begin{figure}[h*]
\plotone{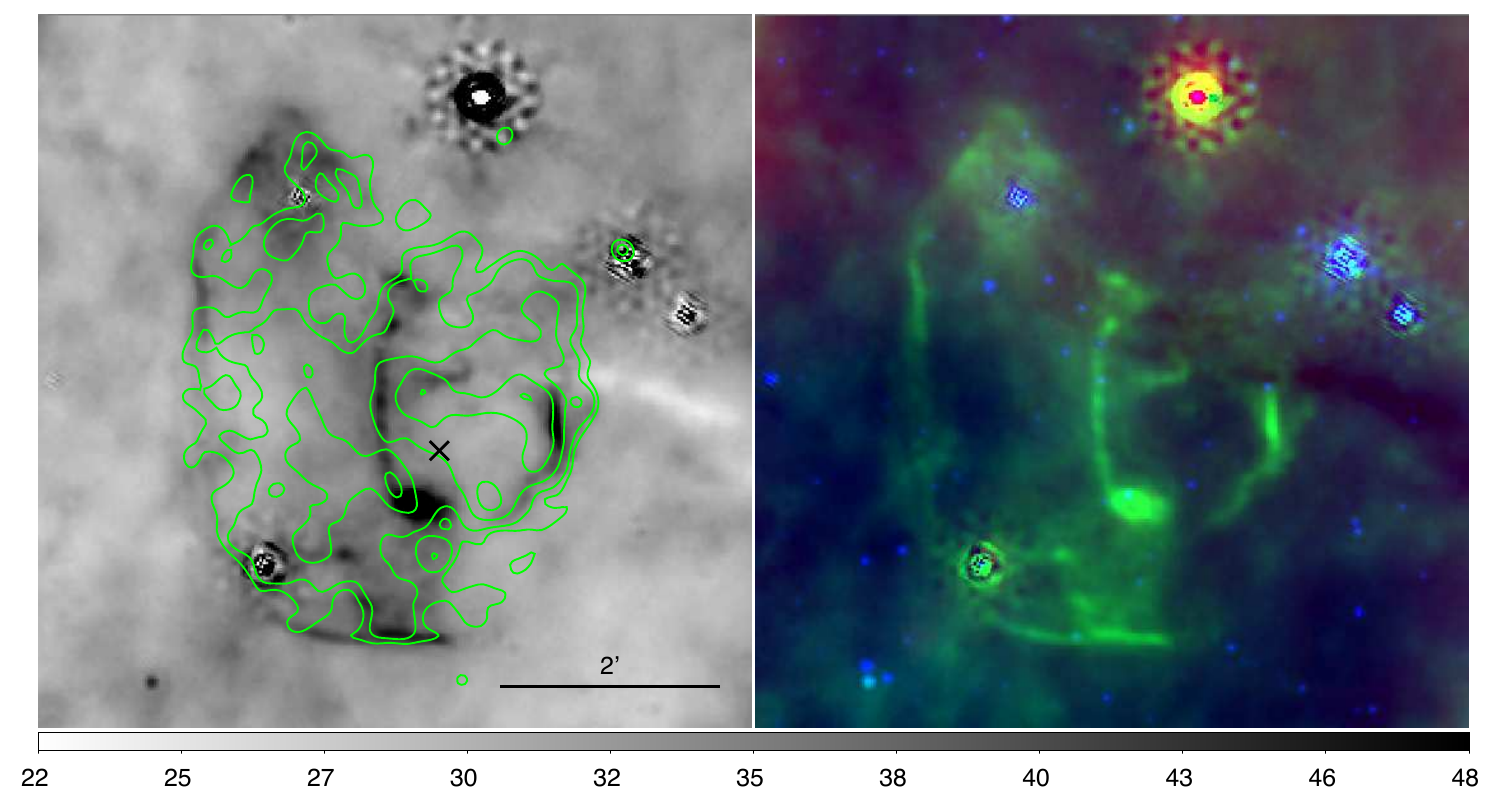}
\epsscale{.9}
\figcaption[f39.pdf]{SNR G337.2-0.7. Contours from Chandra observations: levels are 0.8, 1, 1.6, 2.6 and $4\times10^{-7}$ photons/cm$^2$/sec/pixel. The cross shows the location of the X-ray source J163931.4-475019.\label{snrG337202}}  
\end{figure}

\begin{figure}[h*]
\plotone{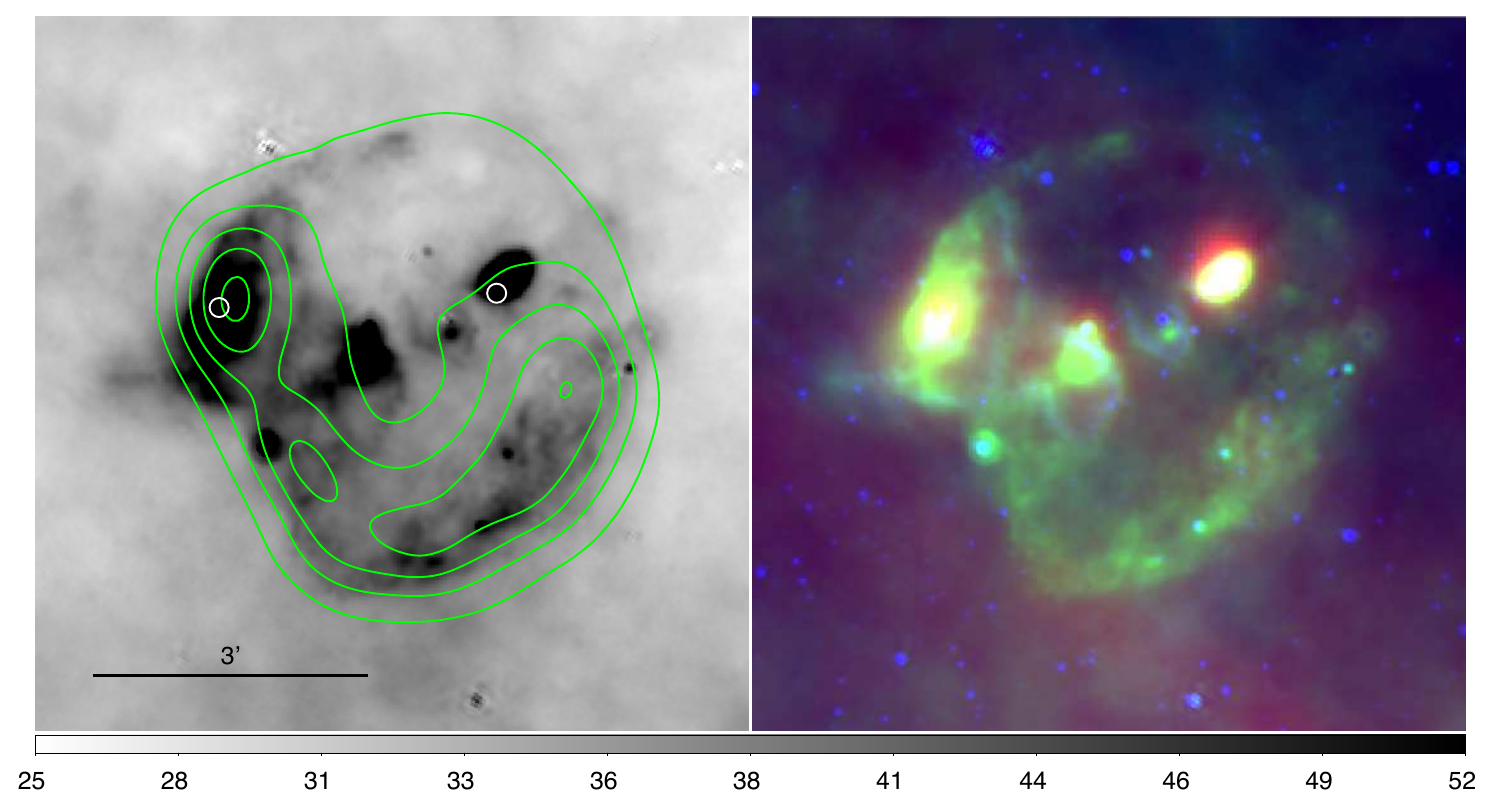}
\epsscale{.9}
\figcaption[f40.pdf]{SNR G340.6-0.3. Contours from MOST observations: levels are from 50 to 350, in steps of 75 mJy/beam. Circles represent the location of two IRAS point sources.\label{snrG340603}}  
\end{figure}

\begin{figure}[h*]
\plotone{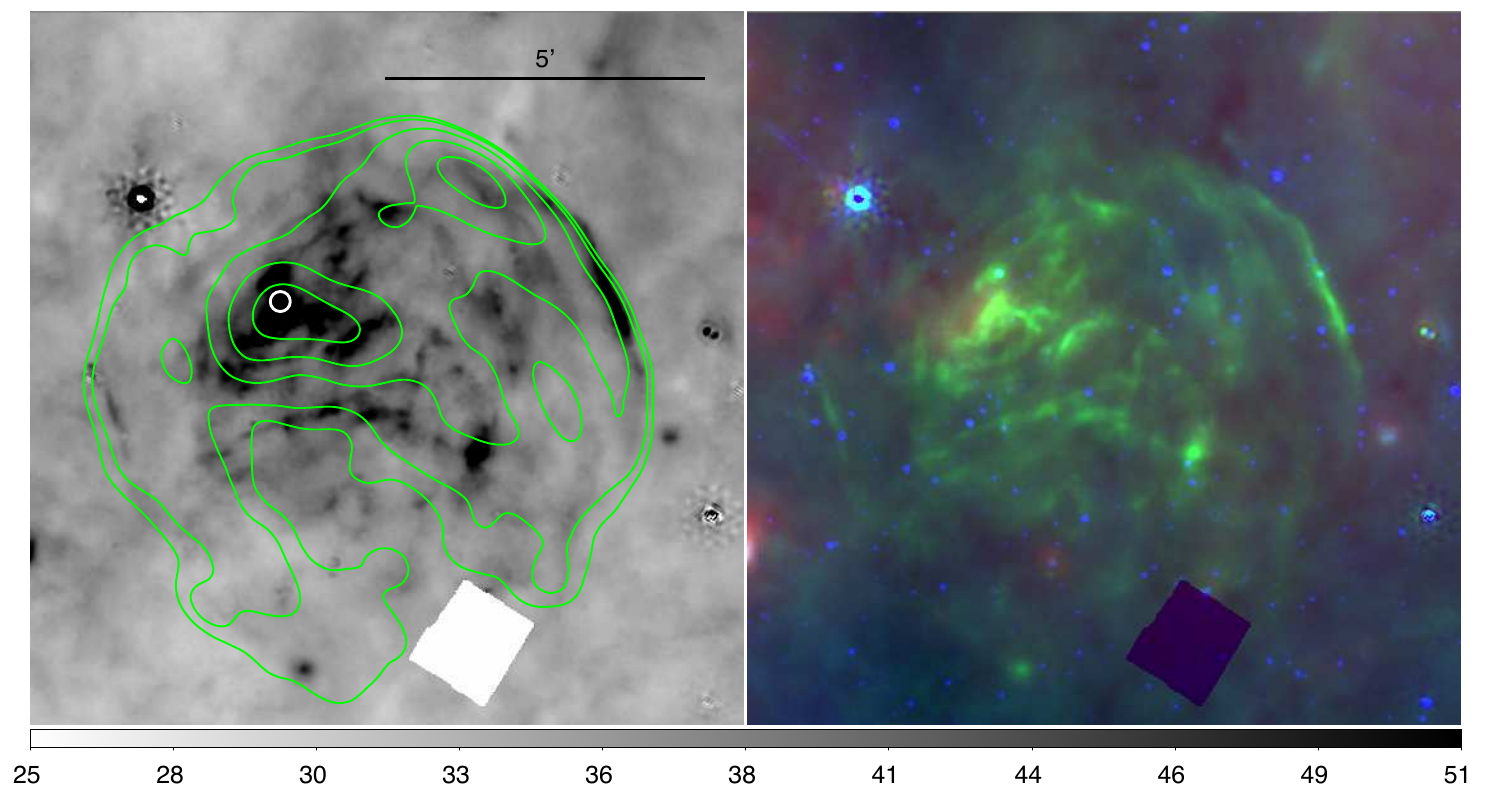}
\epsscale{.9}
\figcaption[f41.ps]{SNR G344.7-0.1. Contours from MOST observations: levels are 16, 21, 35, 58 and 90 mJy/beam. Circle represents the location of a catalogued IRAS point source corresponding to the diffuse emission peak. \label{snrG344701}}  
\end{figure}

\begin{figure}[h*]
\plotone{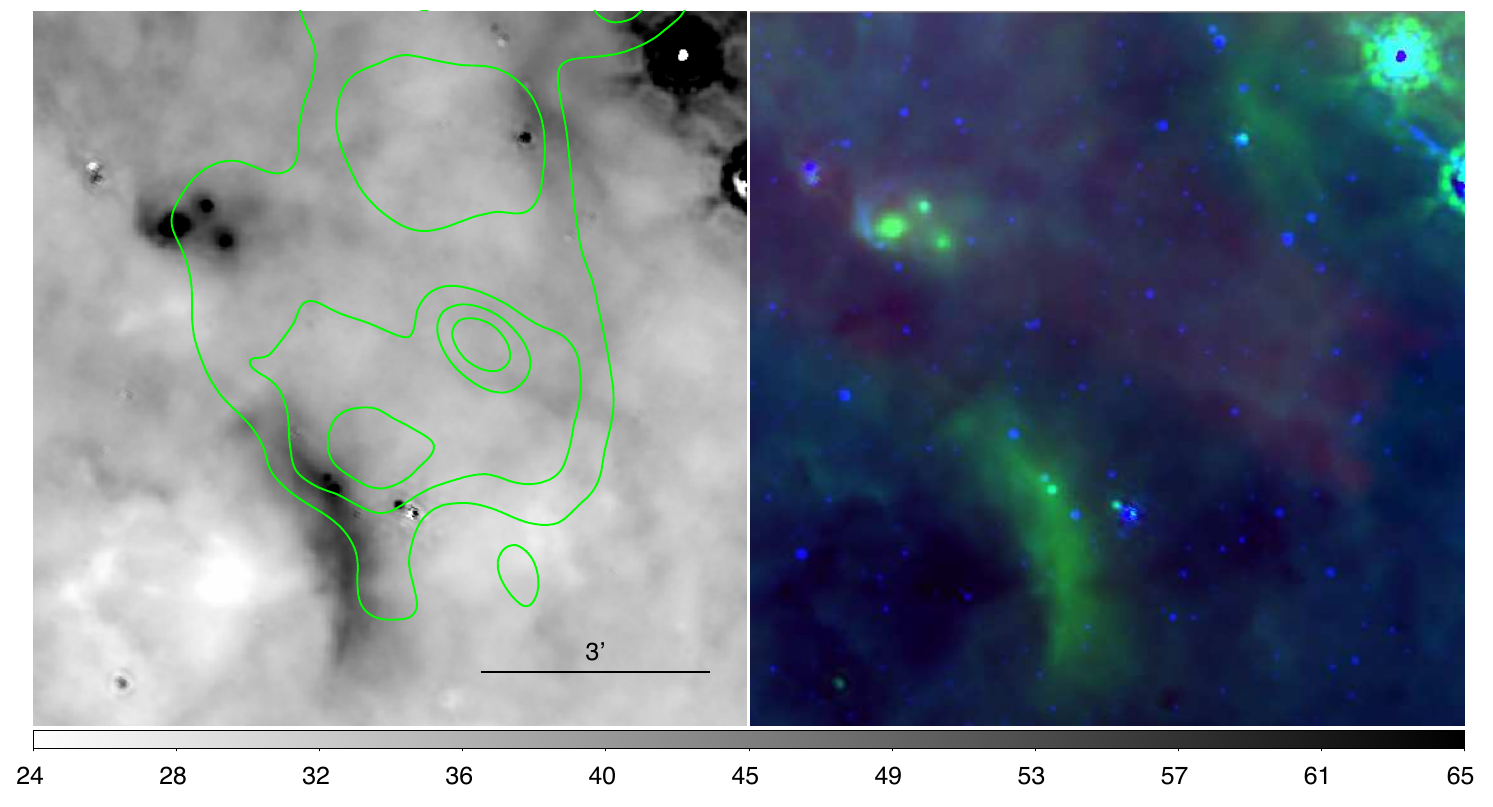}
\epsscale{.9}
\figcaption[f42.ps]{SNR G345.7-0.2. Contours from MOST observations: levels are from 10 to 40, in steps of 10 mJy/beam. PSR B1703-40 is located at brightest radio peak.\label{snrG345702}}  
\end{figure}

\begin{figure}[h*]
\plotone{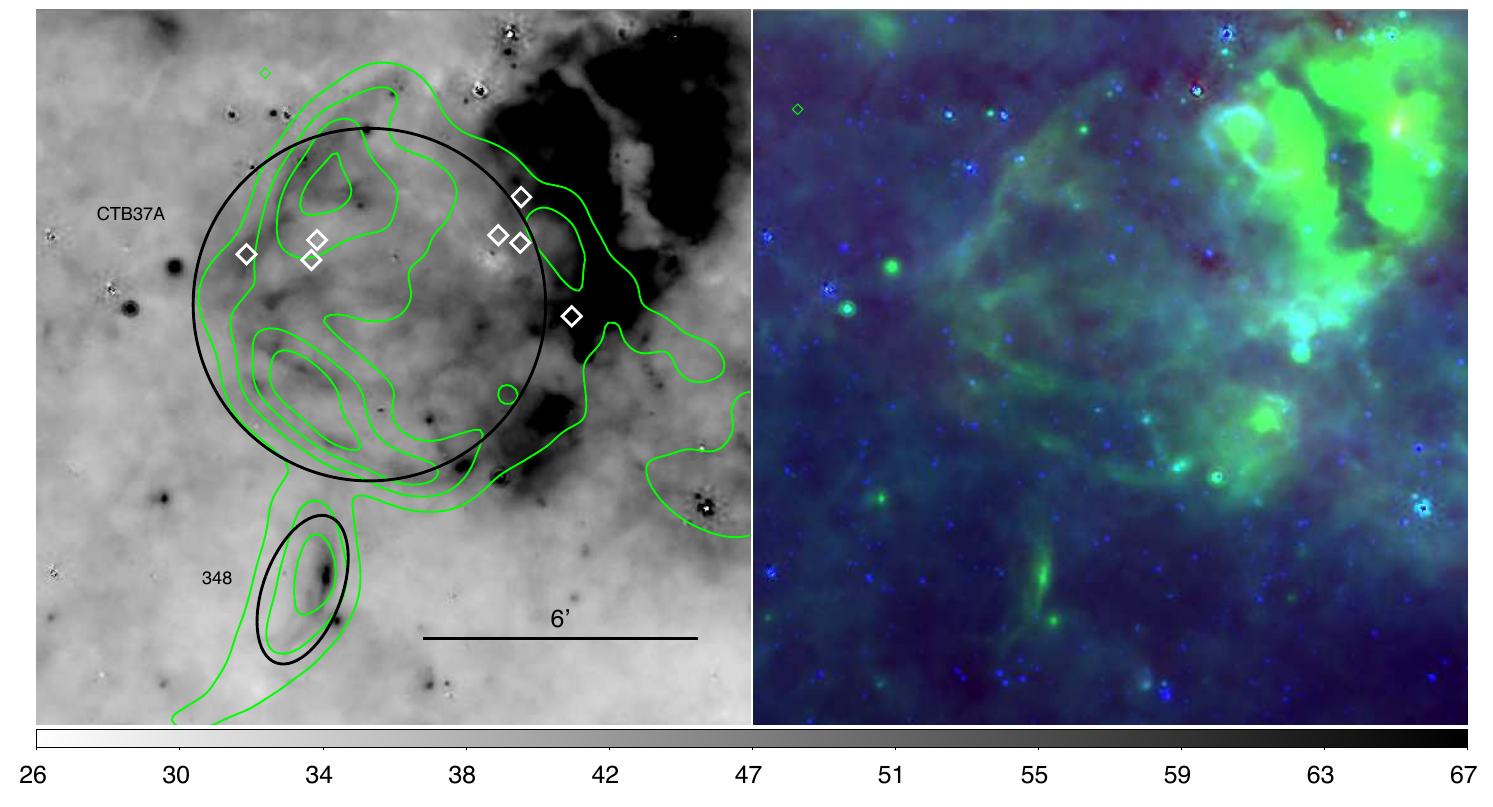}
\epsscale{.9}
\figcaption[f43.pdf]{SNR G348.5-0.0 (south-central countours) and the larger SNR G348.5+0.1. Contours from MOST observations: levels are 0.3, 0.7, 1.1, 1.6 and 2 Jy/beam. Diamonds represent maser locations. Regions used for partial photometry are also indicated in the figure (in black).\label{snrG348db}}  
\end{figure}

\begin{figure}[h*]
\plotone{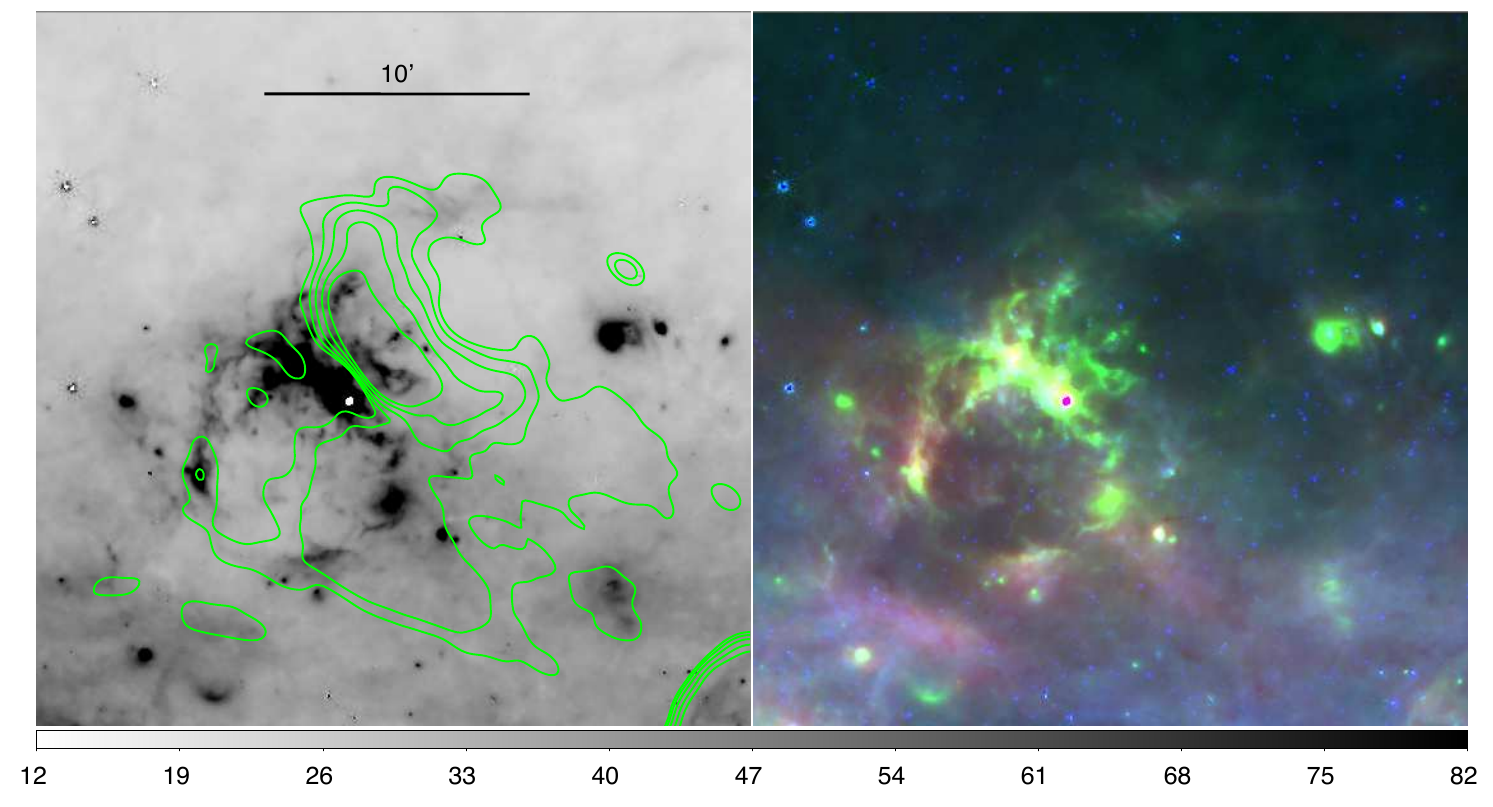}
\epsscale{.9}
\figcaption[f44.pdf]{SNR G348.7+0.3. Contours from MOST observations: levels are 0.03, 0.08, 0.16, 0.32 and 0.55 Jy/beam. \label{snr3487}}  
\end{figure}

\begin{figure}[h*]
\plotone{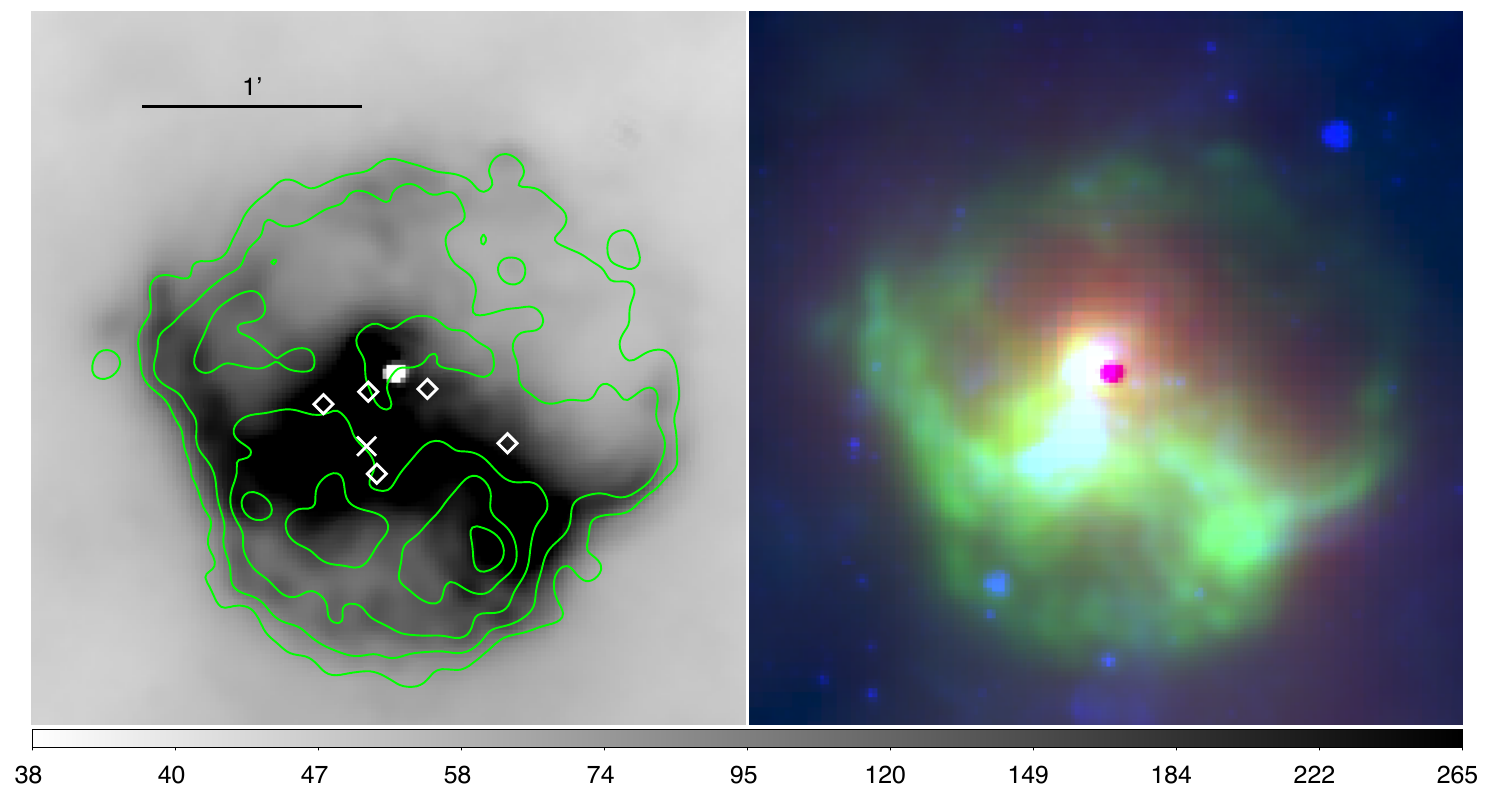}
\epsscale{.9}
\figcaption[f45.pdf]{SNR G349.7-0.2. Contours from Chandra observations: levels are 0.3, 0.7, 1.9, 3.8, 6.5 and $10\times10^{-7}$ photons/cm$^2$/sec/pixel. Masers are represented by diamonds and the cross marks the location of X-ray source J171801.0-372617. \label{snrG349702}}  
\end{figure}

\clearpage

\small{

}


\begin{thebibliography}{110}
\expandafter\ifx\csname natexlab\endcsname\relax\def\natexlab#1{#1}\fi

\bibitem[{{Aharonian} {et~al.}(2008){Aharonian}, {Akhperjanian}, {Barres de
  Almeida}, {Bazer-Bachi}, {Behera}, {Beilicke}, {Benbow}, {Bernl{\"o}hr},
  {Boisson}, {Borrel}, {Braun}, {Brion}, {Brucker}, {B{\"u}hler}, {Bulik},
  {B{\"u}sching}, {Boutelier}, {Carrigan}, {Chadwick}, {Chaves}, {Chounet},
  {Clapson}, {Coignet}, {Cornils}, {Costamante}, {Dalton}, {Degrange},
  {Dickinson}, {Djannati-Ata{\"i}}, {Domainko}, {O'C.~Drury}, {Dubois},
  {Dubus}, {Dyks}, {Egberts}, {Emmanoulopoulos}, {Espigat}, {Farnier},
  {Feinstein}, {Fiasson}, {F{\"o}rster}, {Fontaine}, {Funk}, {F{\"u}{\ss}ling},
  {Gabici}, {Gallant}, {Giebels}, {Glicenstein}, {Gl{\"u}ck}, {Goret},
  {Hadjichristidis}, {Hauser}, {Hauser}, {Heinzelmann}, {Henri}, {Hermann},
  {Hinton}, {Hoffmann}, {Hofmann}, {Holleran}, {Hoppe}, {Horns},
  {Jacholkowska}, {de Jager}, {Jung}, {Katarzy{\'n}ski}, {Kaufmann},
  {Kendziorra}, {Kerschhaggl}, {Khangulyan}, {Kh{\'e}lifi}, {Keogh}, {Komin},
  {Kosack}, {Lamanna}, {Latham}, {Lemoine-Goumard}, {Lenain}, {Lohse},
  {Martin}, {Martineau-Huynh}, {Marcowith}, {Masterson}, {Maurin}, {McComb},
  {Moderski}, {Moulin}, {Naumann-Godo}, {de Naurois}, {Nedbal}, {Nekrassov},
  {Nolan}, {Ohm}, {Olive}, {de O{\~n}a Wilhelmi}, {Orford}, {Osborne},
  {Ostrowski}, {Panter}, {Pedaletti}, {Pelletier}, {Petrucci}, {Pita},
  {P{\"u}hlhofer}, {Punch}, {Quirrenbach}, {Raubenheimer}, {Raue}, {Rayner},
  {Renaud}, {Rieger}, {Reimer}, {Ripken}, {Rob}, {Rosier-Lees}, {Rowell},
  {Rudak}, {Ruppel}, {Sahakian}, {Santangelo}, {Schlickeiser}, {Sch{\"o}ck},
  {Schr{\"o}der}, {Schwanke}, {Schwarzburg}, {Schwemmer}, {Shalchi}, {Skilton},
  {Sol}, {Spangler}, {Stawarz}, {Steenkamp}, {Stegmann}, {Superina}, {Tam},
  {Tavernet}, {Terrier}, {van Eldik}, {Vasileiadis}, {Venter}, {Vialle},
  {Vincent}, {Vivier}, {V{\"o}lk}, {Volpe}, {Wagner}, {Ward}, {Zdziarski}, \&
  {Zech}}]{2008A&A...486..829A}
{Aharonian}, F., {Akhperjanian}, A.~G., {Barres de Almeida}, U., {Bazer-Bachi},
  A.~R., {Behera}, B., {Beilicke}, M., {Benbow}, W., {Bernl{\"o}hr}, K.,
  {Boisson}, C., {Borrel}, V., {Braun}, I., {Brion}, E., {Brucker}, J.,
  {B{\"u}hler}, R., {Bulik}, T., {B{\"u}sching}, I., {Boutelier}, T.,
  {Carrigan}, S., {Chadwick}, P.~M., {Chaves}, R.~C.~G., {Chounet}, L.,
  {Clapson}, A.~C., {Coignet}, G., {Cornils}, R., {Costamante}, L., {Dalton},
  M., {Degrange}, B., {Dickinson}, H.~J., {Djannati-Ata{\"i}}, A., {Domainko},
  W., {O'C.~Drury}, L., {Dubois}, F., {Dubus}, G., {Dyks}, J., {Egberts}, K.,
  {Emmanoulopoulos}, D., {Espigat}, P., {Farnier}, C., {Feinstein}, F.,
  {Fiasson}, A., {F{\"o}rster}, A., {Fontaine}, G., {Funk}, S.,
  {F{\"u}{\ss}ling}, M., {Gabici}, S., {Gallant}, Y.~A., {Giebels}, B.,
  {Glicenstein}, J.~F., {Gl{\"u}ck}, B., {Goret}, P., {Hadjichristidis}, C.,
  {Hauser}, D., {Hauser}, M., {Heinzelmann}, G., {Henri}, G., {Hermann}, G.,
  {Hinton}, J.~A., {Hoffmann}, A., {Hofmann}, W., {Holleran}, M., {Hoppe}, S.,
  {Horns}, D., {Jacholkowska}, A., {de Jager}, O.~C., {Jung}, I.,
  {Katarzy{\'n}ski}, K., {Kaufmann}, S., {Kendziorra}, E., {Kerschhaggl}, M.,
  {Khangulyan}, D., {Kh{\'e}lifi}, B., {Keogh}, D., {Komin}, N., {Kosack}, K.,
  {Lamanna}, G., {Latham}, I.~J., {Lemoine-Goumard}, M., {Lenain}, J., {Lohse},
  T., {Martin}, J.~M., {Martineau-Huynh}, O., {Marcowith}, A., {Masterson}, C.,
  {Maurin}, D., {McComb}, T.~J.~L., {Moderski}, R., {Moulin}, E.,
  {Naumann-Godo}, M., {de Naurois}, M., {Nedbal}, D., {Nekrassov}, D., {Nolan},
  S.~J., {Ohm}, S., {Olive}, J., {de O{\~n}a Wilhelmi}, E., {Orford}, K.~J.,
  {Osborne}, J.~L., {Ostrowski}, M., {Panter}, M., {Pedaletti}, G.,
  {Pelletier}, G., {Petrucci}, P., {Pita}, S., {P{\"u}hlhofer}, G., {Punch},
  M., {Quirrenbach}, A., {Raubenheimer}, B.~C., {Raue}, M., {Rayner}, S.~M.,
  {Renaud}, M., {Rieger}, F., {Reimer}, O., {Ripken}, J., {Rob}, L.,
  {Rosier-Lees}, S., {Rowell}, G., {Rudak}, B., {Ruppel}, J., {Sahakian}, V.,
  {Santangelo}, A., {Schlickeiser}, R., {Sch{\"o}ck}, F.~M., {Schr{\"o}der},
  R., {Schwanke}, U., {Schwarzburg}, S., {Schwemmer}, S., {Shalchi}, A.,
  {Skilton}, J.~L., {Sol}, H., {Spangler}, D., {Stawarz}, {\L}., {Steenkamp},
  R., {Stegmann}, C., {Superina}, G., {Tam}, P.~H., {Tavernet}, J., {Terrier},
  R., {van Eldik}, C., {Vasileiadis}, G., {Venter}, C., {Vialle}, J.~P.,
  {Vincent}, P., {Vivier}, M., {V{\"o}lk}, H.~J., {Volpe}, F., {Wagner}, S.~J.,
  {Ward}, M., {Zdziarski}, A.~A., \& {Zech}, A. 2008, \aap, 486, 829

\bibitem[{{Appleton} {et~al.}(2004){Appleton}, {Fadda}, {Marleau}, {Frayer},
  {Helou}, {Condon}, {Choi}, {Yan}, {Lacy}, {Wilson}, {Armus}, {Chapman},
  {Fang}, {Heinrichson}, {Im}, {Jannuzi}, {Storrie-Lombardi}, {Shupe},
  {Soifer}, {Squires}, \& {Teplitz}}]{2004ApJS..154..147A}
{Appleton}, P.~N., {Fadda}, D.~T., {Marleau}, F.~R., {Frayer}, D.~T., {Helou},
  G., {Condon}, J.~J., {Choi}, P.~I., {Yan}, L., {Lacy}, M., {Wilson}, G.,
  {Armus}, L., {Chapman}, S.~C., {Fang}, F., {Heinrichson}, I., {Im}, M.,
  {Jannuzi}, B.~T., {Storrie-Lombardi}, L.~J., {Shupe}, D., {Soifer}, B.~T.,
  {Squires}, G., \& {Teplitz}, H.~I. 2004, \apjs, 154, 147

\bibitem[{{Arendt}(1989)}]{1989ApJS...70..181A}
{Arendt}, R.~G. 1989, \apjs, 70, 181

\bibitem[{{Arendt} {et~al.}(1999){Arendt}, {Dwek}, \&
  {Moseley}}]{1999ApJ...521..234A}
{Arendt}, R.~G., {Dwek}, E., \& {Moseley}, S.~H. 1999, \apj, 521, 234

\bibitem[{{Barlow} {et~al.}(2010){Barlow}, {Krause}, {Swinyard}, {Sibthorpe},
  {Besel}, {Wesson}, {Ivison}, {Dunne}, {Gear}, {Gomez}, {Hargrave}, {Henning},
  {Leeks}, {Lim}, {Olofsson}, \& {Polehampton}}]{2010A&A...518L.138B}
{Barlow}, M.~J., {Krause}, O., {Swinyard}, B.~M., {Sibthorpe}, B., {Besel}, M.,
  {Wesson}, R., {Ivison}, R.~J., {Dunne}, L., {Gear}, W.~K., {Gomez}, H.~L.,
  {Hargrave}, P.~C., {Henning}, T., {Leeks}, S.~J., {Lim}, T.~L., {Olofsson},
  G., \& {Polehampton}, E.~T. 2010, \aap, 518, L138+

\bibitem[{{Bell}(2003)}]{2003ApJ...586..794B}
{Bell}, E.~F. 2003, \apj, 586, 794

\bibitem[{{Benjamin} {et~al.}(2003){Benjamin}, {Churchwell}, {Babler}, {Bania},
  {Clemens}, {Cohen}, {Dickey}, {Indebetouw}, {Jackson}, {Kobulnicky},
  {Lazarian}, {Marston}, {Mathis}, {Meade}, {Seager}, {Stolovy}, {Watson},
  {Whitney}, {Wolff}, \& {Wolfire}}]{2003PASP..115..953B}
{Benjamin}, R.~A., {Churchwell}, E., {Babler}, B.~L., {Bania}, T.~M.,
  {Clemens}, D.~P., {Cohen}, M., {Dickey}, J.~M., {Indebetouw}, R., {Jackson},
  J.~M., {Kobulnicky}, H.~A., {Lazarian}, A., {Marston}, A.~P., {Mathis},
  J.~S., {Meade}, M.~R., {Seager}, S., {Stolovy}, S.~R., {Watson}, C.,
  {Whitney}, B.~A., {Wolff}, M.~J., \& {Wolfire}, M.~G. 2003, \pasp, 115, 953

\bibitem[{{Blair} {et~al.}(2007){Blair}, {Ghavamian}, {Long}, {Williams},
  {Borkowski}, {Reynolds}, \& {Sankrit}}]{2007ApJ...662..998B}
{Blair}, W.~P., {Ghavamian}, P., {Long}, K.~S., {Williams}, B.~J., {Borkowski},
  K.~J., {Reynolds}, S.~P., \& {Sankrit}, R. 2007, \apj, 662, 998

\bibitem[{{Borkowski} {et~al.}(2006){Borkowski}, {Williams}, {Reynolds},
  {Blair}, {Ghavamian}, {Sankrit}, {Hendrick}, {Long}, {Raymond}, {Smith},
  {Points}, \& {Winkler}}]{2006ApJ...642L.141B}
{Borkowski}, K.~J., {Williams}, B.~J., {Reynolds}, S.~P., {Blair}, W.~P.,
  {Ghavamian}, P., {Sankrit}, R., {Hendrick}, S.~P., {Long}, K.~S., {Raymond},
  J.~C., {Smith}, R.~C., {Points}, S., \& {Winkler}, P.~F. 2006, \apjl, 642,
  L141

\bibitem[{{Boyle} {et~al.}(2007){Boyle}, {Cornwell}, {Middelberg}, {Norris},
  {Appleton}, \& {Smail}}]{2007MNRAS.376.1182B}
{Boyle}, B.~J., {Cornwell}, T.~J., {Middelberg}, E., {Norris}, R.~P.,
  {Appleton}, P.~N., \& {Smail}, I. 2007, \mnras, 376, 1182

\bibitem[{{Broadbent} {et~al.}(1989){Broadbent}, {Osborne}, \&
  {Haslam}}]{1989MNRAS.237..381B}
{Broadbent}, A., {Osborne}, J.~L., \& {Haslam}, C.~G.~T. 1989, \mnras, 237, 381

\bibitem[{{Brogan} {et~al.}(2006){Brogan}, {Gelfand}, {Gaensler}, {Kassim}, \&
  {Lazio}}]{2006ApJ...639L..25B}
{Brogan}, C.~L., {Gelfand}, J.~D., {Gaensler}, B.~M., {Kassim}, N.~E., \&
  {Lazio}, T.~J.~W. 2006, \apjl, 639, L25

\bibitem[{{Camilo} {et~al.}(2006){Camilo}, {Ransom}, {Gaensler}, {Slane},
  {Lorimer}, {Reynolds}, {Manchester}, \& {Murray}}]{2006ApJ...637..456C}
{Camilo}, F., {Ransom}, S.~M., {Gaensler}, B.~M., {Slane}, P.~O., {Lorimer},
  D.~R., {Reynolds}, J., {Manchester}, R.~N., \& {Murray}, S.~S. 2006, \apj,
  637, 456

\bibitem[{{Carey} {et~al.}(2009){Carey}, {Noriega-Crespo}, {Mizuno}, {Shenoy},
  {Paladini}, {Kraemer}, {Price}, {Flagey}, {Ryan}, {Ingalls}, {Kuchar},
  {Pinheiro Gon{\c c}alves}, {Indebetouw}, {Billot}, {Marleau}, {Padgett},
  {Rebull}, {Bressert}, {Ali}, {Molinari}, {Martin}, {Berriman}, {Boulanger},
  {Latter}, {Miville-Deschenes}, {Shipman}, \& {Testi}}]{2009PASP..121...76C}
{Carey}, S.~J., {Noriega-Crespo}, A., {Mizuno}, D.~R., {Shenoy}, S.,
  {Paladini}, R., {Kraemer}, K.~E., {Price}, S.~D., {Flagey}, N., {Ryan}, E.,
  {Ingalls}, J.~G., {Kuchar}, T.~A., {Pinheiro Gon{\c c}alves}, D.,
  {Indebetouw}, R., {Billot}, N., {Marleau}, F.~R., {Padgett}, D.~L., {Rebull},
  L.~M., {Bressert}, E., {Ali}, B., {Molinari}, S., {Martin}, P.~G.,
  {Berriman}, G.~B., {Boulanger}, F., {Latter}, W.~B., {Miville-Deschenes},
  M.~A., {Shipman}, R., \& {Testi}, L. 2009, \pasp, 121, 76

\bibitem[{{Carter} {et~al.}(1997){Carter}, {Dickel}, \&
  {Bomans}}]{1997PASP..109..990C}
{Carter}, L.~M., {Dickel}, J.~R., \& {Bomans}, D.~J. 1997, \pasp, 109, 990

\bibitem[{{Cesarsky} {et~al.}(1999){Cesarsky}, {Cox}, {Pineau des For{\^e}ts},
  {van Dishoeck}, {Boulanger}, \& {Wright}}]{1999A&A...348..945C}
{Cesarsky}, D., {Cox}, P., {Pineau des For{\^e}ts}, G., {van Dishoeck}, E.~F.,
  {Boulanger}, F., \& {Wright}, C.~M. 1999, \aap, 348, 945

\bibitem[{{Chen} {et~al.}(2004){Chen}, {Su}, {Slane}, \&
  {Wang}}]{2004ApJ...616..885C}
{Chen}, Y., {Su}, Y., {Slane}, P.~O., \& {Wang}, Q.~D. 2004, \apj, 616, 885

\bibitem[{{Churchwell} {et~al.}(2009){Churchwell}, {Babler}, {Meade},
  {Whitney}, {Benjamin}, {Indebetouw}, {Cyganowski}, {Robitaille}, {Povich},
  {Watson}, \& {Bracker}}]{2009PASP..121..213C}
{Churchwell}, E., {Babler}, B.~L., {Meade}, M.~R., {Whitney}, B.~A.,
  {Benjamin}, R., {Indebetouw}, R., {Cyganowski}, C., {Robitaille}, T.~P.,
  {Povich}, M., {Watson}, C., \& {Bracker}, S. 2009, \pasp, 121, 213

\bibitem[{{Clark} \& {Stephenson}(1977)}]{1977QB841.C58......}
{Clark}, D.~H. \& {Stephenson}, F.~R. 1977, {The historical supernovae}, ed.
  {Clark, D.~H.~\& Stephenson, F.~R.}

\bibitem[{{Compiegne} {et~al.}(2010){Compiegne}, {Flagey}, {Noriega-Crespo},
  {Martin}, {Bernard}, {Paladini}, \& {Molinari}}]{2010arXiv1010.2774C}
{Compiegne}, M., {Flagey}, N., {Noriega-Crespo}, A., {Martin}, P.~G.,
  {Bernard}, J., {Paladini}, R., \& {Molinari}, S. 2010, ArXiv e-prints

\bibitem[{{Compi{\`e}gne} {et~al.}(2010){Compi{\`e}gne}, {Flagey},
  {Noriega-Crespo}, {Martin}, {Bernard}, {Paladini}, \&
  {Molinari}}]{2010ApJ...724L..44C}
{Compi{\`e}gne}, M., {Flagey}, N., {Noriega-Crespo}, A., {Martin}, P.~G.,
  {Bernard}, J., {Paladini}, R., \& {Molinari}, S. 2010, \apjl, 724, L44

\bibitem[{{Douvion} {et~al.}(2001){Douvion}, {Lagage}, {Cesarsky}, \&
  {Dwek}}]{2001A&A...373..281D}
{Douvion}, T., {Lagage}, P.~O., {Cesarsky}, C.~J., \& {Dwek}, E. 2001, \aap,
  373, 281

\bibitem[{{Draine}(1981)}]{1981ApJ...245..880D}
{Draine}, B.~T. 1981, \apj, 245, 880

\bibitem[{{Draine}(2003)}]{2003ARA&A..41..241D}
---. 2003, \araa, 41, 241

\bibitem[{{Dubner} {et~al.}(1993){Dubner}, {Moffett}, {Goss}, \&
  {Winkler}}]{1993AJ....105.2251D}
{Dubner}, G.~M., {Moffett}, D.~A., {Goss}, W.~M., \& {Winkler}, P.~F. 1993,
  \aj, 105, 2251

\bibitem[{{Dwek}(1987)}]{1987ApJ...322..812D}
{Dwek}, E. 1987, \apj, 322, 812

\bibitem[{{Dwek} \& {Arendt}(1992)}]{1992ARA&A..30...11D}
{Dwek}, E. \& {Arendt}, R.~G. 1992, \araa, 30, 11

\bibitem[{{Dwek} {et~al.}(1987){Dwek}, {Petre}, {Szymkowiak}, \&
  {Rice}}]{1987ApJ...320L..27D}
{Dwek}, E., {Petre}, R., {Szymkowiak}, A., \& {Rice}, W.~L. 1987, \apjl, 320,
  L27

\bibitem[{{Dwek} \& {Werner}(1981)}]{1981ApJ...248..138D}
{Dwek}, E. \& {Werner}, M.~W. 1981, \apj, 248, 138

\bibitem[{{Engelbracht} {et~al.}(2007){Engelbracht}, {Blaylock}, {Su}, {Rho},
  {Rieke}, {Muzerolle}, {Padgett}, {Hines}, {Gordon}, {Fadda},
  {Noriega-Crespo}, {Kelly}, {Latter}, {Hinz}, {Misselt}, {Morrison},
  {Stansberry}, {Shupe}, {Stolovy}, {Wheaton}, {Young}, {Neugebauer},
  {Wachter}, {P{\'e}rez-Gonz{\'a}lez}, {Frayer}, \&
  {Marleau}}]{2007PASP..119..994E}
{Engelbracht}, C.~W., {Blaylock}, M., {Su}, K.~Y.~L., {Rho}, J., {Rieke},
  G.~H., {Muzerolle}, J., {Padgett}, D.~L., {Hines}, D.~C., {Gordon}, K.~D.,
  {Fadda}, D., {Noriega-Crespo}, A., {Kelly}, D.~M., {Latter}, W.~B., {Hinz},
  J.~L., {Misselt}, K.~A., {Morrison}, J.~E., {Stansberry}, J.~A., {Shupe},
  D.~L., {Stolovy}, S., {Wheaton}, W.~A., {Young}, E.~T., {Neugebauer}, G.,
  {Wachter}, S., {P{\'e}rez-Gonz{\'a}lez}, P.~G., {Frayer}, D.~T., \&
  {Marleau}, F.~R. 2007, \pasp, 119, 994

\bibitem[{{Fesen} \& {Kirshner}(1980)}]{1980ApJ...242.1023F}
{Fesen}, R.~A. \& {Kirshner}, R.~P. 1980, \apj, 242, 1023

\bibitem[{{Frail} {et~al.}(1996){Frail}, {Goss}, {Reynoso}, {Giacani}, {Green},
  \& {Otrupcek}}]{1996AJ....111.1651F}
{Frail}, D.~A., {Goss}, W.~M., {Reynoso}, E.~M., {Giacani}, E.~B., {Green},
  A.~J., \& {Otrupcek}, R. 1996, \aj, 111, 1651

\bibitem[{{Fuerst} {et~al.}(1987){Fuerst}, {Reich}, \&
  {Sofue}}]{1987A&AS...71...63F}
{Fuerst}, E., {Reich}, W., \& {Sofue}, Y. 1987, \aaps, 71, 63

\bibitem[{{Gaensler} {et~al.}(1998){Gaensler}, {Manchester}, \&
  {Green}}]{1998MNRAS.296..813G}
{Gaensler}, B.~M., {Manchester}, R.~N., \& {Green}, A.~J. 1998, \mnras, 296,
  813

\bibitem[{{Gordon} {et~al.}(2007){Gordon}, {Engelbracht}, {Fadda},
  {Stansberry}, {Wachter}, {Frayer}, {Rieke}, {Noriega-Crespo}, {Latter},
  {Young}, {Neugebauer}, {Balog}, {Beeman}, {Dole}, {Egami}, {Haller}, {Hines},
  {Kelly}, {Marleau}, {Misselt}, {Morrison}, {P{\'e}rez-Gonz{\'a}lez}, {Rho},
  \& {Wheaton}}]{2007PASP..119.1019G}
{Gordon}, K.~D., {Engelbracht}, C.~W., {Fadda}, D., {Stansberry}, J.,
  {Wachter}, S., {Frayer}, D.~T., {Rieke}, G., {Noriega-Crespo}, A., {Latter},
  W.~B., {Young}, E., {Neugebauer}, G., {Balog}, Z., {Beeman}, J.~W., {Dole},
  H., {Egami}, E., {Haller}, E.~E., {Hines}, D., {Kelly}, D., {Marleau}, F.,
  {Misselt}, K., {Morrison}, J., {P{\'e}rez-Gonz{\'a}lez}, P., {Rho}, J., \&
  {Wheaton}, W.~A. 2007, \pasp, 119, 1019

\bibitem[{{Gotthelf} {et~al.}(2000){Gotthelf}, {Vasisht}, {Boylan-Kolchin}, \&
  {Torii}}]{2000ApJ...542L..37G}
{Gotthelf}, E.~V., {Vasisht}, G., {Boylan-Kolchin}, M., \& {Torii}, K. 2000,
  \apjl, 542, L37

\bibitem[{{Green} {et~al.}(1997){Green}, {Frail}, {Goss}, \&
  {Otrupcek}}]{1997AJ....114.2058G}
{Green}, A.~J., {Frail}, D.~A., {Goss}, W.~M., \& {Otrupcek}, R. 1997, \aj,
  114, 2058

\bibitem[{{Green}(2009{\natexlab{a}})}]{2009BASI...37...45G}
{Green}, D.~A. 2009{\natexlab{a}}, Bulletin of the Astronomical Society of
  India, 37, 45

\bibitem[{{Green}(2009{\natexlab{b}})}]{2009MNRAS.399..177G}
---. 2009{\natexlab{b}}, \mnras, 399, 177

\bibitem[{{Green} \& {Dewdney}(1992)}]{1992MNRAS.254..686G}
{Green}, D.~A. \& {Dewdney}, P.~E. 1992, \mnras, 254, 686

\bibitem[{{Guillard} {et~al.}(2010){Guillard}, {Boulanger}, {Cluver},
  {Appleton}, {Pineau des Forets}, \& {Ogle}}]{2010arXiv1004.0677G}
{Guillard}, P., {Boulanger}, F., {Cluver}, M.~E., {Appleton}, P.~N., {Pineau
  des Forets}, G., \& {Ogle}, P. 2010, ArXiv e-prints

\bibitem[{{Harrus} \& {Slane}(1999)}]{1999ApJ...516..811H}
{Harrus}, I.~M. \& {Slane}, P.~O. 1999, \apj, 516, 811

\bibitem[{{Haslam} \& {Osborne}(1987)}]{1987Natur.327..211H}
{Haslam}, C.~G.~T. \& {Osborne}, J.~L. 1987, \nat, 327, 211

\bibitem[{{Helfand} {et~al.}(2006){Helfand}, {Becker}, {White}, {Fallon}, \&
  {Tuttle}}]{2006AJ....131.2525H}
{Helfand}, D.~J., {Becker}, R.~H., {White}, R.~L., {Fallon}, A., \& {Tuttle},
  S. 2006, \aj, 131, 2525

\bibitem[{{Helfand} {et~al.}(1989){Helfand}, {Velusamy}, {Becker}, \&
  {Lockman}}]{1989ApJ...341..151H}
{Helfand}, D.~J., {Velusamy}, T., {Becker}, R.~H., \& {Lockman}, F.~J. 1989,
  \apj, 341, 151

\bibitem[{{Helou} {et~al.}(1985){Helou}, {Soifer}, \&
  {Rowan-Robinson}}]{1985ApJ...298L...7H}
{Helou}, G., {Soifer}, B.~T., \& {Rowan-Robinson}, M. 1985, \apjl, 298, L7

\bibitem[{{Hewitt} {et~al.}(2008){Hewitt}, {Yusef-Zadeh}, \&
  {Wardle}}]{2008ApJ...683..189H}
{Hewitt}, J.~W., {Yusef-Zadeh}, F., \& {Wardle}, M. 2008, \apj, 683, 189

\bibitem[{{Hines} {et~al.}(2004){Hines}, {Rieke}, {Gordon}, {Rho}, {Misselt},
  {Woodward}, {Werner}, {Krause}, {Latter}, {Engelbracht}, {Egami}, {Kelly},
  {Muzerolle}, {Stansberry}, {Su}, {Morrison}, {Young}, {Noriega-Crespo},
  {Padgett}, {Gehrz}, {Polomski}, {Beeman}, \& {Haller}}]{2004ApJS..154..290H}
{Hines}, D.~C., {Rieke}, G.~H., {Gordon}, K.~D., {Rho}, J., {Misselt}, K.~A.,
  {Woodward}, C.~E., {Werner}, M.~W., {Krause}, O., {Latter}, W.~B.,
  {Engelbracht}, C.~W., {Egami}, E., {Kelly}, D.~M., {Muzerolle}, J.,
  {Stansberry}, J.~A., {Su}, K.~Y.~L., {Morrison}, J.~E., {Young}, E.~T.,
  {Noriega-Crespo}, A., {Padgett}, D.~L., {Gehrz}, R.~D., {Polomski}, E.,
  {Beeman}, J.~W., \& {Haller}, E.~E. 2004, \apjs, 154, 290

\bibitem[{{Hwang} {et~al.}(2000){Hwang}, {Petre}, \&
  {Hughes}}]{2000ApJ...532..970H}
{Hwang}, U., {Petre}, R., \& {Hughes}, J.~P. 2000, \apj, 532, 970

\bibitem[{{Kaspi} {et~al.}(2001){Kaspi}, {Roberts}, {Vasisht}, {Gotthelf},
  {Pivovaroff}, \& {Kawai}}]{2001ApJ...560..371K}
{Kaspi}, V.~M., {Roberts}, M.~E., {Vasisht}, G., {Gotthelf}, E.~V.,
  {Pivovaroff}, M., \& {Kawai}, N. 2001, \apj, 560, 371

\bibitem[{{Kassim} {et~al.}(1991){Kassim}, {Weiler}, \&
  {Baum}}]{1991ApJ...374..212K}
{Kassim}, N.~E., {Weiler}, K.~W., \& {Baum}, S.~A. 1991, \apj, 374, 212

\bibitem[{{Keohane} {et~al.}(2007){Keohane}, {Reach}, {Rho}, \&
  {Jarrett}}]{2007ApJ...654..938K}
{Keohane}, J.~W., {Reach}, W.~T., {Rho}, J., \& {Jarrett}, T.~H. 2007, \apj,
  654, 938

\bibitem[{{Koo} {et~al.}(2007){Koo}, {Moon}, {Lee}, {Lee}, \&
  {Matthews}}]{2007ApJ...657..308K}
{Koo}, B., {Moon}, D., {Lee}, H., {Lee}, J., \& {Matthews}, K. 2007, \apj, 657,
  308

\bibitem[{{Kothes} \& {Dougherty}(2007)}]{2007A&A...468..993K}
{Kothes}, R. \& {Dougherty}, S.~M. 2007, \aap, 468, 993

\bibitem[{{Lazendic} {et~al.}(2005){Lazendic}, {Slane}, {Hughes}, {Chen}, \&
  {Dame}}]{2005ApJ...618..733L}
{Lazendic}, J.~S., {Slane}, P.~O., {Hughes}, J.~P., {Chen}, Y., \& {Dame},
  T.~M. 2005, \apj, 618, 733

\bibitem[{{Leahy} \& {Tian}(2008)}]{2008A&A...480L..25L}
{Leahy}, D.~A. \& {Tian}, W.~W. 2008, \aap, 480, L25

\bibitem[{{Lee} {et~al.}(2009){Lee}, {Moon}, {Koo}, {Lee}, \&
  {Matthews}}]{2009ApJ...691.1042L}
{Lee}, H., {Moon}, D., {Koo}, B., {Lee}, J., \& {Matthews}, K. 2009, \apj, 691,
  1042

\bibitem[{{Li} \& {Draine}(2001)}]{2001ApJ...554..778L}
{Li}, A. \& {Draine}, B.~T. 2001, \apj, 554, 778

\bibitem[{{Livingstone} {et~al.}(2006){Livingstone}, {Kaspi}, {Gotthelf}, \&
  {Kuiper}}]{2006ApJ...647.1286L}
{Livingstone}, M.~A., {Kaspi}, V.~M., {Gotthelf}, E.~V., \& {Kuiper}, L. 2006,
  \apj, 647, 1286

\bibitem[{{Mizuno} {et~al.}(2008){Mizuno}, {Carey}, {Noriega-Crespo},
  {Paladini}, {Padgett}, {Shenoy}, {Kuchar}, {Kraemer}, \&
  {Price}}]{2008PASP..120.1028M}
{Mizuno}, D.~R., {Carey}, S.~J., {Noriega-Crespo}, A., {Paladini}, R.,
  {Padgett}, D., {Shenoy}, S., {Kuchar}, T.~A., {Kraemer}, K.~E., \& {Price},
  S.~D. 2008, \pasp, 120, 1028

\bibitem[{{Moffett} \& {Reynolds}(1994)}]{1994ApJ...437..705M}
{Moffett}, D.~A. \& {Reynolds}, S.~P. 1994, \apj, 437, 705

\bibitem[{{Moon} {et~al.}(2009){Moon}, {Koo}, {Lee}, {Matthews}, {Lee}, {Pyo},
  {Seok}, \& {Hayashi}}]{2009ApJ...703L..81M}
{Moon}, D., {Koo}, B., {Lee}, H., {Matthews}, K., {Lee}, J., {Pyo}, T., {Seok},
  J.~Y., \& {Hayashi}, M. 2009, \apjl, 703, L81

\bibitem[{{Morton} {et~al.}(2007){Morton}, {Slane}, {Borkowski}, {Reynolds},
  {Helfand}, {Gaensler}, \& {Hughes}}]{2007ApJ...667..219M}
{Morton}, T.~D., {Slane}, P., {Borkowski}, K.~J., {Reynolds}, S.~P., {Helfand},
  D.~J., {Gaensler}, B.~M., \& {Hughes}, J.~P. 2007, \apj, 667, 219

\bibitem[{{Murphy}(2009)}]{2009ApJ...706..482M}
{Murphy}, E.~J. 2009, \apj, 706, 482

\bibitem[{{Neufeld} {et~al.}(2007){Neufeld}, {Hollenbach}, {Kaufman}, {Snell},
  {Melnick}, {Bergin}, \& {Sonnentrucker}}]{2007ApJ...664..890N}
{Neufeld}, D.~A., {Hollenbach}, D.~J., {Kaufman}, M.~J., {Snell}, R.~L.,
  {Melnick}, G.~J., {Bergin}, E.~A., \& {Sonnentrucker}, P. 2007, \apj, 664,
  890

\bibitem[{{Neufeld} \& {Yuan}(2008)}]{2008ApJ...678..974N}
{Neufeld}, D.~A. \& {Yuan}, Y. 2008, \apj, 678, 974

\bibitem[{{Noriega-Crespo} {et~al.}(2009){Noriega-Crespo}, {Hines}, {Gordon},
  {Marleau}, {Rieke}, {Rho}, \& {Latter}}]{2009eimw.confE..46N}
{Noriega-Crespo}, A., {Hines}, D.~C., {Gordon}, K., {Marleau}, F.~R., {Rieke},
  G.~H., {Rho}, J., \& {Latter}, W.~B. 2009, in The Evolving ISM in the Milky
  Way and Nearby Galaxies

\bibitem[{{Oliva} {et~al.}(1999{\natexlab{a}}){Oliva}, {Lutz}, {Drapatz}, \&
  {Moorwood}}]{1999A&A...341L..75O}
{Oliva}, E., {Lutz}, D., {Drapatz}, S., \& {Moorwood}, A.~F.~M.
  1999{\natexlab{a}}, \aap, 341, L75

\bibitem[{{Oliva} {et~al.}(1999{\natexlab{b}}){Oliva}, {Moorwood}, {Drapatz},
  {Lutz}, \& {Sturm}}]{1999A&A...343..943O}
{Oliva}, E., {Moorwood}, A.~F.~M., {Drapatz}, S., {Lutz}, D., \& {Sturm}, E.
  1999{\natexlab{b}}, \aap, 343, 943

\bibitem[{{Ostriker} \& {Silk}(1973)}]{1973ApJ...184L.113O}
{Ostriker}, J. \& {Silk}, J. 1973, \apjl, 184, L113+

\bibitem[{{Patnaik} {et~al.}(1990){Patnaik}, {Hunt}, {Salter}, {Shaver}, \&
  {Velusamy}}]{1990A&A...232..467P}
{Patnaik}, A.~R., {Hunt}, G.~C., {Salter}, C.~J., {Shaver}, P.~A., \&
  {Velusamy}, T. 1990, \aap, 232, 467

\bibitem[{{Rakowski} {et~al.}(2006){Rakowski}, {Badenes}, {Gaensler},
  {Gelfand}, {Hughes}, \& {Slane}}]{2006ApJ...646..982R}
{Rakowski}, C.~E., {Badenes}, C., {Gaensler}, B.~M., {Gelfand}, J.~D.,
  {Hughes}, J.~P., \& {Slane}, P.~O. 2006, \apj, 646, 982

\bibitem[{{Raymond} {et~al.}(1976){Raymond}, {Cox}, \&
  {Smith}}]{1976ApJ...204..290R}
{Raymond}, J.~C., {Cox}, D.~P., \& {Smith}, B.~W. 1976, \apj, 204, 290

\bibitem[{{Reach} {et~al.}(2005{\natexlab{a}}){Reach}, {Megeath}, {Cohen},
  {Hora}, {Carey}, {Surace}, {Willner}, {Barmby}, {Wilson}, {Glaccum},
  {Lowrance}, {Marengo}, \& {Fazio}}]{2005PASP..117..978R}
{Reach}, W.~T., {Megeath}, S.~T., {Cohen}, M., {Hora}, J., {Carey}, S.,
  {Surace}, J., {Willner}, S.~P., {Barmby}, P., {Wilson}, G., {Glaccum}, W.,
  {Lowrance}, P., {Marengo}, M., \& {Fazio}, G.~G. 2005{\natexlab{a}}, \pasp,
  117, 978

\bibitem[{{Reach} \& {Rho}(2000)}]{2000ApJ...544..843R}
{Reach}, W.~T. \& {Rho}, J. 2000, \apj, 544, 843

\bibitem[{{Reach} {et~al.}(2005{\natexlab{b}}){Reach}, {Rho}, \&
  {Jarrett}}]{2005ApJ...618..297R}
{Reach}, W.~T., {Rho}, J., \& {Jarrett}, T.~H. 2005{\natexlab{b}}, \apj, 618,
  297

\bibitem[{{Reach} {et~al.}(2002){Reach}, {Rho}, {Jarrett}, \&
  {Lagage}}]{2002ApJ...564..302R}
{Reach}, W.~T., {Rho}, J., {Jarrett}, T.~H., \& {Lagage}, P. 2002, \apj, 564,
  302

\bibitem[{{Reach} {et~al.}(2006){Reach}, {Rho}, {Tappe}, {Pannuti}, {Brogan},
  {Churchwell}, {Meade}, {Babler}, {Indebetouw}, \&
  {Whitney}}]{2006AJ....131.1479R}
{Reach}, W.~T., {Rho}, J., {Tappe}, A., {Pannuti}, T.~G., {Brogan}, C.~L.,
  {Churchwell}, E.~B., {Meade}, M.~R., {Babler}, B., {Indebetouw}, R., \&
  {Whitney}, B.~A. 2006, \aj, 131, 1479

\bibitem[{{Reynolds} {et~al.}(2006){Reynolds}, {Borkowski}, {Hwang}, {Harrus},
  {Petre}, \& {Dubner}}]{2006ApJ...652L..45R}
{Reynolds}, S.~P., {Borkowski}, K.~J., {Hwang}, U., {Harrus}, I., {Petre}, R.,
  \& {Dubner}, G. 2006, \apjl, 652, L45

\bibitem[{{Reynoso} \& {Mangum}(2000)}]{2000ApJ...545..874R}
{Reynoso}, E.~M. \& {Mangum}, J.~G. 2000, \apj, 545, 874

\bibitem[{{Reynoso} \& {Mangum}(2001)}]{2001AJ....121..347R}
---. 2001, \aj, 121, 347

\bibitem[{{Rho} {et~al.}(2001){Rho}, {Jarrett}, {Cutri}, \&
  {Reach}}]{2001ApJ...547..885R}
{Rho}, J., {Jarrett}, T.~H., {Cutri}, R.~M., \& {Reach}, W.~T. 2001, \apj, 547,
  885

\bibitem[{{Rho} {et~al.}(2008){Rho}, {Kozasa}, {Reach}, {Smith}, {Rudnick},
  {DeLaney}, {Ennis}, {Gomez}, \& {Tappe}}]{2008ApJ...673..271R}
{Rho}, J., {Kozasa}, T., {Reach}, W.~T., {Smith}, J.~D., {Rudnick}, L.,
  {DeLaney}, T., {Ennis}, J.~A., {Gomez}, H., \& {Tappe}, A. 2008, \apj, 673,
  271

\bibitem[{{Rho} \& {Petre}(1998)}]{1998ApJ...503L.167R}
{Rho}, J. \& {Petre}, R. 1998, \apjl, 503, L167+

\bibitem[{{Rho} {et~al.}(2009){Rho}, {Reach}, {Tappe}, {Rudnick}, {Kozasa},
  {Hwang}, {Andersen}, {Gomez}, {Delaney}, {Dunne}, \&
  {Slavin}}]{2009ASPC..414...22R}
{Rho}, J., {Reach}, W.~T., {Tappe}, A., {Rudnick}, L., {Kozasa}, T., {Hwang},
  U., {Andersen}, M., {Gomez}, H., {Delaney}, T., {Dunne}, L., \& {Slavin}, J.
  2009, in Astronomical Society of the Pacific Conference Series, Vol. 414,
  Astronomical Society of the Pacific Conference Series, ed. {T.~Henning,
  E.~Gr{\"u}n, \& J.~Steinacker}, 22--+

\bibitem[{{Safi-Harb} {et~al.}(2005){Safi-Harb}, {Dubner}, {Petre}, {Holt}, \&
  {Durouchoux}}]{2005ApJ...618..321S}
{Safi-Harb}, S., {Dubner}, G., {Petre}, R., {Holt}, S.~S., \& {Durouchoux}, P.
  2005, \apj, 618, 321

\bibitem[{{Saken} {et~al.}(1992){Saken}, {Fesen}, \&
  {Shull}}]{1992ApJS...81..715S}
{Saken}, J.~M., {Fesen}, R.~A., \& {Shull}, J.~M. 1992, \apjs, 81, 715

\bibitem[{{Seok} {et~al.}(2008){Seok}, {Koo}, {Onaka}, {Ita}, {Lee}, {Lee},
  {Moon}, {Sakon}, {Kaneda}, {Lee}, {Lee}, \& {Kim}}]{2008PASJ...60S.453S}
{Seok}, J.~Y., {Koo}, B.-C., {Onaka}, T., {Ita}, Y., {Lee}, H.-G., {Lee},
  J.-J., {Moon}, D.-S., {Sakon}, I., {Kaneda}, H., {Lee}, H.~M., {Lee}, M.~G.,
  \& {Kim}, S.~E. 2008, \pasj, 60, 453

\bibitem[{{Seward} {et~al.}(2003){Seward}, {Slane}, {Smith}, \&
  {Sun}}]{2003ApJ...584..414S}
{Seward}, F.~D., {Slane}, P.~O., {Smith}, R.~K., \& {Sun}, M. 2003, \apj, 584,
  414

\bibitem[{{Shaver} \& {Goss}(1970)}]{1970AuJPA..14..133S}
{Shaver}, P.~A. \& {Goss}, W.~M. 1970, Australian Journal of Physics
  Astrophysical Supplement, 14, 133

\bibitem[{{Sibthorpe} {et~al.}(2010){Sibthorpe}, {Ade}, {Bock}, {Chapin},
  {Devlin}, {Dicker}, {Griffin}, {Gundersen}, {Halpern}, {Hargrave}, {Hughes},
  {Jeong}, {Kaneda}, {Klein}, {Koo}, {Lee}, {Marsden}, {Martin}, {Mauskopf},
  {Moon}, {Netterfield}, {Olmi}, {Pascale}, {Patanchon}, {Rex}, {Roy}, {Scott},
  {Semisch}, {Truch}, {Tucker}, {Tucker}, {Viero}, \&
  {Wiebe}}]{2010ApJ...719.1553S}
{Sibthorpe}, B., {Ade}, P.~A.~R., {Bock}, J.~J., {Chapin}, E.~L., {Devlin},
  M.~J., {Dicker}, S., {Griffin}, M., {Gundersen}, J.~O., {Halpern}, M.,
  {Hargrave}, P.~C., {Hughes}, D.~H., {Jeong}, W., {Kaneda}, H., {Klein}, J.,
  {Koo}, B., {Lee}, H., {Marsden}, G., {Martin}, P.~G., {Mauskopf}, P., {Moon},
  D., {Netterfield}, C.~B., {Olmi}, L., {Pascale}, E., {Patanchon}, G., {Rex},
  M., {Roy}, A., {Scott}, D., {Semisch}, C., {Truch}, M.~D.~P., {Tucker}, C.,
  {Tucker}, G.~S., {Viero}, M.~P., \& {Wiebe}, D.~V. 2010, \apj, 719, 1553

\bibitem[{{Slane} {et~al.}(2002){Slane}, {Chen}, {Lazendic}, \&
  {Hughes}}]{2002ApJ...580..904S}
{Slane}, P., {Chen}, Y., {Lazendic}, J.~S., \& {Hughes}, J.~P. 2002, \apj, 580,
  904

\bibitem[{{Stil} {et~al.}(2006){Stil}, {Taylor}, {Dickey}, {Kavars}, {Martin},
  {Rothwell}, {Boothroyd}, {Lockman}, \&
  {McClure-Griffiths}}]{2006AJ....132.1158S}
{Stil}, J.~M., {Taylor}, A.~R., {Dickey}, J.~M., {Kavars}, D.~W., {Martin},
  P.~G., {Rothwell}, T.~A., {Boothroyd}, A.~I., {Lockman}, F.~J., \&
  {McClure-Griffiths}, N.~M. 2006, \aj, 132, 1158

\bibitem[{{Strom} \& {Greidanus}(1992)}]{1992Natur.358..654S}
{Strom}, R.~G. \& {Greidanus}, H. 1992, \nat, 358, 654

\bibitem[{{Sugizaki} {et~al.}(2001){Sugizaki}, {Mitsuda}, {Kaneda},
  {Matsuzaki}, {Yamauchi}, \& {Koyama}}]{2001ApJS..134...77S}
{Sugizaki}, M., {Mitsuda}, K., {Kaneda}, H., {Matsuzaki}, K., {Yamauchi}, S.,
  \& {Koyama}, K. 2001, \apjs, 134, 77

\bibitem[{{Sun} {et~al.}(2004){Sun}, {Seward}, {Smith}, \&
  {Slane}}]{2004ApJ...605..742S}
{Sun}, M., {Seward}, F.~D., {Smith}, R.~K., \& {Slane}, P.~O. 2004, \apj, 605,
  742

\bibitem[{{Tappe} {et~al.}(2006){Tappe}, {Rho}, \&
  {Reach}}]{2006ApJ...653..267T}
{Tappe}, A., {Rho}, J., \& {Reach}, W.~T. 2006, \apj, 653, 267

\bibitem[{{Tian} \& {Leahy}(2008{\natexlab{a}})}]{2008ApJ...677..292T}
{Tian}, W.~W. \& {Leahy}, D.~A. 2008{\natexlab{a}}, \apj, 677, 292

\bibitem[{{Tian} \& {Leahy}(2008{\natexlab{b}})}]{2008MNRAS.391L..54T}
---. 2008{\natexlab{b}}, \mnras, 391, L54

\bibitem[{{Troja} {et~al.}(2006){Troja}, {Bocchino}, \&
  {Reale}}]{2006ApJ...649..258T}
{Troja}, E., {Bocchino}, F., \& {Reale}, F. 2006, \apj, 649, 258

\bibitem[{{Tuohy} \& {Garmire}(1980)}]{1980ApJ...239L.107T}
{Tuohy}, I. \& {Garmire}, G. 1980, \apjl, 239, L107

\bibitem[{{Vasisht} \& {Gotthelf}(1997)}]{1997ApJ...486L.129V}
{Vasisht}, G. \& {Gotthelf}, E.~V. 1997, \apjl, 486, L129+

\bibitem[{{Vink}(2004)}]{2004NuPhS.132...21V}
{Vink}, J. 2004, Nuclear Physics B Proceedings Supplements, 132, 21

\bibitem[{{Wardle} \& {Yusef-Zadeh}(2002)}]{2002Sci...296.2350W}
{Wardle}, M. \& {Yusef-Zadeh}, F. 2002, Science, 296, 2350

\bibitem[{{Whiteoak} \& {Green}(1996)}]{1996A&AS..118..329W}
{Whiteoak}, J.~B.~Z. \& {Green}, A.~J. 1996, \aaps, 118, 329

\bibitem[{{Williams} {et~al.}(2006{\natexlab{a}}){Williams}, {Borkowski},
  {Reynolds}, {Blair}, {Ghavamian}, {Hendrick}, {Long}, {Points}, {Raymond},
  {Sankrit}, {Smith}, \& {Winkler}}]{2006ApJ...652L..33W}
{Williams}, B.~J., {Borkowski}, K.~J., {Reynolds}, S.~P., {Blair}, W.~P.,
  {Ghavamian}, P., {Hendrick}, S.~P., {Long}, K.~S., {Points}, S., {Raymond},
  J.~C., {Sankrit}, R., {Smith}, R.~C., \& {Winkler}, P.~F. 2006{\natexlab{a}},
  \apjl, 652, L33

\bibitem[{{Williams} {et~al.}(2006{\natexlab{b}}){Williams}, {Chu}, \&
  {Gruendl}}]{2006AJ....132.1877W}
{Williams}, R.~M., {Chu}, Y., \& {Gruendl}, R. 2006{\natexlab{b}}, \aj, 132,
  1877

\bibitem[{{Wilner} {et~al.}(1998){Wilner}, {Reynolds}, \&
  {Moffett}}]{1998AJ....115..247W}
{Wilner}, D.~J., {Reynolds}, S.~P., \& {Moffett}, D.~A. 1998, \aj, 115, 247

\bibitem[{{Wolszczan} {et~al.}(1991){Wolszczan}, {Cordes}, \&
  {Dewey}}]{1991ApJ...372L..99W}
{Wolszczan}, A., {Cordes}, J.~M., \& {Dewey}, R.~J. 1991, \apjl, 372, L99

\bibitem[{{Yamauchi} {et~al.}(2005){Yamauchi}, {Ueno}, {Koyama}, \&
  {Bamba}}]{2005PASJ...57..459Y}
{Yamauchi}, S., {Ueno}, M., {Koyama}, K., \& {Bamba}, A. 2005, \pasj, 57, 459

\end{thebibliography}
\end{document}